\documentclass[11pt]{article}
\pdfoutput=1 

\usepackage{jheppub}
\usepackage{graphicx}
\usepackage{amsmath}
\usepackage{amssymb}
\usepackage{amsfonts}
\usepackage{mathtools}
\usepackage{hyperref}
\usepackage{braket}
\usepackage[normalem]{ulem}
\usepackage{tikz}
\usepackage{comment}  
 
\hypersetup{colorlinks=true,urlcolor=[rgb]{0,0,0.5},citecolor=[rgb]{0.5,0,0},linkcolor=[rgb]{0,0,0.4}}

\def\Tr{{\rm tr}}
\def\tr{{\rm tr}}
\def\op{{\cal O}}
\def\C{\mathbb{C}}

\def\Z{\mathbb{Z}}

\def\vev#1{\langle{#1}\rangle}
\def\ketbra#1{ |{#1}\rangle\!\langle{#1}| }

\def\1den{\hbox{$1\hskip -1.2pt\vrule depth 0pt height 1.53ex width 0.7pt
                  \vrule depth 0pt height 0.3pt width 0.12em$}}

\def\where{\quad {\rm where} \quad}

\def\and{\quad {\rm and} \quad}

\def\nn{\nonumber\\}

\def\ra{\rightarrow}
\def\ie{{\rm i.e.\ }}
\def\eg{{\rm e.g.\ }}

\def\CA{{\cal A}}
\def\CB{{\cal B}}

\def\CE{{\cal E}}
\def\CF{{\cal F}}
\def\CH{{\cal H}}

\def\CR{{\cal R}}

\def\Wg{{\cal W}\! g}

\def\ktitle{\ensuremath{k}}

\title{Spectral decoupling in many-body quantum chaos}

\author[a]{Jordan Cotler}
\author[b]{and Nicholas Hunter-Jones}

\affiliation[a]{Stanford Institute for Theoretical Physics,\\ Stanford University, Stanford, California 94305}
\affiliation[b]{Perimeter Institute for Theoretical Physics,\\ Waterloo, Ontario N2L 2Y5}

\emailAdd{jcotler@stanford.edu}
\emailAdd{nickrhj@pitp.ca}

\abstract{We argue that in a large class of disordered quantum many-body systems, the late time dynamics of time-dependent correlation functions is captured by random matrix theory, specifically the energy eigenvalue statistics of the corresponding ensemble of disordered Hamiltonians.
We find that late time correlation functions approximately factorize into a time-dependent piece, which only depends on spectral statistics of the Hamiltonian ensemble, and a time-independent piece, which only depends on the data of the constituent operators of the correlation function.  We call this phenomenon ``spectral decoupling,'' which signifies a dynamical onset of random matrix theory in correlation functions.  A key diagnostic of spectral decoupling is $k$-invariance, which we refine and study in detail.  Particular emphasis is placed on the role of symmetries, and connections between $k$-invariance, scrambling, and OTOCs.  Disordered Pauli spin systems, as well as the SYK model and its variants, provide a rich source of disordered quantum many-body systems with varied symmetries, and we study $k$-invariance in these models with a combination of analytics and numerics.}

\begin{document} 
\maketitle
\flushbottom

\newpage

\section{Introduction}

New connections between many-body quantum chaos and random matrix theory (RMT) have emerged over the last several years \cite{BHRMT16,GarciaSYK16, GarciaSpec17,ChaosRMT,AltlandErgodicity18,gharibyan2018onset,SYKramp18, galitski2019otoc,ProsenQC,ProsenSFF,FRQC17,FRQC18}.  Whereas traditional treatments of single-body or few-body quantum chaos have emphasized characteristic energy level statistics of chaotic Hamiltonians \cite{BGSchaos,RMTphys,HaakeChaos}, recent research has forged connections between RMT and the dynamics of quantum many-body systems, as probed by correlation functions.  Particular emphasis has been placed on disordered systems, which naturally give rise to ensembles of Hamiltonians.  Such ensembles will be the setting for much of the analysis in this paper.

There is an important structural distinction between RMT energy level statistics and correlation functions of local quantum many-body systems.  Energy level statistics of an ensemble of local Hamiltonians are mostly agnostic to local physics, since energy eigenvalues are basis-independent data of a Hamiltonian.  By contrast, the behavior of time-dependent correlation functions crucially depends on whether the constituent operators are local.  Said simply, local correlation functions behave differently than non-local correlation functions, if the Hamiltonian dynamics is local.

Interestingly, the dynamics of time-dependent correlation functions of highly \textit{non}-local operators is \textit{not} sensitive to the specific choice of constituent operators \cite{ChaosRMT}.  Accordingly, the time-dependence of such correlation functions is completely described by basis-independent data of the Hamiltonian dynamics, i.e.\! energy eigenvalue correlation functions of the Hamiltonian ensemble which are the core objects of study in RMT.  For instance, if $A,B$ are highly non-local operators and $\CE_H$ is a Hamiltonian ensemble, we schematically have approximate equalities like
\begin{equation}
\label{eq:schematic1}
\int_{\CE_H} dH \, \text{tr}(e^{i H t} A e^{- i H t} B) \approx \text{[spectral quantity]}(t) \, \text{tr}(AB)\,,
\end{equation}
where the integral over $\CE_H$ performs a disorder average with respect to the Hamiltonian ensemble.  Note that the function $\text{[spectral quantity]}(t)$ will only depend on the joint eigenvalue statistics of the Hamiltonian ensemble $\CE_H$, and not on the highly non-local operators $A,B$.  In fact, it often suffices that only some of the operators in the correlation function be highly non-local.  Eqn.~\eqref{eq:schematic1} exemplifies ``spectral decoupling,'' namely that after a certain timescale, the dynamics of correlation functions (in a certain class of systems) only depend on spectral quantities (i.e., the joint eigenvalue statistics of the Hamiltonian ensemble), schematically denoted by $\text{[spectral quantity]}(t)$.  This dynamical data is decoupled from the data of the operators $A,B$, via an approximate factorization as per Eqn.~\eqref{eq:schematic1}.  We will elaborate precisely on many details in the sections below.  

Thus, we can predict the time dependence of disorder-averaged, highly non-local correlation functions simply by calculating energy eigenvalue correlations of the corresponding Hamiltonian ensemble.  Conversely, we can calculate energy eigenvalue correlations of a Hamiltonian ensemble by studying the dynamics of disorder-averaged, highly non-local correlation functions.

This bridge between non-local correlation functions and RMT energy eigenvalue statistics is attractive, but it is often more natural to consider correlation functions of local operators.  However, consider the following: take a local operator $\op$, and evolve it with a Hamiltonian in the Heisenberg picture out to a time $t$ as $\op(t) = e^{i H t} \, \op \, e^{- i H t}$.  After a sufficiently long time, $\op(t)$ will be a highly non-local operator.  Then it is possible that correlation functions containing $\op(t)$ will satisfy an equation like Eqn.~\eqref{eq:schematic1} after sufficiently long times.  In this way, we can imagine relating RMT energy eigenvalue statistics to the late time dynamics of correlation functions of initially local operators.

In this paper we make the above idea precise, and study how late time dynamics of correlation functions of disordered systems are captured by RMT energy eigenvalue statistics via spectral decoupling.  A quantitative diagnostic of the dynamical onset of spectral decoupling is $k$-invariance, introduced in \cite{ChaosRMT}, which is central to our analysis.  We find an intimate relationship between $k$-invariance, scrambling, and out-of-time-order correlators (OTOCs) \cite{LarkinOv69,Kitaev15,MSSbound}, thus drawing on recent diagnostics of many-body quantum chaos.   A subtle issue is the role of symmetry, which we study in great detail.  We generalize $k$-invariance to systems with symmetries, including time reversal symmetries, particle-hole symmetries, etc.  Disordered Pauli spin systems and the SYK model \cite{SachdevYe, Kitaev15, MS_SYK} will be our analytical and numerical testing ground for these ideas.  These models have quenched disorder, and come with a wide variety of symmetries.  We also analyze $k$-invariance in random circuits, leveraging recent technical results to perform explicit computations \cite{HaarRQC}.

An outline of the paper is as follows:
\begin{itemize}
\item In Section \ref{sec:corrRMT}, we review the essentials of random matrix theory, and discuss time-dependent correlation functions with respect to random matrix ensembles with different symmetries.
\item In Section \ref{sec:kinv}, we define $k$-invariance, explain its relation to the dynamical onset of chaos in time-dependent correlation functions, and make connections to scrambling and operator growth.  We also give a generalization of $k$-invariance to finite temperature, and explore $k$-invariance in random quantum circuits.
\item In Section \ref{sec:kinvsym}, we introduce and study a refinement of $k$-invariance that accounts for symmetries of a Hamiltonian ensemble.  As examples, we treat in explicit detail manifestations of time-reversal symmetry and describe a general construction.
\item In Section \ref{sec:kinvex}, we apply our techniques to various disordered Pauli spin models and several versions of the SYK model, numerically confirming our results, and explore the role of approximate symmetries.  Connections to quantum gravity and Jackiw-Teitelboim matrix models in particular are also explained.
\item In Section \ref{sec:Discussion}, we discuss our results and future directions.
\item The Appendices contain various formulas and derivations, including the first two moments of the Haar unitary, Haar orthogonal, and Haar symplectic ensembles.
\end{itemize}

\section{Correlation functions in RMT}
\label{sec:corrRMT}

Consider a Hilbert $\mathcal{H}$ space of dimension $d$.  We also consider a Hamiltonian ensemble $\CE_H$ which may be viewed as subset of the space of Hamiltonians on $\mathcal{H}$,  equipped with a probability measure $dH$ such that
\begin{equation}
\int_{\CE_H} dH = 1\,.
\end{equation}
Physically, this ensemble $\CE_H$ may be viewed as probabilistic instances of some disordered system.  Before discussing the behavior of correlation functions in disordered quantum many-body systems, we will first review the behavior of correlation functions in conventional random matrix ensembles.  The standard random matrix theory ensembles are the Gaussian Unitary Ensemble (GUE), Gaussian Orthogonal Ensemble (GOE), and Gaussian Symplectic Ensemble (GSE).  We will not require the precise measures of these ensembles for our analysis here, and so refer the reader to reviews in 
\cite{MehtaRMT,RMTphys} for details.  The GUE, GOE, and GSE each belong to different symmetry classes; for instance, Hamiltonians in the GOE or GSE class represent different realizations of time-reversal symmetry.  The GUE has no symmetries, meaning that there is no symmetry possessed by all Hamiltonians in the GUE.

We start by focusing on the Gaussian Unitary Ensemble (GUE), an ensemble of $d\times d$ random Hermitian matrices, defined as having a Gaussian probability density $P(H) \propto e^{-\frac{d}{2} \Tr(H^2)}$ and a unitarily invariant measure $dH = d(UHU^\dagger)$.  Therefore, if $U$ is any unitary on the Hilbert space $\mathcal{H}$, then for any function $f(H)$ we have
\begin{equation}
\label{eq:unitaryInvariance1}
\int_{\mathcal{E}_{\text{GUE}}} d(UHU^\dagger) \, f(H) = \int_{\mathcal{E}_{\text{GUE}}} dH \, f(H)\,.
\end{equation}
Equivalently, consider the probability density function of the GUE, $P(H)$.  Since a Hamiltonian can be specified by its eigenvectors $\{e_1,...,e_d\}$ and eigenvalues $\{\lambda_1,...,\lambda_d\}$, we may write the probability density as $P(\{e_1,...,e_d\}, \{\lambda_1,...,\lambda_d\})$.  Then Eqn.~\eqref{eq:unitaryInvariance1} is equivalent to
\begin{equation}
P(\{U\,e_1,...,U\,e_d\}, \{\lambda_1,...,\lambda_d\}) = P(\{e_1,...,e_d\}, \{\lambda_1,...,\lambda_d\})\,,
\end{equation}
which holds for any unitary $U$.  This means that in the GUE, for a fixed choice of eigenvalues, all eigenbases are equally likely.  As a result, $P$ factorizes as
\begin{equation}
\begin{split}
P(\{e_1,...,e_d\}, \{\lambda_1,...,\lambda_d\}) &= P_{\text{eigvec}}(e_1,...,e_d) \, P_{\text{eigval}}(\lambda_1,...,\lambda_d) \\
&= \frac{1}{\text{vol}(U(d))} \, P_{\text{eigval}}(\lambda_1,...,\lambda_d) \,,
\end{split}
\end{equation}
where in the second line we have used $P_{\text{eigvec}}(e_1,...,e_d) = 1/\text{vol}(U(d))$ since all eigenbases are equally likely.  We see that the probability density over the GUE depends only on eigenvalues.  This feature is also present in both the GOE and GSE.\footnote{The GOE and GSE do not have unitarily invariant measures.  Instead, their measures are invariant under the orthogonal or symplectic groups, subgroups of the unitary group which are compatible with time-reversal symmetry.}

Accordingly, random samples from the GUE are rather non-physical.  Suppose that $d = 2^N$ for some integer $N$.  If we fix a tensor factor decomposition of our Hilbert space into qubits as $\mathcal{H} = \bigotimes_{i=1}^N \mathbb{C}^2$, then a random sample of the GUE will have $N$-body interactions.  Indeed, these interactions are as non-local as possible.\footnote{For a more detailed analysis along these lines, see \cite{cotler2019locality}.}  The simple reason is that locality of interactions with respect to our subsystems of qubits is basis-dependent.  However, the GUE ensemble treats all bases equitably, and so random samples are agnostic to locality.  As we will see shortly, this will have interesting implications for time-dependent correlation functions with respect to evolution by the GUE, namely that the locality of constituent operators will not affect the dynamics in any way. 

\begin{figure}
\centering{\footnotesize
\begin{tikzpicture}[thick,scale=0.5,baseline=-0.4cm]
\node at (0,0) {\includegraphics[width=0.6\linewidth]{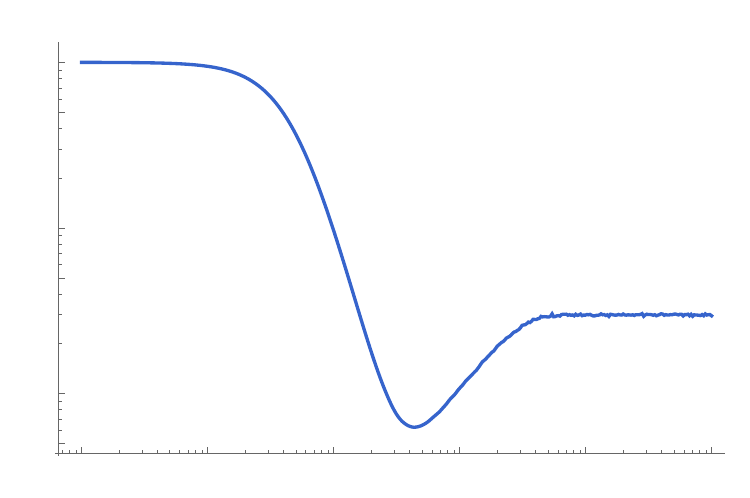}};
\node at (-7.8,5.6) {$\CR_2(t)$};
\node at (9,-5.1) {$t$};
\node at (-8.3,4.7) {$d^2$};
\node at (-8.3,-1.6) {$d$};
\node at (-8.5,-4.3) {$\sqrt{d}$};
\node at (-1.4,4.5) {exp decay};
\node at (0.8,2.8) {power-law decay};
\node at (0.2,2) {($\sim 1/t^3$)};
\node at (-0.3,-4.3) {dip};
\node at (4.4,-3.2) {linear ramp};
\node at (3.8,-4) {($\sim t$)};
\node at (6.6,-1.2) {plateau};
\node at (-3,-5.5) {$1$};
\node at (0.7,-5.5) {$\sqrt{d}$};
\node at (4.5,-5.5) {$2d$};
\end{tikzpicture}}
\caption{The spectral form factor $\CR_2(t)$ for the GUE, depicted on log-log scale.  Various features and their corresponding time scales are labelled.  The dip time is at $\sim \sqrt{d}$ followed by a linear ramp which ends at a time of $2d$ and turns into the plateau.}
\label{fig:R2func}
\end{figure}
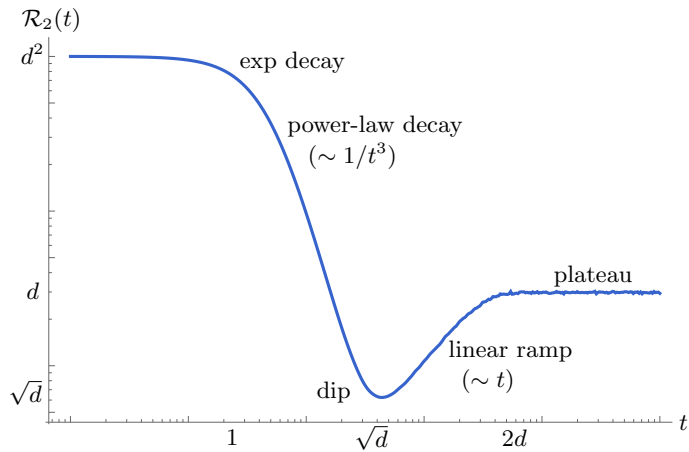

To analyze time-dependent correlation functions with respect to the GUE, we leverage that a key corollary of unitary invariance is Haar-unitary invariance.  That is, Eqn.~\eqref{eq:unitaryInvariance1} implies
\begin{equation}
\label{eq:HaarInvariance1}
\int_{U(d)} dU \, \int_{\CE_{\text{GUE}}} d(UHU^\dagger) \, f(H) = \int_{\CE_{\text{GUE}}} dH \, f(H)\,,
\end{equation}
where on the left-hand side, the outer integral is over the Haar measure on $U(d)$.  Now consider two operators $A$ and $B$ acting on $\mathcal{H}$, and their infinite-temperature time-dependent 2-point function
\begin{equation}
\langle A(t) B(0) \rangle_{\CE_{\text{GUE}}} := \int_{\CE_\text{GUE}} dH \, \frac{1}{d}\,\text{tr}\left(e^{i H t} A e^{- i H t} B \right)\,.
\end{equation}
By Eqn.~\eqref{eq:HaarInvariance1} this is equivalent to
\begin{equation}
\begin{split}
&\int_{U(d)} dU \, \int_{\CE_\text{GUE}} d(UHU^\dagger) \, \frac{1}{d}\,\tr\left(e^{i H t} A e^{- i H t} B \right)\\
& \qquad = \int_{U(d)} dU \, \int_{\CE_\text{GUE}} dH\, \frac{1}{d}\,\tr\left(e^{i (U^\dagger H U) t} A e^{- i (U^\dagger H U) t} B \right)\,,
\end{split}
\end{equation}
where in the first line we have used the change of variables $H \to U H U^\dagger$.  We can compute the Haar-unitary integral to obtain \cite{ChaosRMT}
\begin{align}
\label{eq:2ptfn1}
&\langle A(t) B(0) \rangle_{\CE_{\text{GUE}}}\\
&\qquad = \frac{1}{d}\,\tr(A) \cdot \frac{1}{d} \, \tr(B) + \frac{\int_{\CE_{\text{GUE}}} dH \, \left|\tr(e^{- i H t})\right|^2-1}{d^2 - 1} \left(\frac{1}{d}\,\tr(AB) - \frac{1}{d}\,\tr(A) \cdot \frac{1}{d} \, \tr(B)\right)\,. \nonumber
\end{align}
To clean up the answer somewhat, suppose that $A$ and $B$ are traceless, and define $\langle \, \cdot \, \rangle := \frac{1}{d} \, \tr(\, \cdot \,)$.  We further define the $2k$-spectral form factor with respect to a Hamiltonian ensemble $\CE_H$ by
\begin{equation}
\begin{split}
\label{eq:2kspectral1}
\mathcal{R}_{2k}^{\mathcal{E}_H}(t) &:= \int_{\CE_H} dH \, \left|\tr\left(e^{- i H t} \right) \right|^{2k} \\
&= \int_{\CE_H} dH \, \sum_{\substack{i_1,...,i_k = 1 \\ j_1,...,j_k = 1}}^d e^{- i\,(\lambda_{i_1} +\, \cdots\, + \lambda_{i_k} - \lambda_{j_1} - \,\cdots\, - \lambda_{j_k})\,t}
\end{split}
\end{equation}
which depends solely on eigenvalue statistics for any ensemble $\CE_H$.  When the context is clear, we will often write $\mathcal{R}_{2k}(t)$ without a superscript.  Spectral form factors have been studied extensively in the context of random matrix theory (for instance, see \cite{MehtaRMT,BrezinZee,BrezinHikami1}). The GUE spectral form factor, and its features and time scales, are depicted in Figure~\ref{fig:R2func}. For the GUE, we rewrite Eqn.~\eqref{eq:2ptfn1} as
\begin{align}
\label{eq:2ptfn2}
\langle A(t) B(0) \rangle_{\mathcal{E}_{\text{GUE}}}  = \frac{\mathcal{R}_2^{\mathcal{E}_{\text{GUE}}}(t) - 1}{d^2 - 1} \, \langle AB \rangle\,.
\end{align}
Notice that the dynamics of $\langle A(t) B(0) \rangle_{\mathcal{E}_{\text{GUE}}}$ decouple from the details of the operators $A$ and $B$.  In particular, the only time dependence is due to the spectral form factor $\mathcal{R}_2^{\mathcal{E}_{\text{GUE}}}(t) $ which is an RMT quantity that diagnoses eigenvalue statistics.  Indeed, Eqn.~\eqref{eq:2ptfn1} has precisely the form that was advertised in Eqn.~\eqref{eq:schematic1} in the introduction.

We can generalize Eqn.~\eqref{eq:2ptfn2} in many ways, including to higher point functions at multiple times, to finite temperature, and so on.  As an example, letting $A$ and $B$ be distinct Pauli operators, the infinite-temperature $2k$-OTOC (i.e., out-of-time-order correlator) is, to leading order in large $d$ \cite{ChaosRMT} given by the $2k$-spectral form factor as
\begin{equation}
\langle \underbrace{A(t) B(0) A(t) B(0) \cdots A(t) B(0)}_{k\text{ of }A(t) B(0)} \rangle_{\mathcal{E}_{\text{GUE}}} \simeq \frac{\mathcal{R}_{2k}^{\mathcal{E}_{\text{GUE}}}(t)}{d^{2k}}\,\langle \underbrace{A B A B \cdots A B}_{k\text{ of }A B} \rangle \,.
\end{equation}
The time-dependence of the lower-order terms suppressed above similarly factorize as
$$[\text{spectral form factor}](t) \times [\text{trace of operators}]\,.$$

As emphasized in \cite{ChaosRMT}, the only essential feature in the vast simplication of correlation functions seen above is the Haar-unitary invariance of the Hamiltonian ensemble $\mathcal{E}_{H}$ controlling time evolution.  The GUE possesses an exact Haar-unitary invariance, and accordingly its time-dependent correlation functions solely depend on basis-invariant data of consistuent operators, and the joint eigenvalue distribution of the GUE. Unfortunately, the GUE is not a physical ensemble, since generic samples have $N$-body interactions and so are completely non-local.  In fact, it appears that Haar-unitary invariance is at odds with the locality of interactions of a Hamiltonian ensemble.  Indeed, if we consider a Hamiltonian ensemble comprising solely of Hamiltonians with, say, $k$-body interactions for $k < N$, then this ensemble cannot possibly be exactly Haar-unitary invariant.  Thus it seems difficult to make contact with more physical ensembles Hamiltonians which possess few-body interactions.

To make progress, we need to change our perspective slightly.  Instead of consider ensembles of Hamiltonians, we instead consider ensembles of unitaries generated by Hamiltonian time evolution.  More precisely, consider an ensemble of Hamiltonians $\CE_H$ with measure $dH$.  Now we construct an associated $1$-parameter family of \textit{unitary} ensembles $\CE_t = \{e^{- i H t},\,H \in \CE_H\}$ (parameterized by $t$) with measure $dU$, such that for any function $f(U)$ we have
\begin{equation}
\label{eq:dUmeasure1}
\int_{\mathcal{E}_t} dU \, f(U) := \int_{\mathcal{E}_t} dH \, f(e^{- i H t})\,.
\end{equation}
Equivalently, the probability density of a unitary $e^{- i H t}$ in $\mathcal{E}_t$ is the same as the probability density of $H$ in $\mathcal{E}$.  At first, this appears to be merely a change of notation.  For instance, we can equivalently rewrite the $2k$-spectral form factor $\mathcal{R}_{2k}^{\mathcal{E}_H}(t)$ in Eqn.~\eqref{eq:2kspectral1} as
\begin{align}
\label{eq:2kspectral2}
\mathcal{R}_{2k}^{\mathcal{E}_t} := \int_{\mathcal{E}_t} dU \, \left|\tr\left(U\right) \right|^{2k} 
\end{align}
and equivalently rewrite $\langle A(t) B(0) \rangle_{\mathcal{E}_{\text{GUE}}}$ in Eqn.~\eqref{eq:2ptfn2} as
\begin{align}
\label{eq:2ptfn3}
\langle A(t) B(0) \rangle_{\mathcal{E}_t^{\text{GUE}}}  = \frac{\mathcal{R}_2^{\mathcal{E}_t^{\text{GUE}}} - 1}{d^2 - 1} \, \langle AB \rangle\,.
\end{align}

New notation aside, the unitary ensemble $\mathcal{E}_t$ can have more interesting properties than the Hamiltonian ensemble $\mathcal{E}_H$.  First, we note that if a Hamiltonian ensemble $\mathcal{E}_H$ is Haar-unitary invariant, then so is the corresponding unitary ensemble $\mathcal{E}_t$ for any $t$. Likewise we expect that if $\mathcal{E}_H$ is \textit{approximately} Haar-unitary invariant (with a notion of approximate we make precise) then $\mathcal{E}_t$ should also be approximately Haar-unitary invariant.  But, if $\mathcal{E}_H$ is \textit{not} approximately Haar-unitary invariant, then it is still possible that $\mathcal{E}_t$ can become Haar-unitary invariant for some range of $t$.

The crucial point is that the simplification of correlation functions explained above is only contingent on the (approximate) Haar-unitary invariance of $\mathcal{E}_t$.  We will find that ensembles $\mathcal{E}_H$ of \textit{local} Hamiltonians which themselves are \textit{not} (approximately) Haar-unitary invariant can still have their corresponding $\mathcal{E}_t$ become (approximately) Haar-invariant at sufficiently late times, leading to a dynamical simplification of late time correlation functions.  Spectral decoupling is due in part to the physics of scrambling.

To make the above ideas precise, we must: 
\begin{enumerate}
\item Provide a suitable and useful definition of approximate unitary invariance.
\item Analyze how symmetries of $\CE_H$ affect the dynamical onset of approximate unitary invariance of $\CE_t$.
\item Study which kinds of ensembles $\CE_H$ have the property that $\CE_t$ becomes approximately unitary invariant for sufficiently large $t$, and understand what controls the corresponding timescale of $t$.
\end{enumerate}
In the next section, we provide an answer to the first of these points, by reviewing and refining $k$-invariance \cite{ChaosRMT}.  The remaining points will be discussed throughout the remainder of the paper.

\section{\ktitle-invariance and many-body chaos}
\label{sec:kinv}

In this section we introduce and discuss $k$-invariance as a measure of spectral decoupling in disordered quantum many-body systems. The goal will be to understand and quantify the ergodic nature of chaotic Hamiltonian evolution. Consider the ensemble of unitary time evolutions at a fixed time $t$, generated by an ensemble of Hamiltonians $\CE_H$ as
\begin{equation}
\CE_t = \big\lbrace e^{-iHt}\,, ~ H\in \CE_H \big\rbrace
\label{eq:Htens}
\end{equation}
with measure $dU$, as defined by Eqn.~\eqref{eq:dUmeasure1}.  The ensemble of Hamiltonians $\mathcal{E}_H$ might be a spin system with quenched disorder, the SYK model, a random matrix ensemble, etc. One question that we can ask about the ensemble of unitaries $\CE_t$ is how rapidly and uniformly it spreads out over the unitary group $U(d)$.  A precise measure of the distance between $\mathcal{E}_t$ and $U(d)$ is provided by unitary $k$-designs.  In words, if a unitary ensemble forms a $k$-design, its first $k$ moments agree with the corresponding moments of the Haar unitary ensemble $U(d)$.  But as discussed in \cite{ChaosRMT}, $k$-designs fail to capture aspects of late time ergodicity even for random matrix ensembles (i.e., the GUE, GOE and GSE). An alternative to $k$-designs which encapsulates properties of random matrices is $k$-invariance. Next we will define $k$-designs and $k$-invariance and discuss their connection to correlation functions of a disordered theory.

\subsubsection*{$k$-designs}
We begin with unitary $k$-designs, which are subsets of $U(d)$ which reproduce moments of the full unitary group.  The Haar ensemble consists of the Haar measure on the unitary group $U(d)$, where the Haar measure is the unique left- and right-invariant measure on $U(d)$. We will want to consider moments of this ensemble, and averages of operators over the unitary group. Consider an operator $\op$ acting on the $k$-fold Hilbert space $\CH^{\otimes k}$. The $k$-fold channel with respect to the Haar ensemble is
\begin{equation}
\label{eq:Phikchannel1}
\Phi^{(k)}_{\mathcal{E}_{\text{Haar}}}(\op) = \int dU\, U^{\otimes k} \,\op\, U^\dagger{}^{\otimes k}\,.
\end{equation}
If we instead consider a weighted subset of the unitary group, \ie an ensemble of unitaries $\CE_U = \{U_i\}$ where each $U_i$ has probability $p_i$, the $k$-fold channel with respect to $\CE_U$ is 
\begin{equation}
\Phi^{(k)}_{\CE_U}(\op) = \sum_i p_i \, U_i^{\otimes k} \,\op\, U_i^\dagger{}^{\otimes k}\,.
\label{eq:kfold}
\end{equation}
Such an ensemble $\CE_U \subset U(d)$ might be discrete or continuous.  In the latter case, Eqn.~\eqref{eq:kfold} is upgraded to an integral with an appropriate probability density $P(U)$ as
\begin{equation}
\label{eq:Phikchannel2}
\Phi^{(k)}_{\CE_U}(\op) = \int_{\CE_U} \! dU\,P(U)\, U^{\otimes k} \,\op\, U^\dagger{}^{\otimes k}\,.   
\end{equation}
One might ask when averages over the ensemble $\CE_U$ look like averages over the full unitary group. An {\it exact unitary $k$-design} is an ensemble $\CE_U$ for which the $k$-fold channels are equal for all operators $\op$ acting on $\CH^{\otimes k}$
\begin{equation}
\label{eq:kdesigndef1}
\text{$k$-design}:\quad \Phi^{(k)}_{\CE_U}(\op) = \Phi^{(k)}_{\mathcal{E}_{\text{Haar}}}(\op) \,.
\end{equation}
For the first moment, Pauli operators (or any basis of operators of $\CB(\CH)$) form an exact 1-design. There are examples of exact 2 and 3-designs \cite{Dankert09,Zhu15,Kueng15,Webb15}, but for higher $k$, exact constructions of unitary $k$-designs are not known.

One might instead ask when an ensemble is merely \textit{close} to replicating moments of the Haar measure. An approximate $k$-design is an ensemble of unitaries $\CE_U$ for which the distance between $k$-fold channels in the diamond norm is small,
\begin{equation}
\text{approximate $k$-design}:\quad \big\|\Phi^{(k)}_{\CE_U} - \Phi^{(k)}_{\mathcal{E}_{\text{Haar}}}\big\|_\diamond \leq \epsilon\,,
\label{eq:approxk}
\end{equation}
where the diamond norm is defined in Appendix~\ref{app:designs}. We define approximate designs in terms of the diamond norm, but note that there are other definitions of an approximate design in the literature involving different norms.  However, different norms bound each other up to factors of $d$\,; see \cite{LowThesis} for details in the context of $k$-designs.

\subsubsection*{Frame potential}

A more tractable measure of approximate designs and the Haar randomness of an ensemble is the 2-norm on quantum channels, namely
\begin{equation}
\big\|\Phi^{(k)}_{\CE_U} - \Phi^{(k)}_{\mathcal{E}_{\text{Haar}}}\big\|_2^2 \,.
\label{eq:HSnorm1}
\end{equation}
Usually, we consider the 2-norm on the space of operators $\mathcal{B}(\mathcal{H})$ acting on a Hilbert space $\mathcal{H}$. However, in the present setting we are considering superoperators acting on $\mathcal{B}(\mathcal{H}^{\otimes k})$.  These superoperators also have a natural 2-norm (more precisely, the 2-norm of the moment operators), which we discuss in Appendix~\ref{app:designs}.

A convenient way of representing Eqn.~\eqref{eq:HSnorm1} is in terms of {\it frame potentials}. The $k$-th frame potential for an ensemble of unitaries $\CE_U$ is defined as \cite{Gross07,Scott08}
\begin{equation}
\CF^{(k)}_{\CE_U} = \int_{U,V\in \CE_U} dU dV\, \big| \Tr(U^\dagger V) \big|^{2k}\,,
\label{eq:FPdef}
\end{equation}
written here for a continuous ensemble equipped with a probability measure.  The key fact is that we have
\begin{equation}
\label{eq:HSnorm2}
\big\|\Phi^{(k)}_{\CE_U} - \Phi^{(k)}_{\mathcal{E}_{\text{Haar}}}\big\|_2^2  = \CF^{(k)}_{\CE_U} - \CF^{(k)}_{\mathcal{E}_{\text{Haar}}}\,.
\end{equation}
As a result, we see that the frame potential $\CF^{(k)}_{\CE_U}$ is lower bounded by the corresponding frame potential computed for the Haar measure \cite{Scott08}
\begin{equation}
\CF^{(k)}_{\CE_U} - \CF^{(k)}_{\mathcal{E}_{\text{Haar}}}  \geq 0\,,
\end{equation}
where the Haar value is $\CF^{(k)}_{\mathcal{E}_{\text{Haar}}} = k!$ for $k \leq d$. For higher moments $k>d$, the Haar value is given by a known combinatorial expression. Closeness in the 2-norm, as seen by the frame potential, is a weaker notion of approximate $k$-design as defined in Eqn.~\eqref{eq:approxk}. Nevertheless, the difference in frame potentials bounds the difference in $k$-fold channels up to factors of the dimension of the Hilbert space, as shown in Appendix~\ref{app:designs}. 

The frame potential has more recently been understood as a diagnostic of chaotic dynamics \cite{ChaosDesign,ChaosRMT,ChaosSUSY,CVFP2019,Chenu2019}.
Specifically, the $k$-th frame potential was related to operator-averaged OTOCs, making precise the connection between the chaotic decay of correlation functions and the approach to Haar randomness. We define the $2k$-OTOCs averaged over an ensemble $\CE_t$ in Eqn.~\eqref{eq:Htens} by
\begin{equation}
\vev{A_1B_1(t)\ldots A_k B_k(t)}_{\CE_t} := \int_{\CE_H} \!\!dH\, \frac{1}{d} \,\Tr(A_1 B_1(t) \ldots A_k B_k(t))\,,
\end{equation}
where $dH$ is the measure over $\CE_H$, $B(t) = e^{iHt}B e^{-iHt}$, and we take the expectation in the infinite temperature state. Note that these OTOCs can be naturally generalized to an ensemble of unitaries $\mathcal{E}_U$ instead of just $\mathcal{E}_t$.  The $k$-th frame potential is then related to operator-averaged $2k$-OTOCs as \cite{ChaosDesign}
\begin{equation}
\CF^{(k)}_{\CE_t} = \frac{1}{d^{2k-2}} \sum_{\{A_i\}, \{B_i\}} \big|\vev{A_1B_1(t)\ldots A_k B_k(t)}_{\CE_t} \big|^2\,,
\label{eq:opavgOTOCs}
\end{equation}
where we take the sum of each operator over an orthonormal basis of operators $\{\mathcal{O}_i\}_{i=1}^{d^2}$, i.e. satisfying $\frac{1}{d}\,\tr(\mathcal{O}_i^\dagger \mathcal{O}_j) = \delta_{ij}$.  Since the frame potential is lower bounded by its Haar value $\CF^{(k)}_{\mathcal{E}_{\text{Haar}}} = k!$, the chaotic decay of OTOCs signifies the approach to randomness.

\subsubsection*{\ktitle-designs are not enough}

In some models of stochastic, time-dependent evolution up to time $t$, such as random quantum circuits or Brownian circuits, it is known that for late enough times $t$ the ensembles converge to approximate unitary $k$-designs \cite{HL08,BHH12,HaarRQC,Nakata16,Onorati17}.
But for time-independent evolution by a disordered Hamiltonian this is not the case.  Considering the ensemble $\CE_t = \{e^{- i H t} \, , \, H \in \CE_H\}$ of evolution by disordered Hamiltonians up to time $t$, there is a simple argument for why $\mathcal{E}_t$ will never become Haar random at very late times \cite{ChaosDesign}.  The eigenvalues of a sample from the Haar unitary ensemble have the form $e^{- i \theta_j}$ for $j = 1,...,d$, which experience level repulsion on the complex circle.  However, sampling a Hamiltonian $H$ from $\CE_H$ with eigenvalues $\lambda_j$ for $j=1,...,d$, the eigenvalues of $e^{- i H t}$ are $e^{- i \lambda_j t}$, which become Poisson distributed on the complex circle as $t$ approaches infinity.  Thus $e^{- i H t}$ for large $t$ cannot look like a Haar random unitary.
 
 However, at intermediate time scales, $e^{- i H t}$ can look approximately Haar random.  Since the eigenvalues $\lambda_j$ of random Hamiltonians experience level repulsion, $e^{- i \lambda_j t}$ still may experience level repulsion on the complex circle so long as $t$ is not too large.  Considering Hamiltonian evolution by a standard random matrix ensemble such as the GUE, we can analytically compute the frame potentials in terms of the spectral functions \cite{ChaosRMT}. Hamiltonians drawn from the GUE become close to Haar random in 2-norm at an intermediate time scale called the ``dip time'' \cite{ChaosRMT}, see Figure~\ref{fig:R2func}.  We believe that this phenomenon of becoming close to Haar random at an intermediate time holds more generally for chaotic quantum many-body systems, once the symmetries of the system are taken into account. Nevertheless, as evolution by random matrix Hamiltonians is not Haar random at late times, we require a new quantity to measure ergodicity and the onset of random matrix theory.

\subsubsection*{\ktitle-invariance}

We now introduce $k$-invariance as a measure of the onset of random matrix behavior, and specifically spectral decoupling.
We are interested in the ensemble of unitary time evolutions generated by disordered Hamiltonians, $\CE_t = \{e^{-iHt}, ~ H\in \CE_H\}$. For systems that break all symmetries, we expect that the ensemble will achieve (approximate) unitary invariance.  This is motivated by a defining property of the GUE, namely that the measure on the space of Hermitian matrices is invariant under unitary conjugation, as discussed above.  Given the ensemble $\CE_t$, we define an ensemble which is unitarily invariant: $\widetilde \CE_t = \{e^{i(U H U^\dagger) t} \,, \, H \in \CE_H\text{ and }U \in U(d)\}$ where the $U$'s are Haar distributed and the $H$'s are still distributed according to $dH$.  We will often refer to $\widetilde{\mathcal{E}}_t$ as the ``Haar'ed'' version of $\mathcal{E}_t$.
The distance between the ensembles $\CE_t$ and $\widetilde \CE_t$ captures how basis-invariant the dynamics of $\CE_t$ are at the time $t$.  We say that an ensemble $\CE_t$ is $k$-invariant if the $k$-fold channels of the two ensembles are equal
\begin{equation}
\label{eq:kinvdef1}
\text{$k$-invariant}:\quad \Phi^{(k)}_{\CE_t}(\op) = \Phi^{(k)}_{\widetilde \CE_t}(\op) \,.
\end{equation}
Given the time evolution of a chaotic quantum many-body system, we expect that the ensemble of unitary time-evolutions become approximately $k$-invariant at late times as information is scrambled and the dynamics become invariant under an arbitrary change of basis. Thus, we want to quantify an how close the two ensembles $\mathcal{E}_t$ and $\widetilde{\mathcal{E}}_t$ are at a given time $t$. 
We might consider a strong notion of $k$-invariance in terms of the diamond distance between the ensembles
\begin{equation}
\text{approximate $k$-invariance}:\quad \big\|\Phi^{(k)}_{\CE_t} - \Phi^{(k)}_{\widetilde \CE_t}\big\|_\diamond \leq \epsilon\,,
\label{eq:approxkinv}
\end{equation}
which captures the distinguishability of the ensembles. In the interest of working with more tractable quantities, we will instead focus on the difference in frame potentials to understand $k$-invariance
\begin{equation}
\big\|\Phi^{(k)}_{\CE_t} - \Phi^{(k)}_{\widetilde \CE_t}\big\|_2^2 = \CF^{(k)}_{\CE_t} - \CF^{(k)}_{\widetilde \CE_t}\geq 0\,.
\label{eq:kinv}
\end{equation}
Again, the two frame potentials quantify a 2-norm distance between the ensembles and the difference in frame potentials bounds the diamond distance up to factors of the dimension; see Appendix~\ref{app:designs}.  We will refer to $\CF^{(k)}_{\widetilde \CE_t}$ as a Haar'ed frame potential.  As we see, $k$-invariance as measured by the frame potentials defines a distance to unitary-invariance and will signify the onset of random matrix behavior in late time quantum many-body dynamics.  In particular, an ensemble $\CE_t$ can become (approximately) $k$-invariant at a timescale $t_{k\text{-inv}}$ such that $t_{1\text{-inv}} \leq t_{2\text{-inv}} \leq \cdots$, thus introducing a new hierarchy of timescales into quantum many-body chaos.  Intuitively, $t_{k\text{-inv}}$ is the timescale when $2k$-OTOCs with adjacent operators inserted $t_{k\text{-inv}}$ apart undergo spectral decoupling. More explicitly, a $2k$-OTOC
\begin{equation}
\langle A_1(t) \,A_2(0) \,A_3(t) \cdots\, A_{2k}(0) \rangle_{\CE_t}
\end{equation}
with $t \geq t_{k\text{-inv}}$ will be approximately spectrally decoupled.  Similarly, a $2k$-point Keldysh-ordered correlation function of the form
\begin{equation}
\langle A_k(kt)\,\cdots\,A_2(2t)\,A_1(t) \, \rho_0 \, A_{k+1}(t)^\dagger \, A_{k+2}(2t)^\dagger \, \cdots A_{2k}(k t)^\dagger \rangle_{\CE_t}
\end{equation}
will be spectrally decoupled for $t \geq t_{k\text{-inv}}$.  Here we have assumed that $\rho_0$ does not depend on the ensemble and so $\rho_0$ is not, for instance, a finite-temperature Gibbs state.  It is possible to generalize the above discussion to finite-temperature OTOCs and finite-temperature Keldysh-ordered correlation functions.  This utilizes a finite-temperature version of $k$-invariance, which we discuss later in this section.

The definition of $k$-invariance in Eqn.~\eqref{eq:kinvdef1} is weaker than the definition of $k$-designs in Eqn.~\eqref{eq:kdesigndef1}.  In fact, if a unitary ensemble forms a $k$-design, it is necessarily $k$-invariant.  This is readily seen since if $\Phi_{\CE_U}^{(k)} = \Phi_{\mathcal{E}_{\text{Haar}}}^{(k)}$, then $\Phi_{\CE_U}^{(k)}$ must be $k$-invariant since $\Phi_{\mathcal{E}_{\text{Haar}}}^{(k)}$ is $k$-invariant. Lastly, if an ensemble is $k$-invariant it is also $(k-1)$-invariant.

\subsubsection*{1-invariant frame potentials}
Given some ensemble of Hamiltonians time-evolutions $\CE_t$, the question of whether the ensemble is $k$-invariant amounts to comparing the frame potential of $\CE_t$ to that of the invariant ensemble $\widetilde \CE_t$. By construction, the invariant frame potential can be computed in terms of spectral functions of the Hamiltonian ensemble. 

Here we quickly review the calculation of the 1-invariant frame potential \cite{ChaosRMT}, which will lower bound $\CF^{(1)}_{\CE_t}$ with equality if and only if the ensemble is 1-invariant. We have
\begin{equation}
\CF^{(1)}_{\widetilde \CE_t} = \int_{U(d)} dU \int_{\CE_H} dH_1 dH_2 \,\tr(U \,e^{i H_1 t}\, U^\dagger e^{-i H_2 t}) \tr(U e^{-i H_1 t} U^\dagger e^{i H_2 t})\,.
\end{equation}
Integrating using the second Haar moment (Eqn.~\eqref{eq:U2ndmom}), we compute the 1-invariant frame potential in terms of the 2-point form factor of $\CE_H$ as
\begin{equation}
\CF^{(1)}_{\widetilde \CE_t} = \frac{\CR_2(t)^2 + d^2 - 2\CR_2(t)^2}{d^2-1}\,.
\label{eq:FP1inv}
\end{equation}
The early time decay of the function is $\approx \CR_2(t)^2/d^2$, and for a chaotic system we approach a late time value of $\CF^{(1)}_{\widetilde \CE_t} \approx 2$ when $t\sim d$. We derive the reproduce the 2-invariant frame potential in terms of spectral functions in Appendix~\ref{app:OTOCs}.

\subsection{Relation to correlation functions}
\label{sec:kinvcorr}

We have given a general definition of $k$-invariance in Eqn.~\eqref{eq:kinv}, and in order to make the definition less abstract we now relate it to correlation functions of observables. Recall that we can exactly relate the frame potential of an ensemble to operator-averaged $2k$-OTOCs as in Eqn.~\eqref{eq:opavgOTOCs}. From the definition of $k$-invariance, we have
\begin{equation}
\label{eq:2kOTOCkinv1}
\CF^{(k)}_{\CE_t} - \CF^{(k)}_{\widetilde \CE_t} = \frac{1}{d^{2k-2}} \sum_{\{A_i\},\{B_i\}} \bigg( \big|\vev{A_1 B_1(t)\ldots A_k B_k(t)}_{\CE_t}\big|^2 -  \big|\vev{A_1 B_1(t)\ldots A_k B_k(t)}_{\widetilde\CE_t}\big|^2\bigg)\,,
\end{equation}
where we sum each $A_i$ and $B_i$ over a complete orthonormal basis of operators $\{\mathcal{O}_i\}_{i=1}^{d^2}$, i.e. satisfying $\frac{1}{d}\,\tr(\mathcal{O}_i^\dagger \mathcal{O}_j) = \delta_{ij}$.  We can also write (see Appendix~\ref{app:OTOCs})
\begin{equation}
\label{eq:2kOTOCkinv2}
\CF^{(k)}_{\CE_t} - \CF^{(k)}_{\widetilde \CE_t} = \frac{1}{d^{2k-2}} \sum_{\{A_i\},\{B_i\}} \bigg| \vev{A_1 B_1(t)\ldots A_k B_k(t)}_{\CE_t} -  \vev{A_1 B_1(t)\ldots A_k B_k(t)}_{\widetilde\CE_t}\bigg|^2\,.
\end{equation}
As such, approximate $k$-invariance can be interpreted in terms of the distance between correlation functions of the two ensembles $\mathcal{E}_t$ and $\widetilde{\mathcal{E}}_t$.  We will now explore Eqn.~\eqref{eq:2kOTOCkinv1} in detail in the case of 2-point functions and the onset of 1-invariance.

\subsubsection*{1-invariance and variance of correlators}
The condition for 1-invariance may be written as the following difference of 2-point functions
\begin{equation}
\CF^{(1)}_{\CE_t} - \CF^{(1)}_{\widetilde \CE_t} = \sum_{A,B \in P} \big| \vev{A(t) B}_{\CE_t} \big|^2 - \sum_{A,B \in P} \big| \vev{A(t) B}_{\widetilde \CE_t} \big|^2\,,
\end{equation}
where $P$ denotes the basis of generalized Pauli operators.  Also, as in the previous section, $\vev{\,\cdot\,}_{\CE_t}$ is the average of the correlator over the ensemble
\begin{equation}
\vev{A(t) B}_{\CE_t} = \int_{\CE_H} dH\, \frac{1}{d} \,\Tr(A(t) B) \where A(t) = e^{iHt}Ae^{-iHt}\,.
\end{equation}
The brackets $\vev{\,\cdot\,}_{\widetilde{\CE}_t}$ are defined similarly, but with respect to $\widetilde{\mathcal{E}}_t$.  We can compute the averaged correlation functions $\vev{A(t) B}_{\widetilde \CE_t}$ over the unitarily invariant ensemble explicitly. For Pauli operators $A$ and $B$, $\vev{A(t) B}_{\widetilde \CE_t}$ is only nonzero for $A = B$. For non-identity Paulis, $\vev{A(t) A}_{\widetilde \CE_t} = (\CR_2(t)-1)/(d^2-1)$, and thus
\begin{equation}
\CF^{(1)}_{\CE_t} - \CF^{(1)}_{\widetilde \CE_t} = \sum_{A,B \in P'} \big| \vev{A(t) B}_{\mathcal{E}_t} \big|^2 - \sum_{A \in P'} \bigg( \frac{\CR_2(t)-1}{d^2-1}\bigg)^2\,,
\end{equation}
where have canceled the identity contributions from $\CF^{(1)}_{\CE_t}$ and $\CF^{(1)}_{\widetilde{\CE}_t}$ and summed over non-identity Paulis $P'$.  Note that the second term on the right-hand size is independent of $A$, and so the sum over $A$ will just produce a multiplicative prefactor.  As a sanity check, we note that
\begin{equation}
\CF^{(1)}_{\widetilde \CE_t} = \sum_{A,B \in P} \big| \vev{A(t) B}_{\widetilde \CE_t} \big|^2 = \sum_{A \in P'} \bigg( \frac{\CR_2(t)-1}{d^2-1}\bigg)^2 + 1 = \frac{\CR_2(t)^2 + d^2 - 2\CR_2(t)}{d^2-1}
\end{equation}
which is our previously derived expression for the unitarily invariant frame potential in terms of spectral form factors.

Using the above expressions, we can understand the decay of the difference in frame potentials, and thus the onset of approximate 1-invariance, in terms of 2-point functions.  First, note that the 2-point spectral form factor can be written as an average over non-identity Paulis \cite{ChaosRMT}
\begin{equation}
\CR_2(t) = \sum_{A\in P'} \vev{A(t) A}_{\mathcal{E}_t} + 1\,.
\end{equation}
We define notation for the average over non-identity Pauli operators as
\begin{equation}
\big\langle \cdot \big\rangle_A := \frac{1}{d^2-1} \sum_{A\in P'} (\,\cdot\,)\,.
\end{equation}
If we assume that the 2-point functions with $A\neq B$ are approximately zero at all times $\vev{A(t) B}_{\mathcal{E}_t} \approx 0$, then the $\CF^{(1)}_{\CE_t} - \CF^{(1)}_{\widetilde \CE_t}$ can be written as the variance of 2-point functions
\begin{equation}
\frac{\CF^{(1)}_{\CE_t} - \CF^{(1)}_{\widetilde \CE_t}}{d^2-1} \approx \Big\langle\big| \vev{A(t) A}_{\CE_t} \big|^2 \Big\rangle_A - \Big|\big\langle \vev{A(t) A}_{\CE_t} \big\rangle_A \Big|^2 = {\rm Var}\big(\vev{A(t) A}_{\mathcal{E}_t}\big)\,.
\end{equation}
Therefore, if the difference in frame potentials becomes small and $\vev{A(t) B}_{\mathcal{E}_t} \approx 0$, then the variance of 2-point functions ${\rm Var}\big(\vev{A(t) A}_{\mathcal{E}_t}\big)$ is small. 

\subsection{Chaos from approximate 2-invariance}

Here, we explain how the onset of $2$-invariance gives rise to scrambling and late time operator growth.  We will characterize scrambling from the decay of the 4-point OTOCs as well as quantum mutual information of subsystems of the time evolution operator.  Additionally, we show that $2$-invariance implies universal spectral behavior of late time operator dynamics. We note that thermalization by random Hamiltonian evolution was studied in \cite{Cramer12,BrandaoThermal12,Vinayak12,Masanes13,NakataThermal}, and these results hold for 2-invariant ensembles.

\subsubsection{OTOC decay with 2-invariant Hamiltonians}

To explore why 2-invariance gives rise to scrambling, we first look at OTOCs as a probe of chaotic dynamics. At infinite temperature, we consider the OTOCs $\vev{A(t)BC(t)D}_{\mathcal{E}_t}$. We want to show that if the ensemble $\CE_t$ is approximately 2-invariant, then OTOCs necessarily decay. We compute the full OTOC for the invariant ensemble $\widetilde\CE_t$ in Appendix~\ref{app:OTOCs}, and for traceless operators we find to leading order in $1/d$
\begin{align}
&\vev{A(t)BC(t)D}_{\widetilde \CE_t}\\
&= \frac{\CR_4(t)}{d^4} \vev{ABCD} + \left(\frac{\CR_{4,1}(t)}{d^3}-\frac{\CR_4(t)}{d^4}\right)\vev{AB}\vev{CD} + \left(\frac{\CR^*_{4,1}(t)}{d^3}-\frac{\CR_4(t)}{d^4}\right) \vev{AD}\vev{BC} + \op\Big(\frac{1}{d^2}\Big)\nonumber \,.
\end{align}
The spectral functions $\mathcal{R}_{4}(t)$ and $\mathcal{R}_{4,1}(t)$ are defined by
\begin{align}
\CR_4(t) &:= \big\langle \Tr(e^{-iHt})^2\Tr(e^{iHt})^2\big\rangle_{\CE_H} = \int D\lambda \sum_{i,j,k,\ell} e^{i(\lambda_i+\lambda_j -\lambda_k-\lambda_\ell) t} \\
\CR_{4,1}(t) &:= \big\langle \Tr(e^{iHt})^2\Tr(e^{-2iHt})\big\rangle_{\CE_H} = \int D\lambda \sum_{i,j,\ell} e^{i(\lambda_i+\lambda_j -2\lambda_\ell)t}\,,
\end{align}
where we note that $\CR_{4,1}(t)$ is generically complex.

For non-identity Pauli operators with $A$ and $C$ not equal to $B$ or $D$, the decay of the 2-invariant OTOC is entirely governed by the spectral 4-point form factor $\CR_4(t)$, as shown in \cite{ChaosRMT}. Moreover, for more conventional OTOCs of the form $\vev{A(t)BA(t)B}_{\mathcal{E}_t}$, again assuming non-identity Pauli operators, we find
\begin{equation}
\vev{A(t)BA(t)B}_{\widetilde \CE_t} \approx \frac{\CR_4(t)}{d^4} \,\langle A B A B \rangle\,.
\end{equation}
Simply assuming a non-degenerate spectrum, the long-time average of the 4-point form factor is $\CR_4(t) \approx 2d^2$.  Then for 2-invariant ensembles, these OTOCs decay to a late time value of $1/d^2$. Moreover, in many examples we can explicitly compute the early time decay of the spectral functions. As such, 2-invariance of a disordered system implies the decay of OTOCs.

\subsubsection{Information scrambling with 2-invariant Hamiltonians}

Now we turn to study the scrambling of quantum information under evolution for 2-invariant Hamiltonians.  In the following, we assume that the ensemble $\CE_t$ has reached a time where it is approximately $2$-invariant.

The setup we consider will be to study the scrambling by looking at the mutual information of the unitary time-evolution operator \cite{ChaosChannels,ChaosDesign,HaydenPreskill}. More precisely, we bipartition the input and output of $e^{-iHt}$ into $\bar A \bar B$ and $CD$ respectively, and take $\bar A$ and $\bar B$ to be entangled with reference systems $A$ and $B$, respectively. We then look at the 4-partite state on regions $ABCD$. By studying the mutual information between various subregions of the state, we probe the delocalization and scrambling of quantum information under $e^{-iHt}$. This setup is similar to studying the entanglement of the Choi state of the operator \cite{ChaosChannels}, and has an operational interpretation in terms of the Hayden-Preskill decoding protocol \cite{HaydenPreskill,YoshidaKitaev17}.

Consider an initial state $\rho = \ketbra{\psi}$, where $\ket\psi_{A\bar A \bar B B}$ is a pure state on a 4-partite system. The state can be depicted by \vspace*{-10pt}
\begin{center}
\begin{tikzpicture}[line width=1,scale=0.7]
\node at (0,0) {$\rho$};
\draw[rounded corners] (-0.4,-1) rectangle (0.4,1);
\foreach \x in {-0.9,0.4}
{\draw (\x,0.25) -- (\x+0.5,0.25);
\draw (\x,-0.25) -- (\x+0.5,-0.25);
\draw (\x,0.8) -- (\x+0.5,0.8);
\draw (\x,-0.8) -- (\x+0.5,-0.8);}
\node[anchor=east] at (-0.9,0.9) {{\small $A$}};
\node[anchor=east] at (-0.9,0.3) {{\small $\bar A$}};
\node[anchor=east] at (-0.9,-0.3) {{\small $\bar B$}};
\node[anchor=east] at (-0.9,-0.9) {{\small $B$}};
\end{tikzpicture}
\end{center}
We evolve the $\bar A\bar B$ subsystems as $\rho_{\bar A \bar B}(t) = e^{-iHt}\,\rho_{\bar A \bar B}\, e^{iHt}$, and partition the output of the time-evolution into $CD$, such that the evolved state $\rho(t)$ is
\begin{center}
\begin{tikzpicture}[line width=1,scale=0.75]
\node at (-2,0.06) {$e^{-iHt}$};
\node at (0,0) {$\rho$};
\node at (2,0.06) {$e^{iHt}$};
\draw[rounded corners] (-0.4,-1) rectangle (0.4,1);
\draw[rounded corners] (-3,-0.5) rectangle (-0.9,0.5);
\draw[rounded corners] (0.9,-0.5) rectangle (3,0.5);
\foreach \x in {-0.9,0.4,3,-3.5}
{\draw (\x,0.2) -- (\x+0.5,0.2);
\draw (\x,-0.2) -- (\x+0.5,-0.2);}
\foreach \x in {-3.5,0.4}
{\draw (\x,0.8) -- (\x+3.1,0.8);
\draw (\x,-0.8) -- (\x+3.1,-0.8);}
\node[anchor=east] at (-3.4,0.9) {{\small $A$}};
\node[anchor=east] at (-3.4,0.3) {{\small $C$}};
\node[anchor=east] at (-3.4,-0.3) {{\small $D$}};
\node[anchor=east] at (-3.4,-0.9) {{\small $B$}};
\end{tikzpicture}
\end{center}
To be clear, the unitary $e^{- i H t}$ is expressed as a map $\mathcal{H}_{\bar{A}}\otimes\mathcal{H}_{\bar{B}} \to \mathcal{H}_{C} \otimes \mathcal{H}_D$
\begin{center}
\begin{tikzpicture}[line width=1,scale=0.70]
\node at (2,0.06) {$e^{-iHt}$};
\draw[rounded corners] (0.9,-0.5) rectangle (3,0.5);
\foreach \x in {0.4,3}
{\draw (\x,0.2) -- (\x+0.5,0.2);
\draw (\x,-0.2) -- (\x+0.5,-0.2);}
\node[anchor=east] at (0.4,0.3) {{\small $C$}};
\node[anchor=west] at (3.4,0.3) {{\small $\bar A$}};
\node[anchor=west] at (3.4,-0.3) {{\small $\bar B$}};
\node[anchor=east] at (0.4,-0.3) {{\small $D$}};
\end{tikzpicture}
\end{center}
where $\mathcal{H}_{\bar{A}}\otimes\mathcal{H}_{\bar{B}}$ and $\mathcal{H}_{C} \otimes \mathcal{H}_D$ are different bipartitions of the same Hilbert space (i.e., $\mathcal{H}_{\bar{A}}\otimes\mathcal{H}_{\bar{B}} \simeq \mathcal{H}_{C} \otimes \mathcal{H}_D$).

To determine whether information has delocalized and scrambled over the entire system, we want to compute the mutual informations $I(A,BD)$ and $I(A,C)$ in the state $\rho(t)$. If $I(A,C)$ is small for any small input region $A$, then information has delocalized and we cannot learn about the state on $A$ by performing measurements on $C$. When $I(A,BD)$ becomes large, the information in $A$ has scrambled and we can reconstruct the state by acting on the $BD$ systems alone.

To make the computation more tractable, we instead consider the R\'{e}nyi-2 mutual information
\begin{equation}
I^{(2)}(A,BD) = S^{(2)}_A+S^{(2)}_{BD}-S^{(2)}_{ABD}\,,
\end{equation}
where $S^{(2)}_A = -\log \Tr (\rho_A^2)$ and the other terms are defined similarly.  We compute the R\'{e}nyi-2 entropies by calculating the purities of subsystems of $\rho(t)$ averaged over 2-invariant ensembles. For instance, considering $\Tr (\rho_{BD}(t)^2)$, we use the invariance of the Hamiltonians to Haar-conjugate the expression as
\begin{center}
\begin{tikzpicture}[line width=1,scale=0.55]
\node at (-3.5,0) {$\Tr(\rho_{BD} (t)^2) = $};
\node at (0,0) {$U$};
\node at (1.1,0) {$\Lambda$};
\node at (2.4,0.05) {$U^\dagger$};
\node at (3.7,-0.08) {$\rho$};
\node at (5,0) {$U$};
\node at (6.2,0.05) {$\Lambda^\dagger$};
\node at (7.6,0.05) {$U^\dagger$};
\node at (9.9,0) {$U$};
\node at (11,0) {$\Lambda$};
\node at (12.3,0.05) {$U^\dagger$};
\node at (13.6,-0.08) {$\rho$};
\node at (14.9,0) {$U$};
\node at (16.1,0.05) {$\Lambda^\dagger$};
\node at (17.5,0.05) {$U^\dagger$};
\foreach \x in {2.9,4.1,12.8,14}
{\draw (\x,0.1) -- (\x+0.4,0.1);
\draw (\x,-0.1) -- (\x+0.4,-0.1);}
\foreach \x in {0.35,1.4,5.3,6.6,10.2,11.3,15.2,16.5}
{\draw (\x,0) -- (\x+0.4,0);}
\draw[rounded corners] (-0.4,-0.1) -- (-0.8,-0.1) -- (-0.8,-0.6) -- (8.6,-0.6) -- (8.6,-0.1) -- (8.2,-0.1);
\draw[rounded corners] (9.4,-0.1) -- (9,-0.1) -- (9,-0.6) -- (18.4,-0.6) -- (18.4,-0.1) -- (18,-0.1);
\draw (8.2, 0.1) -- (9.4,0.1);
\draw[rounded corners] (-0.4,0.1) -- (-0.8,0.1) -- (-0.8,0.6) --  (18.4,0.6) -- (18.4,0.1) -- (18,0.1);
\end{tikzpicture}\,,
\end{center}
where $\Lambda$ is the unitary time evolution operator $e^{-iHt}$ written in the diagonal basis and $U$ is a Haar random unitary. We then average over $U$ using the 4-th moment of Haar random unitaries.

Computing the quantity, we find that in terms of the purities of the initial state, we have to leading order in $1/d$
\begin{equation}
\big\langle \Tr(\rho_{BD} (t)^2)\big\rangle_{\CE_t} = \frac{\CR_4}{d^4} \Tr(\rho_{BD}^2) + \frac{1}{d_D} \bigg(1-\frac{\CR_4}{d^4}\bigg) \Tr(\rho_{A}^2) + \frac{1}{d_C} \bigg(1-\frac{\CR_4}{d^4}\bigg) \Tr(\rho_{B}^2) + \cdots
\end{equation}
and similarly mutual information
\begin{equation}
\big\langle\Tr(\rho_{ABD} (t)^2)\big\rangle_{\CE_t} = \frac{\CR_4}{d^4} \Tr(\rho_{ABD}^2) + \frac{1}{d_D} \bigg(1-\frac{\CR_4}{d^4}\bigg) + \frac{1}{d_C} \bigg(1-\frac{\CR_4}{d^4}\bigg) \Tr(\rho_{AB}^2) + \cdots
\end{equation}
and trivially we also have $\vev{\Tr(\rho_{A} (t)^2)}_{\CE_t} = \Tr(\rho_{A}^2)$.

Combining these, the full expression for the mutual information $I^{(2)}(A,BD)$ at early times is
\begin{equation}
\text{Early times}:\qquad I^{(2)}(A,BD) \approx \log \left( \frac{\Tr(\rho_{ABD}^2)}{\Tr(\rho_{A}^2)\Tr(\rho_{BD}^2)}\right)\,,
\end{equation}
which is simply the mutual information of the initial state.  At late times we find
\begin{equation}
\text{Late times}:\qquad I^{(2)}(A,BD) \approx \log \left( \frac{\frac{1}{d_C}\Tr(\rho_{AB}^2)+\frac{1}{d_D}}{\Tr(\rho_A^2)(\frac{1}{d_C}\Tr(\rho_{B}^2)+\frac{1}{d_D}\Tr(\rho_{A}^2))}\right)\,.
\end{equation}
Now assume that we have maximally entangled inputs, $\rho_{A\bar A \bar B B} \approx \rho_{A\bar A}\otimes \rho_{\bar B B}$, where we take $A$ to be maximally entangled with $\bar A$ and take $B$ to be nearly maximally entangled with $\bar B$. Then $\Tr (\rho_{A}^2) = 1/d_A$ and $\Tr (\rho_{B}^2) \approx 1/d_B$, and at late times we find the mutual information is nearly equal to its maximal value
\begin{equation}
\text{Late times}:\qquad I^{(2)}(A,BD) \approx 2\log(d_A) = 2n_A\,,
\end{equation}
where $n_A$ is the number of qubits in $A$, indicating that reconstruction of the state $A$ is possible by only acting on $BD$. Here we have assumed that $d\gg1$ and that the $B$ and $C$ systems are larger than their complements, $d_C\gg d_D$ and $d_B\gg d_A$. We note that if the system is 2-invariant, we only really need that $t\gg 1$ so that the spectral function $\CR_4(t)$ has decayed. We do not necessarily need to be probing exponentially long times.

Similarly, we can compute the mutual information between $A$ and $C$
\begin{equation}
I^{(2)}(A,C) = S^{(2)}_A+S^{(2)}_{C}-S^{(2)}_{AC}\,,
\end{equation}
and at early times the quantity is large, but at late times becomes small $I^{(2)}(A,C) \approx 0$, indicating that $A$ and $C$ are no longer entangled.  We have
\begin{equation}
\text{Early times}:\qquad I^{(2)}(A,C) \approx \log \left( \frac{\Tr(\rho_{AC}^2)}{\Tr(\rho_{A}^2)\Tr(\rho_{C}^2)}\right)\,.
\end{equation}
and at late times
\begin{equation}
\text{Late times}:\qquad I^{(2)}(A,C) \approx \log \left( \frac{\frac{1}{d_D}\Tr(\rho_{A}^2)+\frac{1}{d_C}\Tr(\rho_{B}^2)}{\Tr(\rho_A^2)(\frac{1}{d_C}\Tr(\rho_{AB}^2)+\frac{1}{d_D})}\right)\,.
\end{equation}
Again assuming $A$ is  maximally entangled with $\bar A$ and $B$ is nearly maximally entangled with $\bar B$, we find that at late times $I^{(2)}(A,C) \approx 0$. 

In summary, for ensembles $\mathcal{E}_t$ which become 2-invariant, we have that at time scales greater than $t \sim 1$,
\begin{equation}
I^{(2)}(A,BD) \approx \text{maximal} \and I^{(2)}(A,C) \approx 0\,,
\end{equation}
indicating that 2-invariance is sufficient for the delocalization and scrambling of information.

\subsubsection{Late time operator growth}
Chaotic many-body systems exhibit a ballistic growth in the support of an initially local operator \cite{LocalizedShocks,Aleiner16,NahumRQC17,VRPS17,SYKopgrowth,SYKopbeta}. Here we show that many-body systems which become approximately $2$-invariant exhibit a universal behavior in the late time growth of operators.  In particular, suppose we have a traceless operator $\op$ in the Heisenberg picture which evolves by
\begin{equation}
\op(t) = \sum_{A\in P} \gamma_A(t) \, A
\end{equation}
which is summed over an orthonormal basis of operators.  Also, $\gamma_A(t)$ is the support of the growing operator on a particular basis element $A$, and $\sum_{A \in P} |\gamma_A(t)|^2 = 1$.

We further denote $\op(t=0) = \op_0$\,, which is taken to be a non-identity Pauli operator.  Then we can express the coefficients $\gamma_A(t)$ by
\begin{equation}
\gamma_A(t) = \frac{1}{d} \text{tr}(\op_0(t) A) = \frac{1}{d} \text{tr}\left(e^{-i H t} \op_0 e^{i H t} A \right)
\end{equation}
and
\begin{equation}
|\gamma_A(t)|^2 = \frac{1}{d^2} \, \text{tr}\left(e^{i H t} \op_0 e^{-i H t} A \right) \text{tr}\left(e^{i H t} \op_0 e^{-i H t} A \right) \,.
\end{equation}

Following the operator growth calculation in Appendix E of \cite{RQCsym}, suppose we disorder average over an ensemble $\mathcal{E}_{t}$ where the constituent Hamiltonian ensemble $\CE_H$ has no symmetry.  For $t$ greater than the approximate $2$-invariance time, we have
\begin{equation}
|\gamma_{\op_0}(t)|^2 \simeq \frac{\mathcal{R}_4(t)}{d^4} \quad\and\quad |\gamma_{A\neq \op_0}(t)|^2 \simeq \frac{1}{d^2}\,.
\label{eq:latetimeop1}
\end{equation}
This captures a type of \textit{ergodicity} of late time operator growth -- namely, how an operator spreads itself uniformly across operator space.  For a chaotic Hamiltonian ensemble with no symmetries, we expect that at late times of $t\sim d$, we have $\mathcal{R}_4(t) \sim d^2$ so that $|\gamma_{\op_0}(t)|^2 \sim |\gamma_{A \neq \op_0}(t)|^2 \sim 1/d^2$.  This is the timescale when the initial operator $\op_0$ has evenly spread itself around operator space.  So before times of $t\sim d$ but after the approximate $2$-invariance time, the spectral statistics of the Hamiltonian ensemble evidently capture the approach to ergodicity. We note that in the approximation for $|\gamma_{\op_0}(t)|^2$ in Eqn.~\eqref{eq:latetimeop1}, which captures the decay of the support of $\mathcal{O}_0(t)$ on the initial operator $\mathcal{O}_0$, we have assumed that the approximate 2-invariance time is shorter than the time scale where the support on all operators becomes uniform ($t\sim \sqrt{d}$).

\subsection{Finite temperature}

We can generalize our discussion of $k$-invariance to finite temperature by using the thermal frame potential defined and computed in \cite{ChaosDesign,ChaosRMT}, as well as an alternative definition better suitable to ``physical'' correlation functions. We begin with the former definition first.  Recalling that the $k$-th frame potential can be expressed as the operator-averaged $2k$-point OTOCs, we can also consider a thermal $2k$-point OTOC
\begin{equation}
\begin{split}
\label{thermalOTOC}
&\langle A_1(0) B_1(t) \cdots A_k(0) B_k(t) \rangle_{\beta, 1} \\
& \qquad \qquad \qquad = \frac{1}{\tr \big(e^{- \beta H}\big)}\,\tr\left(e^{- \beta H/2k} A e^{- \beta H /2k} B(t) \cdots e^{- \beta H/2k} A e^{- \beta H/2k} B(t) \right)\,.
\end{split}
\end{equation}
Here, we have equitably distributed $2k$ interstitial factors of $e^{- \beta H/2k}$, and then normalized by $\text{tr}(e^{- \beta H})$.  This convention is packaged in the notation for the correlator, $\langle \, \cdot \, \rangle_{\beta, 1}$\,, where the ``$1$'' subscript denotes that this is a thermal correlator of the ``first kind.''  (A thermal correlator of the ``second kind'' will follow shortly.)  Taking the square of the correlator and averaging over orthonormal bases of the $A$'s and $B$'s, we find the finite-temperature frame potential of the first kind:
\begin{equation}
\CF_{\CE_{t,\beta,1}}^{(k)} = \int dH_1 \, dH_2 \, \frac{\left| \text{tr}\left(e^{-(\beta/2k - it)H_1 } e^{-(\beta/2k + it)H_2} \right)\right|^{2k}}{\text{tr}\big(e^{- \beta H_1}\big) \text{tr}\big(e^{- \beta H_2}\big)/d^2}\,.
\end{equation}
This frame potential has a pleasing form, which is why it is so-defined.  Note that as $\beta \to 0$, we recover the infinite-temperature frame potential.

Of course, thermal correlators that arise in the study of physical systems\footnote{We are referring to the placement of the $e^{- \beta H/2k}$ factors as being unphysical, not the lack of time order (which is also unphysical).} do not have the form of Eqn.~\eqref{thermalOTOC}.  Rather, there is a single Gibbs state in the correlator so that it takes the form
\begin{equation}
\label{thermalOTOC2}
\langle A_1(0) B_1(t) \cdots A_k(0) B_k(t) \rangle_{\beta, 2} = \frac{1}{\tr\big(e^{- \beta H}\big)}\,\tr\left(e^{- \beta H} A(0) B(t) \cdots A(0)  B(t) \right)\,.
\end{equation}
which we call a thermal correlator of the second kind.  Similarly taking the square of the correlator and averaging over orthonormal bases of the $A$'s and $B$'s, we arrive at the finite-temperature frame potential of the second kind:
\begin{equation}
\CF_{\CE_{t,\beta,2}}^{(k)} = \int dH_1 \, dH_2 \, \frac{\text{tr}\left(e^{-(\beta - it)H_1 } e^{- (\beta + i t) H_2 } \right) \text{tr}\left(e^{i H_1 t} e^{- i H_2 t} \right) \left| \text{tr}\left(e^{i H_1 t} e^{- i H_2 t} \right)\right|^{2k - 2}}{\text{tr}\big(e^{- \beta H_1}\big) \text{tr}\big(e^{- \beta H_2}\big)/d^2}\,.
\end{equation}
Similar to before, as $\beta \to 0$, we recover the infinite-temperature frame potential.

The finite-temperature frame potentials can help characterize $k$-invariance for thermal correlators.  In particular, for $j = 1,2$ (i.e., for thermal correlators of either the first or second kind), we have
\begin{equation}
\CF_{\CE_{t,\beta,j}}^{(k)} - \CF_{\widetilde{\CE}_{t,\beta,j}}^{(k)} = \frac{1}{d^{2k - 2}} \sum_{i_1,...,i_k = 1}^{d^2} \left|\langle A_{i_1} A_{i_2}(t) \cdots A_{i_{k-1}} A_{i_k}(t)\rangle_{\beta,j, \CE}  - \langle A_{i_1} A_{i_2}(t) \cdots A_{i_{k-1}} A_{i_k}(t)\rangle_{\beta,j, \widetilde{\CE}} \right|^2
\end{equation}
which is the finite-temperature version of Eqn.~\eqref{eq:2kOTOCkinv2}.  The proof of this identity is a trivial modification of the infinite-temperature analysis in Appendix~\ref{app:OTOCs}.  Notably, if $\mathcal{F}_{\mathcal{E}_{t,\beta,2}}^{(k)} - \mathcal{F}_{\widetilde{\mathcal{E}}_{t,\beta,2}}^{(k)} \approx 0$ for $t \geq t_{k\text{-inv}}$, then we expect $2k$-point finite-temperature Keldysh-ordered correlation functions
\begin{equation}
\langle A_k(kt)\,\cdots\,A_2(2t)\,A_1(t) \, \frac{e^{- \beta H}}{\text{tr}(e^{- \beta H})} \, A_{k+1}(t)^\dagger \, A_{k+2}(2t)^\dagger \, \cdots A_{2k}(k t)^\dagger \rangle_{\mathcal{E}_t}
\end{equation}
will be spectrally decoupled for $t \geq t_{k\text{-inv}}$.

Naturally, the Haar'ed finite-temperature frame potentials can be expressed solely in terms of spectral statistics.  For the Haar'ed finite-temperature frame potential of the first kind, we have
\begin{equation}
\mathcal{F}_{\mathcal{E}_{t,\beta,1}}^{(1)} = \frac{1}{d^2 - 1}\left(\widetilde{\mathcal{R}}_2^2(t,\beta/2) + d^2 - 2 \widetilde{\mathcal{R}}_2(t,\beta/2) \right),
\end{equation}
where we define \cite{ChaosRMT}
\begin{equation}
\widetilde{\mathcal{R}}_2(t,\beta) := \left\langle \frac{Z(t,\beta) Z^*(t,\beta)}{Z(2\beta)/d}  \right\rangle_{\widetilde{\mathcal{E}}} = \int D\lambda \, \frac{\sum_{ij} e^{it(\lambda_i - \lambda_j)} e^{- \beta(\lambda_i + \lambda_j)}}{\sum_i e^{- 2 \beta \lambda_i}/d}\,. 
\end{equation}
In a similar vein, the Haar'ed finite-temperature frame potential of the second kind can be written as
\begin{equation}
\mathcal{F}_{\mathcal{E}_{t,\beta,2}}^{(1)} = \frac{1}{d^2 - 1}\left(\overline{\mathcal{R}}_2^2(t,\beta) + d^2 - 2 \overline{\mathcal{R}}_2(t,\beta) \right),
\end{equation}
where we now define
\begin{equation}
\overline{\mathcal{R}}_2(t,\beta) := \left\langle \frac{Z(t,\beta) Z^*(t,0)}{Z(\beta)/d}  \right\rangle_{\widetilde{\mathcal{E}}} = \int D\lambda \, \frac{\sum_{ij} e^{it(\lambda_i - \lambda_j)} e^{- \beta \lambda_i}}{\sum_i e^{- \beta \lambda_i}/d}\,.
\end{equation}
One can obtain similar expressions for the higher-order finite-temperature frame potentials as well.  We expect that most of the results in this paper generalize to the finite-temperature settings.

\subsection{\ktitle-invariance in random circuits}
We end the section with an example where the $k$-invariance time is essentially exactly computable. Random quantum circuits (RQCs) are solvable models of strongly-coupled local unitary dynamics, and as such are a valuable resource for understanding many-body chaos. It is known that random circuits are rapid information scramblers \cite{BF12,HaydenPreskill}, decouplers \cite{BF13}, and generators of randomness \cite{HL08,BHH12}. Random circuits are known to form approximate $k$-designs in depth $O(n\, {\rm poly}(k))$, although it is likely that the dependence on $k$ is linear \cite{HaarRQC}, reaching the optimal lower bound $O(n k)$. It has also been shown that random circuits on $D$-dimensional lattices form approximate designs in $O(n^{1/D} {\rm poly}(k))$ depth \cite{HM18}.

As explained earlier in this section, a unitary ensemble forming a $k$-design is a sufficient condition for that ensemble to be $k$-invariant.  The convergence of random circuits to $k$-designs directly implies that they become $k$-invariant. Furthermore, we motivated $k$-invariance as a measure of chaotic dynamics in time-independent Hamiltonian systems, as opposed to the strongly time-dependent dynamics of random circuits. Nevertheless, it is still instructive to compute the $k$-invariance times in a solvable model with local dynamics.

\begin{figure}
\centering
\includegraphics[width=0.4\linewidth]{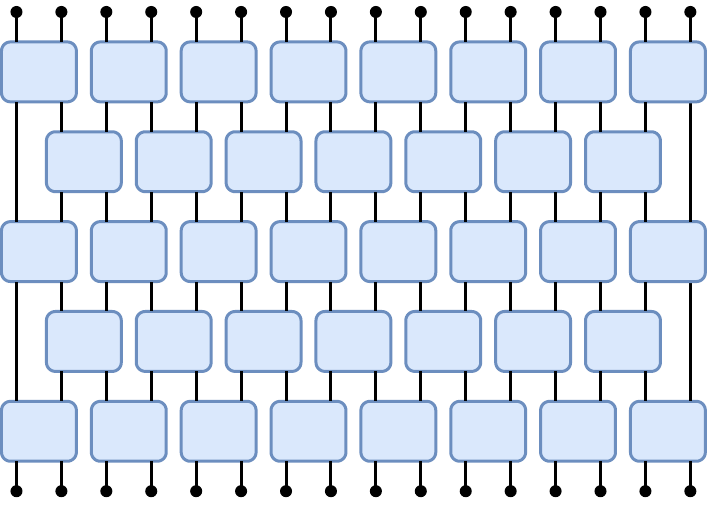}
\begin{tikzpicture}[thick,scale=0.5,baseline=-0.4cm]
\draw[thick,->] (0,0) -- (0,6) node[anchor=south] {$t$};
\end{tikzpicture}
\caption{A random quantum circuit on $n$ qudits of local dimension $q$.  Each layer is comprised of parallelized 2-site unitaries, where each gate is randomly sampled from $U(q^2)$.}
\label{fig:RQCdiag}
\end{figure}
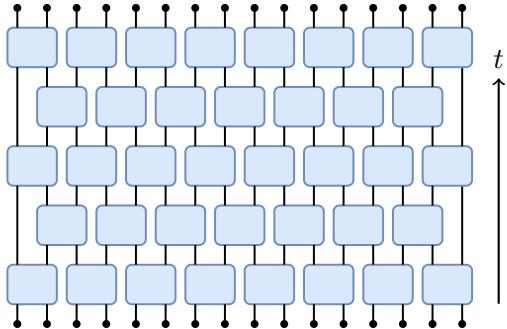

The random circuits we consider act on $n$ qudits, each of local dimension $q$, arranged in a 1-dimensional chain.  The circuits are built out of layers of 2-site random unitaries, acting in parallel on even links at even time steps and odd links at odd time steps.  The 2-site unitaries are each independently samples from $U(q^2)$.  For a diagrammatic depiction of the circuit, see Figure~\ref{fig:RQCdiag}.  Evolution to time $t$ is given by $t$ layers of the circuit, and the total unitary evolution is denoted by $U_t$. The ensemble of the accumulated random unitary circuit up to time $t$ will be notated as $\mathcal{E}_t^{\text{RQC}}$. See \cite{NahumRQC16,NahumRQC17,VRPS17,RQCstatmech} for a recent treatment of entanglement and operator growth in these models.

We start by discussing the first moment of the random circuit ensemble $\mathcal{E}_{t}^{\text{RQC}}$. The $k=1$ frame potential for random circuits is exactly $1$ after a single time step, i.e.~$\CF^{(1)}_{\CE_t^{\text{RQC}}} = 1$.  This is the Haar value of the frame potential, and so RQCs form exact 1-designs. Furthermore, the form factor for random circuit evolution is $\CR_2 = \vev{\Tr(U_t)\Tr(U_t^\dagger)}_{\mathcal{E}_{t}^{\text{RQC}}} = 1$. This means that the invariant frame potential in Eqn.~\eqref{eq:FP1inv} is $\CF^{(1)}_{\mathcal{E}_{t}^{\text{RQC}}} = 1$, and thus $\mathcal{E}_{t}^{\text{RQC}}$ is exactly 1-invariant for any $n$, $q$, and $t>0$. 

The second moment is nontrivial. The frame potential for random quantum circuits was recently computed in \cite{HaarRQC} using a mapping to a statistical mechanics model \cite{NahumRQC17,RQCstatmech}. The $k=2$ frame potential can be written as
\begin{equation}
\CF^{(2)}_{\CE_t^{\text{RQC}}} = 2 + w(t,q,n) \approx 2\bigg( 1+ \bigg(\frac{2q}{q^2+1}\bigg)^{2t}\bigg)^{n} \approx 2 + \frac{n}{q^{2t}}\,,
\end{equation}
where $w(t,q,n)$ is a sum of time-dependent contributions, which exhibits an exponential decay in time since $w(t,q,n) \sim n/q^{2t}$. The function $w(t,q,n)$ can be computed exactly, but the key fact is that the frame potential is exactly expressed as a constant term plus an exponentially decaying function in time. Recalling that $\CF^{(2)}_{\CE_t^{\text{RQC}}}$ is lower bounded by its Haar value of 2, this proves that random circuits form 2-designs.\footnote{We note that although the frame potential decays in $\log(n)$ time, defining an approximate design in terms of the diamond norm and bounding it in terms of the difference in frame potentials we pick up an additional factor of $n$. So the 2-design time is $O(n)$ for local random circuits.}

We can also compute the invariant frame potential by computing the form factors for RQCs. First note that the Haar values for the higher-point spectral form factors that appear in the 2-invariance calculation are $\CR_4 = \vev{\Tr(U)^2\Tr(U^\dagger)^2}_{\mathcal{E}_\text{Haar}} = 2$, $\CR_{4,1} = \vev{\Tr(U^2)\Tr(U^\dagger)^2}_{\mathcal{E}_\text{Haar}} = 0$, and $\CR_{4,2} = \vev{\Tr(U^2)\Tr(U^\dagger{}^2)}_{\mathcal{E}_\text{Haar}} = 2$.  Since random circuits will become $2$-designs and thus become $2$-invariant, the previous Haar values of the spectral correlators will be achieved by the RQC.  In summary, the exact values of the form factors for Haar random unitaries are
\begin{equation}
\text{Haar:}\qquad
\begin{aligned}
\CR_{2} = 1\,, \quad &\CR_{4} = 2\,, \\
\CR_{4,1} = 0\,, \quad &\CR_{4,2} = 2\,,
\end{aligned}
\end{equation}
which are achieved by the RQC at the 2-design time.  It turns out that $\CR_{4,1} = 0$ for all times under random circuit evolution. This fact is transparent for $t=1$ when we have a disjoint tensor product of unitaries. But for general times a proof of this involves expressing $\mathcal{R}_{4,1}$ as the partition function for a statistical mechanics model and seeing that all spin configurations are disallowed. Furthermore, the form factors $\CR_{4}$ and $\CR_{4,2}$ can be computed as the partition function of a lattice model, and the resulting expressions involve the same time dependence as the frame potential. For $t>0$, the exact values of the form factors for local random quantum circuits are
\begin{equation}
\text{RQCs:}\qquad
\begin{aligned}
\CR_{2}(t) = 1\,, \quad &\CR_{4}(t) = 2+w(t,q,n)\,, \\
\CR_{4,1}(t) = 0\,, \quad &\CR_{4,2}(t) = 2+w(t,q,n)\,,
\end{aligned}
\label{eq:RQCsFF}
\end{equation}
where again $w(t,n,q) \approx n/q^{2t}$, which we emphasize is an exactly calculable quantity for random circuits \cite{HaarRQC}.

Using the expressions for the RQC form factors in Eqn.~\eqref{eq:RQCsFF}, we can compute the $2$-invariant frame potential (see Eqn.~\eqref{eq:FP2inv} in Appendix~\ref{app:OTOCs}) to find
\begin{equation}
\CF^{(2)}_{\widetilde \CE^{\rm RQC}_t} = 2+ \frac{2 (d^2 -3)}{d^2(d^2-1) (d^2-9) } \, w(t,q,n)^2\,.
\end{equation}
With these expressions it becomes clear that the decay to 2-invariance precedes the decay to a 2-design, with a very short timescale in between. The decay to 2-invariance and 2-design in random circuits can be summarized as
\begin{align}
\text{2-invariance:}\qquad &\CF^{(2)}_{\CE_t^\text{RQC}}-\CF^{(2)}_{\widetilde{\CE}_t^\text{RQC}} \approx w(t,q,n) - \frac{2}{q^{4n}} w(t,q,n)^2 \sim  \frac{n}{q^{2t}} - \frac{2n}{q^{4t+4n}}\,\nn
\text{2-design:}\qquad &\CF^{(2)}_{\CE_t^{\text{RQC}}}-\CF^{(2)}_{\mathcal{E}_{\text{Haar}}} = w(t,q,n) \sim \frac{n}{q^{2t}}\,.
\end{align}
Unsurprisingly, the time scale that both $\CF^{(2)}_{\CE_t^{\text{RQC}}}-\CF^{(2)}_{\mathcal{E}_{\text{Haar}}}$ and $\CF^{(2)}_{\CE_t^\text{RQC}}-\CF^{(2)}_{\widetilde{\CE}_t^\text{RQC}}$ become small is of order $\log(n)$. Asking when the diamond norm distance between the ensembles is small, the 2-invariance and 2-design times are both $O(n)$. 

We emphasize that approximate $2$-invariance is achieved before the system becomes an approximate $2$-design.  That is, for some small tolerance $\varepsilon$, we have $\mathcal{F}_{\mathcal{E}_t^{\text{RQC}}}^{(2)} - \mathcal{F}_{\widetilde{\mathcal{E}}_t^{\text{RQC}}}^{(2)}  = \varepsilon$ at a time before $\mathcal{F}_{\CE_t^\text{RQC}}^{(2)} - \mathcal{F}_{\CE_\text{Haar}}^{(2)}  = \varepsilon$, even though the difference between the two times is parametrically suppressed.  This is in line with the fact that the $2$-invariance time should come before the $2$-design time, as being a $2$-design implies being $2$-invariant.  Since $2$-invariance implies scrambling, as previously discussed, we expect a hierarchy $t_{\text{scramble}} \leq t_{2\text{-inv}} \leq t_{2\text{-design}}$, where we have shown the latter inequality is a strict inequality for random circuits.

\subsection{Connection to ETH}
A related framework for connecting chaotic quantum systems with random matrix theory is provided by the Eigenstate Thermalization Hypothesis (ETH) \cite{SrednickiETH,DeutschETH,ChaosETH}.  There are many refinements of the original conjectures, but here we will only sketch one of the original formulations, more or less following \cite{Srednicki98}.  Our main aim is to argue that ETH and $k$-invariance are distinct but compatible notions. 

Consider a Hermitian operator $A$.  If we have a single instance of a chaotic Hamiltonian with eigenstates $\{|E_i\rangle\}_{i=1}^d$ with corresponding eigenvalues $\{E_i\}_{i=1}^d$, then ETH proposes that
\begin{equation}
\label{eq:ETH1}
\langle E_m | A | E_n \rangle = \mathcal{A}(E) \, \delta_{mn} +  e^{-S(E)/2}\,f^A(E, \omega) \, R_{mn}\,,
\end{equation}
where $E = \frac{1}{2}(E_m + E_n)$ and $\omega = E_m - E_n$.  Above, $\mathcal{A}(E)$ and $f^A(E, \omega)$ are smooth functions of $E$ and $\omega$, $S(E)$ is an energy-smoothed thermodynamic entropy, and $R_{mn}$ are random numbers with mean zero and unit variance.

To illustrate how the ETH ansatz is applied, consider the infinite-temperature 2-point function $\vev{A(t)A} = \frac{1}{d}\,\tr(A(t) A(0))$ for a fixed Hamiltonian $H$.  Then expanding in an eigenbasis of $H$, we have
\begin{equation}
\frac{1}{d}\,\tr(A(t) A(0)) = \frac{1}{d} \sum_{m,n=1}^d e^{i(E_m - E_n)t} |A_{mn}|^2\,,
\label{eq:2ptETH}
\end{equation}
where $A_{mn} = \braket{E_m|A|E_n}$. Taking an infinite time average, we find that only the diagonal terms in the sum contribute, giving $\frac{1}{d} \sum_n |A_{nn}|^2$. This is consistent with the diagonal ansatz in Eqn.~\eqref{eq:ETH1}, as $\CA(E_n) \approx A_{nn}$. Moreover, for a traceless operator, we expect the ensemble averaged diagonal matrix elements to have $\vev{A_{nn}}_{\CE_H} = 0$ and fluctuations $\vev{A_{nn}^2}_{\CE_H} \approx 1/d$. This is also consistent with the late time form of the 1-invariant 2-point function, $\vev{A(t)A}_{\widetilde \CE_t} \approx \CR_2(t)/d^2$, which goes to $1/d$. 

The first term in Eqn.~\eqref{eq:ETH1}, known as the diagonal ansatz, helps explain features of thermalization and the equilibrium values of observables.  The second term in Eqn.~\eqref{eq:ETH1}, known as the off-diagonal ansatz, aims to describe the dynamical approach to equilibrium. In our present context, the off-diagonal ansatz in Eqn.~\eqref{eq:ETH1} will give a prediction for the decay in time of the 2-point function in Eqn.~\eqref{eq:2ptETH}. We see that this agrees with the 1-invariant form and the decoupling of the matrix elements from the sum over energies in Eqn.~\eqref{eq:2ptETH} if the function $f^A(E,\omega)$ is constant.  Indeed, in energy windows $\omega \lesssim E_{\rm Th}$ where $E_{\rm Th}$ is set by a scale called the Thouless energy, $f^A(E,\omega)$ is constant (although this can be subtle in systems with diffusive transport \cite{Dymarsky18}).  Such behavior in the off-diagonal matrix elements has been observed numerically \cite{dymarsky2019new,RigolOffDiag,ChaosETH}. Then it is consistent with $k$-invariance that in small energy windows (and thus at late time scales) ETH predicts decoupled forms of correlation functions. Aside from this consistency check, it would be interesting to more fully understand the precise interplay between $k$-invariance and ETH, and how the timescales involved in each compare with one another.

\section{Spectral decoupling and symmetry}
\label{sec:kinvsym}

Our discussion of many-body chaos and the onset of spectral decoupling has thus far been centered on systems which break all symmetries.  However, many physical systems do possess symmetries, such as time reversal symmetry, particle-hole symmetry, and so on.  Here, we refine our analysis to accommodate symmetries.  In particular, we analyze the extent to which spectral decoupling can occur in systems with symmetry, and diagnose this with a symmetrized form of $k$-invariance.

\subsection{Symmetries in random matrix theory}

Random matrix theory is meant to capture the universal eigenvalue statistics of `generic' Hamiltonians, constrained by symmetry. For a system exhibiting features of quantum chaos, and which is also constrained by symmetry, a heuristic intuition is that its Hamiltonian behaves as if it was randomly sampled from some universal ensemble of Hamiltonians with appropriate symmetries.  

Here, we emphasize the role of symmetry.  Suppose we have a system with a Hamiltonian $H$, which generates unitary evolution via $U(t) = e^{- i H t}$.  If the system is endowed with symmetries, this means that there is a group $G = \{g\}$ with unitary and anti-unitary representations $V_g$ such that $[H, V_g] = 0$ for all $g \in G$.  The unitary $U(t)$ acts on states in a Hilbert space $\mathcal{H}$, which is itself endowed with an inner product structure.  Indeed, the inner product of two states is invariant under unitary time evolution.  Conversely, this invariance structure is the defining property of unitary evolution.

Then according to Dyson's threefold way \cite{DysonSym}, one can block diagonalize the Hamiltonian generating the time evolution such that the blocks are each either (i) a complex Hermitian matrix, (ii) a real symmetric matrix, or (iii) a real quaternionic matrix.  The latter two possess different manifestations of time-reversal symmetry.  For a review, see \cite{Zirnbauer10}.  If the Hamiltonian is sufficiently `generic,' it is expected that each block, necessarily falling into the categories (i), (ii), or (iii), will behave like a random matrix sampled from a corresponding universal ensemble.  Indeed, Wigner and Dyson constructed such universal random matrix ensembles by considering Gaussian random matrices consistent with (i), (ii) or (iii).  The resulting ensembles are referred to as the Gaussian Unitary Ensemble (GUE), the Gaussian Orthogonal Ensemble (GOE), and the Gaussian Symplectic Ensemble (GSE), respectively.

There are other situations in which the invariance structure of the Hilbert space under Hamiltonian time evolution is enriched.  Such is the case for systems with fermions.  Then Dyson's threefold way is expanded to the tenfold way of Altland and Zirnbauer \cite{AltlandZirnbauer}.  In this paper, we only consider the threefold classification, although our analysis could be generalized to the extended ensembles.

\subsection{Symmetric \ktitle-invariance and time-reversal symmetry}
\label{sub:kinvtsym}

So far, we have only defined $k$-invariance for ensembles $\mathcal{E}_t$ generated by Hamiltonians with no symmetries.  In the presence of symmetry, we need to modify our definition of $k$-invariance.  The idea is to find a modified version of the Haar'ed ensemble $\widetilde{\mathcal{E}}_t$ which we will call $\widetilde{\mathcal{E}}_t^{\text{sym}}$.  The ensemble $\widetilde{\mathcal{E}}_t^{\text{sym}}$ will be compatible with the symmetries of $\widetilde{\mathcal{E}}_t$, and we will use
\begin{equation}
\big\|\Phi_{\mathcal{E}_t}^{(k)} - \Phi_{\widetilde{\mathcal{E}}_t^{\text{sym}}}^{(k)}\big\|_2^2 = \CF^{(k)}_{\CE_t} - \CF^{(k)}_{\widetilde{\CE}_t^{\text{sym}}} \geq 0
\label{eq:symkinv}
\end{equation}
to quantify the distance between the two ensembles.  We begin with the specific examples of time-reversal symmetry and particle-hole symmetry, and then build up to a general construction.

In the spirit of Dyson's classification, we want to consider many-body systems invariant under time-reversal symmetry. For such Hamiltonians, there is an antiunitary operator $T$ which commutes with the Hamiltonian and squares to $\pm 1$, i.e.~$T^2 = \pm 1$. If $T$ squares to unity, then the Hamiltonian can be written as a real symmetric matrix with respect to a class of bases constructed from $T$, and which do not depend on $H$.  For chaotic quantum systems in the $T^2 = 1$ symmetry class, the spectral statistics are expected to be that of the GOE, and thus at late times we expect $\mathcal{E}_t$ to become orthogonally invariant, i.e.~the measure over $\mathcal{E}_t$ becomes invariant under conjugation by an orthogonal matrix.  (This is in contrast to the previous setting with no symmetry, in which we expect unitary invariance.)  For such systems, ``orthogonal $k$-invariance'' is determined by the distance between $\mathcal{E}_t$ and the orthogonally Haar'ed ensemble
\begin{equation}
\widetilde{\mathcal{E}}_t^O = \big\{e^{- i (O H O^T)t}\,~ H \in \CE_H ~\text{and}~ O\in O(d)\big\}\,,
\end{equation}
where the $O$'s are Haar distributed on $O(d)$ and the $H$'s are distributed according to $dH$.  Indeed, the measure on $\widetilde{\mathcal{E}}_t^O$ has the desired orthogonal invariance.

If instead $T^2=-1$ then we expect GSE-type spectral statistics and at late times we expect $\mathcal{E}_t$ to become symplectically invariant, i.e.~the measure over $\mathcal{E}_t$ becomes invariant under conjugation by a symplectic matrix.  Then symplectic $k$-invariance quantifies the distance between $\mathcal{E}_t$ and
\begin{equation}
\widetilde{\mathcal{E}}_t^{Sp} = \big\{e^{- i (S H S^D)t}\,,~ H \in \CE_H ~\text{and}~ S\in Sp(d)\big\}\,,
\end{equation}
where the $S$'s are Haar distributed on $Sp(d)$ (where we assume $d$ is even) and the $H$'s are distributed according to $dH$.  Also, $S^D$ is the symplectic transpose, given by $S^D := J S^T J^{-1}$ such that $J$ is the canonical symplectic form
\begin{equation}
J := \begin{bmatrix}
\phantom{-}\textbf{0} & & \textbf{1} \, \\ - \textbf{1} & & \textbf{0} \,
\end{bmatrix},
\end{equation}
where $\textbf{0}$ is the $d/2 \times d/2$ zero matrix, and $\textbf{1}$ is the $d/2 \times d/2$ identity matrix.

For a given $k$, we compute Eqn.~\eqref{eq:symkinv} using the orthogonally and symplectically invariant frame potentials. Recall that in the case with no symmetries, we compute the $k=1$ unitary invariant frame potential for $\widetilde \CE_t$ in terms of spectral functions using the Haar invariance of the integration measure. We can proceed similarly to compute the invariant frame potentials for time-reversal symmetric systems.

First, consider symmetric $k$-invariance for systems with $T^2=1$, where we look at the distance to orthogonal invariance. The $k=1$ invariant frame potential for $\widetilde \CE^O_t$ is written as
\begin{align}
\CF^{(1)}_{\widetilde \CE_t^O} &= \int_{O(d)} dO \int_{\CE_H} dH_1 dH_2 \,\big|\tr(e^{i(O H_1 O^T) t}e^{-i (O H_2 O^T) t})\big|^{2} \nonumber \\
&= \int_{O(d)} dO \int_{\CE_H} dH_1 dH_2 \,\tr(O \,e^{i H_1 t}\, O^T e^{-i H_2 t}) \tr(O e^{-i H_1 t} O^T e^{i H_2 t})\,.
\end{align}
We can evaluate the above integral by utilizing the second moment of the Haar measure on $O(d)$, with the expression given in Eqn.~\eqref{eq:O2ndmom} in Appendix~\ref{app:designs}.  We find
\begin{equation}
\CF^{(1)}_{\widetilde \CE_t^O} = \frac{1}{d(d+2)(d-1)} \big( (d+1) \CR_2(t)^2 + 2d^3 - 4d \CR_2(t)\big)\,.
\label{eq:OinvFP}
\end{equation}
In the $T^2 = -1$ case, the symplectic invariant frame potential for $k=1$ can also be computed similarly, this time using the second moment of the Haar measure on $Sp(d)$, written explicitly in Eqn.~\eqref{eq:Sp2ndmom} in Appendix~\ref{app:designs}. The result is
\begin{equation}
\CF^{(1)}_{\widetilde \CE_t^{Sp}} = \frac{1}{d(d-2)(d+1)} \big( (d-1) \CR_2(t)^2 + 2d^3 - 4d \CR_2(t)\big)\,.
\label{eq:SpinvFP}
\end{equation}

\subsubsection*{1-invariance and time-reversal symmetry}

Let us explore the relation between approximate 1-invariance and 2-point functions in systems with time-reversal symmetry. Here we assume a multiqubit system, $d=2^n$, and use the basis of Pauli strings. If $T^2=1$, then we expect the late time dynamics to achieve approximate orthogonal invariance, such that $\CF^{(k)}_{\CE_t} - \CF^{(k)}_{\widetilde \CE_t^O} \approx 0$. The difference in $k=1$ frame potentials may be written as the average over 2-point functions
\begin{equation}
\CF^{(1)}_{\CE_t} - \CF^{(1)}_{\widetilde \CE_t^O} = \sum_{A,B \in P'} \big| \vev{A(t) B}_{\CE_t} \big|^2 - \sum_{A,B \in P'} \big| \vev{A(t) B}_{\widetilde \CE_t^O} \big|^2\,.
\end{equation}
The 2-point functions $\vev{A(t) B}_{\widetilde \CE_t^O}$ with respect to the orthogonally invariant ensemble can be derived using Eqn.~\eqref{eq:O2ndmom} in Appendix~\ref{app:designs}. We compute the orthogonally invariant 2-point functions for Pauli strings $A$ to be
\begin{equation}
A~ {\rm even}:~ \vev{A(t) A}_{\widetilde \CE_t^O} = \frac{\CR_2(t)+d-2}{(d+2)(d-1)}\,, \qquad A ~{\rm odd}:~ \vev{A(t) A}_{\widetilde \CE_t^O} = \frac{\CR_2(t)-d}{d(d-1)}\,,
\label{eq:Oinv2pt}
\end{equation}
depending on whether the Pauli string $A$ is even or odd, \ie $A^T = \pm A$.  We will simply refer to the Pauli strings as Pauli operators.  The 2-point functions $\vev{A(t)B}$ with $A\neq B$ vanish identically. 

As a consistency check, we can write out the full expression for the orthogonally invariant frame potential.  The $d^2-1$ non-identity Pauli operators contain $(d+2)(d-1)/2$ even operators and $d(d-1)/2$ odd operators. Therefore,
\begin{equation}
\CF^{(1)}_{\widetilde \CE_t^O} = \sum_{A~{\rm even}} \bigg(\frac{\CR_2(t)+d-2}{(d+2)(d-1)}\bigg)^2 + \sum_{A~{\rm odd}} \bigg(\frac{\CR_2(t)-d}{d(d-1)}\bigg)^2 +1 = \frac{(d+1)\CR_2(t)^2+2d^3-4d\CR_2(t)}{d(d+2)(d-1)}
\end{equation}
which is the same as the expression we derived above in Eqn.~\eqref{eq:OinvFP}.

We would like to understand orthogonal 1-invariance in terms of 2-point functions.  Assuming that the contribution from $A\neq B$ correlation functions are negligible, we can write
\begin{align}
\CF^{(1)}_{\CE_t} - \CF^{(1)}_{\widetilde \CE_t^O} &\approx \sum_{A \in P'} \big| \vev{A(t) A}_{\CE_t} \big|^2 - \sum_{A \in P'} \big| \vev{A(t) A}_{\widetilde \CE_t^O} \big|^2\nn
&= \sum_{A \in P'} \big| \vev{A(t) A}_{\CE_t} \big|^2 - \frac{(\CR_2(t)+d-2)^2}{2(d+2)(d-1)} - \frac{(\CR_2(t)-d)^2}{2d(d-1)} \,.
\end{align}
Since the operator average of the 2-point function $\langle A(t) A \rangle_{\mathcal{E}_t}$ is given by the spectral form factor
\begin{equation}
\big\langle \vev{A(t)A}_{\mathcal{E}_t}\big\rangle_A = \frac{1}{d^2-1} \sum_{A\in P'} \vev{A(t)A} = \frac{\CR_2(t) -1}{d^2-1} \,,
\end{equation}
after some algebra we can write the condition for orthogonal 1-invariance as
\begin{equation}
\frac{\CF^{(1)}_{\CE_t} - \CF^{(1)}_{\widetilde \CE_t^O}}{d^2-1} \approx \Big\langle \big| \vev{A(t) A}_{\CE_t} \big|^2 \Big\rangle_A- \big|\big\langle \vev{A(t) A}_{\CE_t} \big\rangle_A \big|^2 - \frac{\big(\big\langle \vev{A(t) A}_{\CE_t} \big\rangle_A-1\big)^2}{d(d+2)} \,.
\end{equation}
To leading order in $d$ this is simply
\begin{equation}
\CF^{(1)}_{\CE_t} - \CF^{(1)}_{\widetilde \CE_t^O} \approx (d^2-1) {\rm Var}\big(\vev{A(t) A}_{\CE_t} \big) - 1 \,.
\end{equation}
Therefore, achieving orthogonal invariance implies that at late times, there will be a residual variance in the 2-point functions. We will see this explicitly in our numerics for Pauli spin systems in Section~\ref{sec:kinvex}, specifically in Figure~\ref{fig:2ptsR2}.  Although our analysis of this variance is for systems with $T^2 = 1$ time-reversal symmetry, a similar analysis holds for $T^2 = -1$ time-reversal symmetry.

Also, we note that if one did not know the system had time-reversal symmetry and went ahead computing the unitary 1-invariance, the late time value would be off by this additive factor of one, which is indeed what we observe in our numerics for $T$-invariant spin systems.

\subsubsection*{Orthogonal invariance from unitary invariance}
In deriving the expression for orthogonal 1-invariance above, we assumed that the Hamiltonian was real, and thus $e^{-iHt}$ is a symmetric unitary. This glosses over an issue of basis dependence when computing the symmetric frame potentials. In general, making no assumptions about the symmetry of the Hamiltonian ensemble, the expression for the orthogonally-invariant frame potential is
\begin{equation}
\CF^{(1)}_{\widetilde\CE_t^O} = \frac{(d+1)\big(\CR_2^2+d^2+\CR_T^2\big) -2\big(d\CR_2+d\CR_T+\CR_2\CR_T\big)}{d(d+2)(d-1)}\,,
\label{eq:OinvFPfull}
\end{equation}
where
\begin{equation}
\CR_T(t) := \big\langle \tr(e^{iHt}e^{-iH^* t}) \big\rangle_{\CE_H}\,.
\end{equation}
We note that when the Hamiltonian is of GOE class and commutes with an antiunitary operator $T$, there exists a set of bases (constructed independently of $H$) in which $H$ is real (see Appendix~\ref{app:num}). In such bases, $\CR_T = d$ and the expression reduces to the orthogonal 1-invariant frame potential previously derived in Eqn.~\eqref{eq:OinvFP}.

As a consistency check, we can compute $\CR_T(t)$ for the GUE, and find
\begin{equation}
\CR_T(t) = \big\langle \tr(e^{iHt}e^{-iH^* t}) \big\rangle_{\CE_{\rm GUE}} = \frac{\CR_2(t)+d}{d+1} \,.
\end{equation}
Plugging this into the expression for $\CF^{(1)}_{\widetilde\CE_t^O}$ above, we find that the orthogonally invariant frame potential for the GUE reduces to the unitarily invariant expression $\CF^{(1)}_{\mathcal{E}_{\text{GUE}}}$ in Eqn.~\eqref{eq:FP1inv} as expected, meaning that the GUE is exactly orthogonally invariant.

More generally, if we wanted to check orthogonal invariance in the absence of time-reversal symmetry, we should use Eqn.~\eqref{eq:OinvFPfull} where $\CR_T$ no longer reduces to $d$ in a $T$-invariant basis.

\subsection{\ktitle-invariance and a \ensuremath{\Z_2} global symmetry}
Now we consider a system with a $\Z_2$ global symmetry. This example will be a good warm-up for our discussion of SYK in the next section. Suppose we have a Hamiltonian ensemble such that each Hamiltonian is block diagonal with respect to a $\Z_2$ symmetry operator.  In particular, there are two blocks of equal size:
\begin{equation}
H = \begin{pmatrix}H_+&\\ &H_- \end{pmatrix}\,.
\end{equation}
We further suppose that each block $H_+$ and $H_-$ has no further symmetry.  Also, the subscripts $+$ and $-$ are purely for convenience, and do not mean that the eigenvalues of $H_+$ and $H_-$ have a definite sign.  Then the correctly Haar-symmetrized version of the ensemble $\mathcal{E}_t$ is
\begin{align}
\widetilde{\mathcal{E}}_t &= \left\{ \begin{pmatrix} e^{- i (U_+ H_+ U_+^\dagger) t} & \textbf{0} \\
\textbf{0} & e^{- i (U_- H_- U_-^\dagger) t} \end{pmatrix},~ \begin{pmatrix}H_+&\\ &H_- \end{pmatrix} \in \CE_H ~~\text{and}~~ U_+, U_- \in U(d/2)\right\}\,.
\end{align}
We can compute the invariant frame potential by
\begin{align}
\CF_{\widetilde\CE_t}^{(1)} = \int_{U(d/2)} \!\!dU_+ dU_- \int_{\CE_H} \!\!dH dH'  \,  \bigg|  \Tr \bigg[\!\begin{pmatrix}U_+&\textbf{0}\\ \textbf{0}&\hspace*{-6pt}\,\,U_- \end{pmatrix}\!\! \begin{pmatrix}e^{i H_+ t}&\textbf{0}\\ \textbf{0}&\hspace*{-6pt}e^{i H_- t} \end{pmatrix}\!\! \begin{pmatrix}U_+^\dagger&\textbf{0}\\ \textbf{0}&\hspace*{-6pt}\,\,U_-^\dagger \end{pmatrix}\!\! \begin{pmatrix}e^{-i H'_+ t}&\textbf{0}\\ \textbf{0}&\hspace*{-6pt}e^{-i H'_- t} \end{pmatrix}\! \bigg]\bigg|^2\,.
\end{align}
As it will be convenient to refer to later, we write this out explicitly as
\begin{align}
\label{eq:symFP}
\CF_{\widetilde\CE_t}^{(1)} = \int_{U(d/2)} \!\!dU_+ dU_- \int_{\CE_H} \!\!dH dH'  \,  &\Big( \big|\Tr(U_+ e^{i H_+ t} U_+^\dagger e^{-i H'_+ t}) \big|^2 + \big|\Tr(U_- e^{i H_- t} U_-^\dagger e^{-i H'_- t}) \big|^2\nn
&+ \Tr(U_+ e^{i H_+ t} U_+^\dagger e^{-i H'_+ t})\Tr(U_- e^{-i H_- t} U_-^\dagger e^{i H'_- t}) \nn
&+ \Tr(U_+ e^{-i H_+ t} U_+^\dagger e^{i H'_+ t}) \Tr(U_- e^{i H_- t} U_-^\dagger e^{-i H'_- t}) \Big)\,.
\end{align}
We can proceed by computing the first two terms using the second Haar unitary moment and the second two terms using the first Haar unitary moment, which gives
\begin{align}
\CF_{\widetilde\CE_t}^{(1)} = \frac{1}{d_+^2-1} \big( \CR_{2,+}(t)^2 + d_+^2 -2 \CR_{2,+}(t)\big) &+ \frac{1}{d_-^2-1} \big( \CR_{2,-}(t)^2 + d_-^2 -2 \CR_{2,-}(t)\big) + \CF^{(1)}_{\rm int}\,,
\end{align}
where $d_+ = d_- = d/2$ are the dimensions of the two sectors and the form factors $\CR_{2,\pm}$ are those evaluated in each respective sector, \ie involving only the eigenvalues in that part of the spectrum. In the first two terms are two decoupled versions of the normal Haar-invariant frame potential for each sector. We have also defined an interaction term, which is a mixed moment depending on both the $+$ and $-$ sectors
\begin{equation}
\CF^{(1)}_{\rm int} =  \frac{2}{d_+ d_-} \big| \big\langle \Tr(e^{iH_+ t}) \Tr(e^{-iH_- t}) \big\rangle_{\CE_H} \big|^2 = \frac{2}{d_+ d_-} \Big| \int D\lambda \,\Big(\sum_{\lambda_+} e^{i\lambda_i t}\Big) \Big(\sum_{\lambda_-} e^{-i\lambda_i t}\Big)\Big|^2\,,
\end{equation}
where we see that the eigenvalues in each sector are summed over separately.

\subsection{General construction}

Having treated time-reversal symmetry and particle-hole symmetry on a case-by-case basis, we want to generalize our definition of $k$-invariance to accommodate general symmetries.  To appreciate the need for a modification of $k$-invariance in the presence of symmetry, it is worth revisiting the definitions of our main ensembles of study.  Recalling that $\mathcal{E}_t = \{e^{- i H t}\,, \, H \in \CE_H\}$ and
$$\widetilde{\mathcal{E}}_t = \{e^{- i (U H U^\dagger) t}\,, \, H \in \CE_H\text{  and  }U \in U(d)\}\,,$$
$k$-invariance aims to quantify the degree to which
\begin{equation}
\label{eq:approxssatisfy}
\mathcal{E}_t \approx \widetilde{\mathcal{E}}_t
\end{equation}
by comparing the $k$-th moments of each ensemble.  Notice that while $\mathcal{E}_t$ and $\widetilde{\mathcal{E}}_t$ have the same joint eigenvalue distribution, the two ensembles may differ in their eigenvector distribution.  In particular, $\widetilde{\mathcal{E}}_t$ has all possible eigenbases as equally likely.  For $\mathcal{E}_{H} = \mathcal{E}_\text{GUE}$, we have $\mathcal{E}_t = \widetilde{\mathcal{E}}_t$ since $\mathcal{E}_t$ is \textit{already} Haar-invariant (since the GUE is Haar-invariant), i.e.~it has all possible eigenbases as equally likely.  If on the other hand $\mathcal{E}_t$ is not Haar-invariant, and further does not possess any symmetries, then we may still have $\mathcal{E}_t \approx \widetilde{\mathcal{E}}_t$ for sufficiently large $t$ due to spectral decoupling -- in particular, moments of the ensemble $\mathcal{E}_t$ may behave as approximately Haar-invariant at sufficiently late times, and thus approximately equal the corresponding moments of $\widetilde{\mathcal{E}}_t$.

Now we consider the role of symmetry.  If $\CE_H$ has a symmetry group $G$, then for all $g \in G$ there is a unitary (or anti-unitary) representation $V_g$ such that $V_g H V_g^\dagger = H$, or equivalently\footnote{We will restrict our attention to symmetries that commute with the Hamiltonian, and not treat the anti-commuting cases here.}
\begin{equation}
[H, V_g] = 0\,.
\end{equation}
Suppose that the symmetry group is non-trivial, i.e.~contains more than the identity.  A key point is that (non-trivial) symmetry imposes linear constraints on the space of Hamiltonians $H$ (i.e., the space of Hermitian operators). Let $\text{Herm}(\mathcal{H})$ be the vector space of Hermitian operators on the Hilbert space $\mathcal{H}$.  If both of $H_1,H_2 \in \text{Herm}(\mathcal{H})$ satisfy the linear constraints
\begin{equation}
[H_1, V_g] = 0\,, \qquad [H_2, V_g] = 0\,,
\end{equation}
then by linearity
\begin{equation}
[\alpha \, H_1 + \beta \, H_2, V_g] = 0
\end{equation}
for $\alpha, \beta \in \mathbb{R}$.  More generally, we have a system of linear constraints given by $[H, V_g] = 0$ for all $g \in G$.  Let us denote the subspace of Hermitian operators satisfying these symmetry constraints by $S_G$.  In particular,
\begin{equation}
S_G = \{H \in \text{Herm}(\mathcal{H}) \, | \, [H, V_g] = 0\,,\, \forall g \in G\}\,.
\end{equation}
In the trivial case for which $G = \{\textbf{1}\}$, we have $S_{G} = \text{Herm}(\mathcal{H})$.

\begin{figure}
\centering
\includegraphics[width=0.4\linewidth]{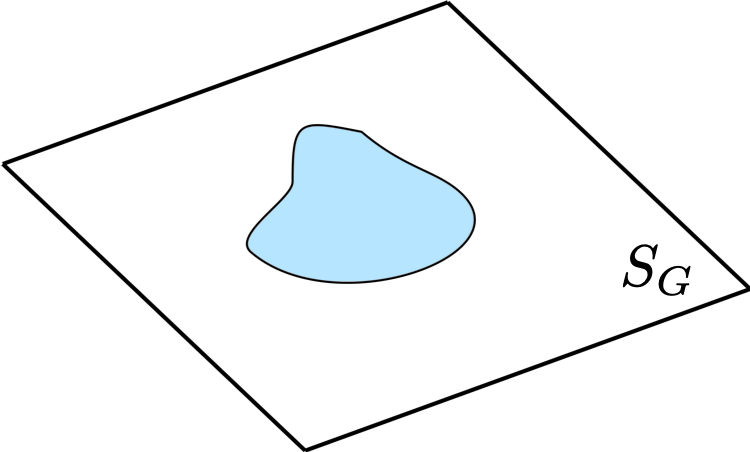}
\caption{A schematic diagram of $\text{Herm}(\mathcal{H})$, which contains the subspace $S_G$ of Hamiltonians invariant under $G$.  The subspace contains the Hamiltonian ensemble $\CE_H$, represented by a light blue blob.}
\label{fig:symkinv1}
\end{figure}

We now have the following picture: the ensemble of Hamiltonians $\CE_H$ with symmetry group $G$ is supported on the subspace $S_G$ of $\text{Herm}(\mathcal{H})$.  This is notated as
\begin{equation}
\label{eq:support1]}
\text{supp}(\CE_H) \subseteq S_G\,.
\end{equation}
A diagram can be seen in Figure~\ref{fig:symkinv1}. 

To make this explicit, recall that the ensemble $\CE_H$ is equivalent to a probability measure $P(H) \, dH$ on $\text{Herm}(\mathcal{H})$.  Let $\pi : \text{Herm}(\mathcal{H}) \to S_G$ be the projection map, and let $\delta_{S_G}(H)$ be the delta function of the subspace $S_G$, satisfying
\begin{equation}
\int_{\text{Herm}(H)} dH \, f(H) \, \delta_{S_G}(H) = \int_{S_G} d(\pi_* H) \, f(H)
\end{equation}
for any function $f(H)$.  In the above equation, $d(\pi_* H)$ is the pushforward of the volume form $dH$ to the subspace $S_G$.  Then we can interpret Eqn.~\eqref{eq:support1]} as saying that the probability measure $P(H) \, dH$ associated to $\CE_H$ has the form
\begin{equation}
\label{eq:PSG1}
P(H) = p(H) \, \delta_{S_G}(H)
\end{equation}
for some $p(H)$.

\begin{figure}
\centering
\includegraphics[width=0.32\linewidth]{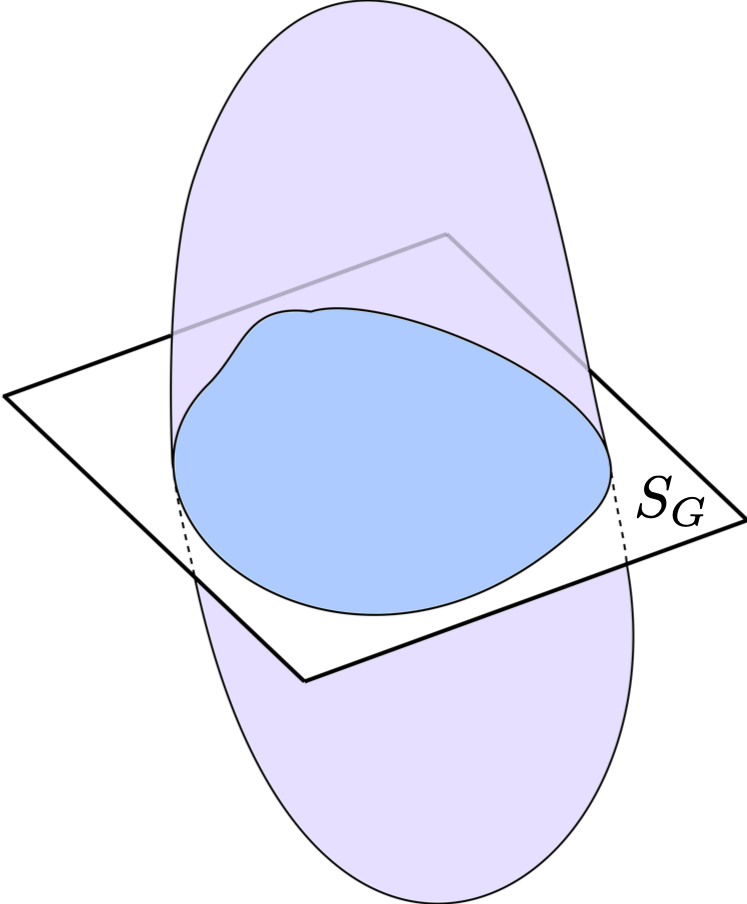}
\caption{The subspace $S_G$ does not completely contain the Haar'ed Hamiltonian ensemble $\widetilde\CE_H$, represented by a purple blob.  The intersection $S_G \cap \widetilde\CE_H$ is in blue.}
\label{fig:symkinv2}
\end{figure}

The Haar'ed ensemble $\CE_H$, denoted by $\widetilde\CE_H$, has an associated probability distribution
\begin{equation}
\label{eq:PHaar1}
\widetilde{P}(H) = \int_{U(d)} dU \, P(U H U^\dagger)\,.
\end{equation}
As a consequence, for any unitary $V$, we have $\widetilde{P}(V H V^\dagger) = \widetilde{P}(H)$, demonstrating that the ensemble $\widetilde\CE_H$ is Haar-invariant (and hence only depends on the joint eigenvalue distribution).  However, if $\CE_H$ has a non-trivial symmetry group, then
\begin{equation}
\text{supp}(\widetilde\CE_H) \not\subseteq S_G\,.
\end{equation}
This is true because putting together Eqn.'s~\eqref{eq:PSG1} and~\eqref{eq:PHaar1}, we have
\begin{equation}
\widetilde{P}(H) = \int_{U(d)} dU \, p(U H U^\dagger) \, \delta_{S_G}(U H U^\dagger) = \int_{U(d)} dU \, p(U H U^\dagger) \, \delta_{U^\dagger S_G U}(H)
\end{equation}
which is clearly no longer supported on $S_G$ alone.  A schematic diagram is shown in Figure~\ref{fig:symkinv2}.  Note that $U^\dagger S_G U$ is the subspace obtained from $S_G$ by conjugating each constituent Hermitian matrix by $U^\dagger$.

So far, we have seen that an ensemble $\CE_H$ with (non-trivial) symmetry group $G$ is supported on a subspace $S_G$ of the space of all Hermitian operators, whereas the Haar'ed version of the ensemble $\widetilde\CE_H$ is supported on the \textit{entire} space of Hermitian operators, and in fact is not supported on \textit{any} proper subspace thereof.  Thus, for ensembles with symmetry, we do not expect $\CE_H \approx \widetilde \CE_H$.

To ameliorate this issue, we appropriately modify the definition of $\widetilde \CE_H$ so that it is consistent with the symmetries of the original ensemble $\CE_H$.  In other words, if $\CE_H$ has symmetry group $G$ so that $\text{supp}(\CE_H)  \subseteq S_G$, we want to find a ``canonical'' definition of $\widetilde\CE_H^G$ such that $\text{supp}(\widetilde\CE_H^G)  \subseteq S_G$.  To do so, consider a symmetry group $G$ and the associated invariant subspace of $\text{Herm}(\mathcal{H})$ which we have denoted by $S_G$.  There is a natural action of the unitary group $U(d)$ on $\text{Herm}(\mathcal{H})$ by $U \cdot H = U H U^\dagger$.  Then we can consider a subgroup of $U(d)$ defined by
\begin{equation}
\text{Invar}_{U(d)}(S_G) = \{U \in U(d) \, | \, U H U^\dagger \in S_G\,,\,\,\forall H \in S_G\}\,.
\end{equation}
This is precisely the subgroup of $U(d)$ which leaves $S_G$ invariant under the group action.  The idea is to ``symmetrize'' the ensemble $\CE_H$ with respect to $\text{Invar}_{U(d)}(S_G)$.

More precisely, we define $\widetilde \CE_H^G$ by the probability distribution
\begin{equation}
\widetilde{P}^G(H) := \int_{\text{Invar}_{U(d)}(S_G)} \!\!dW \, P(W H W^\dagger)\,.
\end{equation}
where $dW$ is the Haar measure on $\text{Invar}_{U(d)}(S_G)$.  Plugging in $P(H) = p(H) \, \delta
_{S_G}(H)$ as per Eqn.~\eqref{eq:PSG1}, we find
\begin{equation}
\widetilde{P}^G(H) = \int_{\text{Invar}_{U(d)}(S_G)} \!\!dW \, p(W H W^\dagger) \, \delta_{S_G}(W H W^\dagger) = \int_{\text{Invar}_{U(d)}(S_G)} \!\!dW \, p(W H W^\dagger) \, \delta_{S_G}(H) 
\end{equation}
since $W^\dagger S_G W = S_G$ for all $W \in \text{Invar}_{U(d)}(S_G)$.  (Note we also have $W^\dagger S_G W = S_G$ for all $W \in \text{Invar}_{U(d)}(S_G)$ because $\text{Invar}_{U(d)}(S_G)$ is itself a group and thus contains both $W$ and $W^\dagger$.)

\begin{figure}
\centering
\includegraphics[width=0.35\linewidth]{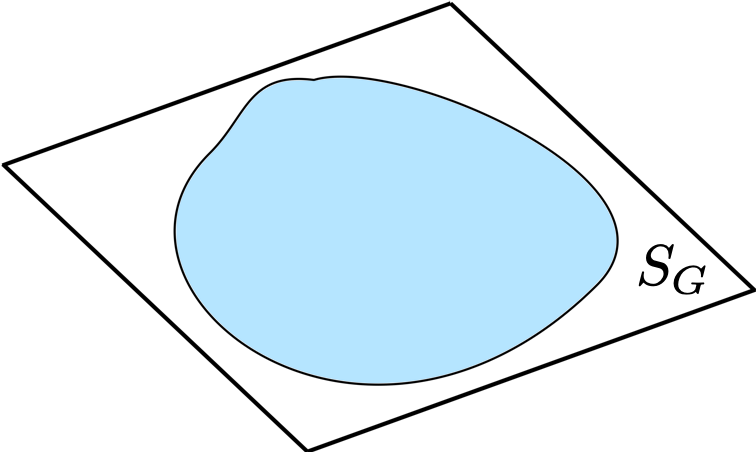}
\caption{$S_G$ completely contains the $G$-Haar'ed Hamiltonian ensemble $\widetilde\CE_H^G$, represented by a light blue blob.}
\label{fig:symkinv3}
\end{figure}

In less mathematical terms, $\widetilde\CE_H^G$ is the most Haar'ed version of $\CE_H$, consistent with its symmetries.  We sketch a schematic diagram in Figure~\ref{fig:symkinv3}.
Now we return to the ensemble $\mathcal{E}_t$.  This ensemble of unitaries has probability measure $P_t(U) \,dU = dP_t$.   Defining $g_t : \text{Herm}(H) \to U(d)$ by
\begin{equation}
g_t(H) = e^{- i H t}\,,
\end{equation}
we have
\begin{equation}
dP_t := d(g_{t\,*} P)\,,
\end{equation}
where $d(g_{t\,*} P)$ is the pushforward measure of $P$ with respect to $g_t$.  More simply, we have
\begin{equation}
\int_{U(d)} dP_t(U) \, f(U) = \int_{\text{Herm}(H)} \!\!dH \, P(H) \, f(e^{- i H t})
\end{equation}
for any function $f(U)$.
Then if $\CE_H$ has symmetry group $G$, the measure $dP_t(U)$ has the form
\begin{equation}
\int_{U(d)} dP_t(U) \, f(U) = \int_{\text{Herm}(H)} \!\!dH \, p(H) \, \delta_{S_G}(H) \,f(e^{- i H t})\,.
\end{equation}
We can naturally define $\widetilde{\mathcal{E}}_t^G$ by the probability distribution
\begin{equation}
\widetilde{P}_t^G(U) := \int_{\text{Invar}_{U(d)}(g_t(S_G))}\!\! dW \, P_t(W U W^\dagger)\,,
\end{equation}
and this probability distribution satisfies
\begin{equation}
\int_{U(d)} dU \, \widetilde{P}_t^G(U) \, f(U) = \int_{\text{Herm}(H)} dH \, \widetilde{P}^G(H) \, f(e^{- i H t}) \,.
\end{equation}
Thus, we have defined an appropriate (and canonical) symmetrically-invariant ensemble $\widetilde{\mathcal{E}}_t^G$.

\subsubsection*{Approximate symmetry}

Suppose we have a Hamiltonian ensemble $\CE_H$ and a symmetry group $G = \{g\}$ with unitary and anti-unitary representations $V_g$ such that for all $H$ in $\CE_H$ and all $g$ in $G$,
\begin{equation}
\|[H, V_g]\|_2 \leq \varepsilon
\end{equation}
for $\varepsilon$ small.  Some other norm may be used instead, if desired.  We say that $\CE_H$ has $G$ as an approximate symmetry.  While $\CE_H$ may not strictly lie in $S_G$, the ensemble will be concentrated near $S_G$. Considering an ensemble $\mathcal{E}_t$ as usual, we can ask if the ensemble becomes $k$-invariant at late times.  We would expect that the ensemble does not become $k$-invariant, but rather may become approximately symmetric $k$-invariant with respect to $G$.  The reason is that if we consider the Haar'ed version of $\CE_H$, namely $\widetilde \CE_H$, then the new ensemble will not be concentrated near $S_G$.  The simple reason is that Haar-averaging with respect to the unitary ensemble does not `know' about the approximate symmetry group $G$.  On the other hand, we expect that the $G$-symmetrized version of $\CE_H$, namely $\widetilde \CE_H^{G}$, is concentrated near $S_G$.

In summary, if $\CE_H$ and thus $\mathcal{E}_t$ has an approximate symmetry group $G$, we may anticipate approximate symmetric $k$-invariance with respect to $G$.  In Section~\ref{sec:ApproxPauliExamples} we will explore this numerically.

\subsection{Summary of symmetric \ktitle-invariance}

We have argued above that for ensembles of Hamiltonians possessing some symmetry, $k$-invariance must be modified.  For instance, a Hamiltonian ensemble with $T^2 = 1$ time-reversal symmetry will not become \textit{unitarily} $k$-invariant (the most general kind of $k$-invariance), but can instead become \textit{orthogonally} $k$-invariant (i.e., $k$-invariance augmented to accommodate $T^2 = 1$ time-reversal symmetry).  We have provided a systematic approach to construct the appropriate form of $k$-invariance for a given symmetry.  In the next section, we will put this approach to work: we will show numerically that ensembles with certain symmetries do not become unitarily $k$-invariant, but do become (approximately) symmetrically $k$-invariant according to the symmetries.

\section{\ktitle-invariance in spin systems and SYK}
\label{sec:kinvex}

In this section we examine $k$-invariance numerically in a number of disordered many-body systems, including disordered Pauli spin systems and various versions of the SYK model. We proceed by numerically computing the frame potential, as defined in Eq.~\eqref{eq:FPdef}, for the ensemble $\CE_t$ of a given disordered system at fixed times.  In particular, we generate a set of Hamiltonians by randomly sampling from the ensemble of disordered Hamiltonians, and use these samples as an approximation to the full Hamiltonian ensemble to compute $\CF^{(k)}_{\CE_t}$. We then compute the invariant frame potentials by numerically evaluating the spectral form factors for the same ensemble of randomly generated Hamiltonians. Some additional details regarding our numerics are given in Appendix~\ref{app:num}. 

The quantity we compute is the difference between $k=1$ frame potentials
\begin{equation*}
\CF^{(1)}_{\CE_t} - \CF^{(1)}_{\widetilde{\CE}_t}\,,
\end{equation*}
which we sometimes refer to informally as the `distance to $1$-invariance.' 
In computing this for systems breaking all symmetries, we take $\widetilde{\CE}_t$ to be the unitarily invariant ensemble and use the 1-invariant frame potential in Eqn.~\eqref{eq:FP1inv}. For time-reversal symmetric systems we use the orthogonal and symplectic invariant expressions for the frame potential derived in Section~\ref{sub:kinvtsym}. For the SYK model, which nontrivially realizes a particle-hole symmetry due to the interplay with charge-parity sectors, we derive the appropriate 1-invariant expressions for $\CF^{(1)}_{\widetilde{\CE}_t}$ below. 

\subsection{Pauli spin models}
The simplest examples of disordered spin systems are Pauli spins with disordered interactions. We consider several versions with nonlocal and geometrically local interactions as well as systems which realize or break time-reversal symmetry. Numerics studying $k$-invariance in Pauli spin systems were first reported in \cite{NHJthesis}.

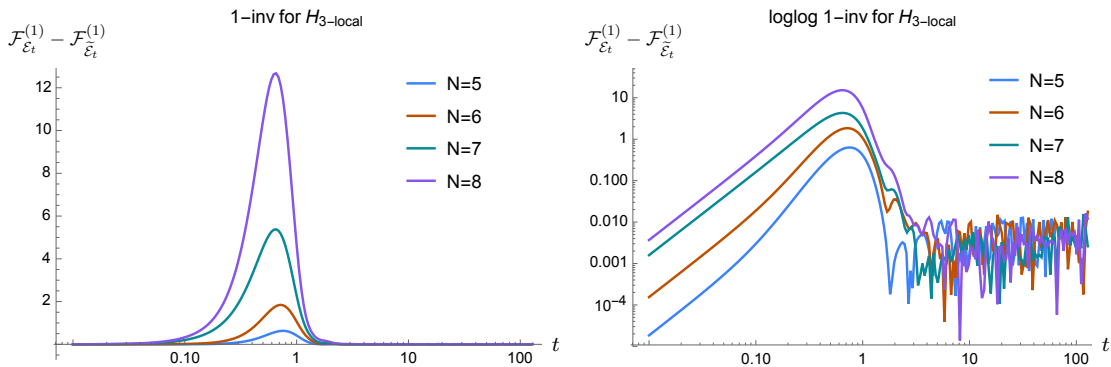
\begin{figure}
\centering
\begin{tikzpicture}[scale=0.8,baseline=-0.4cm]
\node at (0,0) {\includegraphics[width=0.45\linewidth]{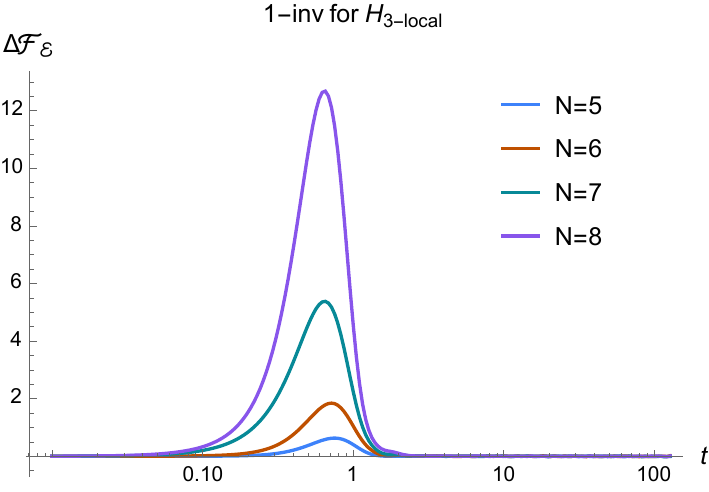}\quad \includegraphics[width=0.45\linewidth]{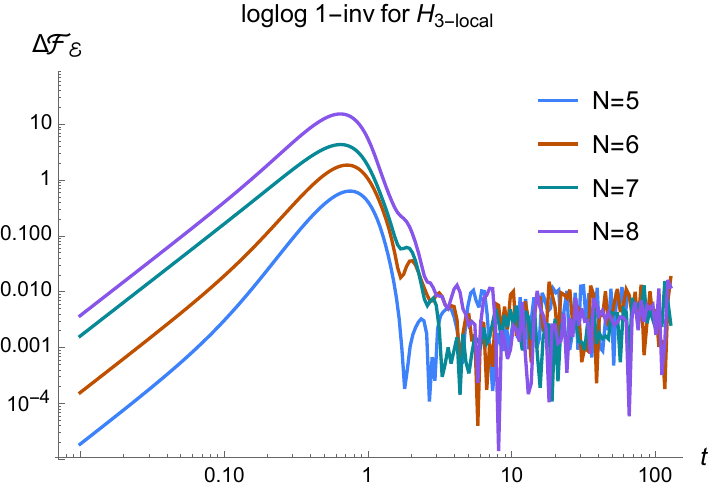}};
\node[fill=white] at (-8.6,2.45) {\scalebox{0.69}{$\CF^{(1)}_{\CE_t}-\CF^{(1)}_{\widetilde \CE_t}$}};
\node[fill=white] at (-0.32,-2.6) {\scriptsize $t$};
\node[fill=white] at (0.98,2.45) {\scalebox{0.69}{$\CF^{(1)}_{\CE_t}-\CF^{(1)}_{\widetilde \CE_t}$}};
\node[fill=white] at (8.82,-2.6) {\scriptsize $t$};
\end{tikzpicture}
\caption{We plot the distance to 1-invariance, the difference between the first frame potential and the unitarily invariant frame potential, for the random 3-local model with $N=5,6,7$ and 8 spins and observe the decay to approximate 1-invariance at late times.}
\label{fig:1invH3L}
\end{figure}

\subsubsection*{Random Nonlocal 3-body}
The first spin model we consider is a system of $N$ all-to-all randomly coupled spins with 3-local interactions
\begin{equation}
H_\text{3-local} = \sum_{\substack{i,j,k \\ \alpha,\beta,\gamma}} J_{ijk,\alpha\beta\gamma}\, \sigma^\alpha_i \sigma^\beta_j \sigma^\gamma_k\,,
\label{eq:HR3L}
\end{equation}
where each $J_{ijk,\alpha\beta\gamma}$ is an i.i.d.\ Gaussian random variable with zero mean and variance $1/N^2$.  In the Hamiltonian, we sum $i,j,k$ over all possible 3-body interactions between all triplets of sites (\ie sum each over all $N$) and sum $\alpha,\beta,\gamma$ over $\{x,y,z\}$, \ie local Paulis at each site. The main feature of this Hamiltonian is that it consists of nonlocal (but still 3-local) random interactions which break all symmetries.

In Figure~\ref{fig:1invH3L}, we plot $\CF^{(1)}_{\CE_t} - \CF^{(1)}_{\widetilde{\CE}_t}$ for $H_\text{3-local}$ and with an increasing number of spins. For each $N = 5,6,7,8$ we numerically sampled 300 3-local Hamiltonians with random couplings as described above. The finite ensemble defined by this set of Hamiltonians with uniform weights was then used to compute the frame potential for different times. As per Eq.~\eqref{eq:FP1inv}, the invariant frame potential was computed by numerically evaluating the spectral form factor at different times for the same set of randomly sampled Hamiltonians.
At early times, we observe growth in the different of frame potentials and a peak at intermediate times with a magnitude which grows with $N$.  The reason that the distance to $1$-invariance is zero at $t = 0$ is because $\mathcal{E}_{t=0} = \{\textbf{1}\}$.  So then the distance to $1$-invariance must increase and then decrease.

At late times, the dynamics appear approximately unitarily invariant as the distance drops to zero. Note that on the log-log plot we take the absolute value of the difference in frame potentials, and so the late time fluctuations are actually around zero. The late time floor appears to be zero within the precision of our numerics for all values of $N$ we checked.  Therefore, the random 3-local system appears to achieve approximate invariance at late times.

\begin{figure}
\centering
\begin{tikzpicture}[scale=0.8,baseline=-0.4cm]
\node at (0,0) {\includegraphics[width=0.45\linewidth]{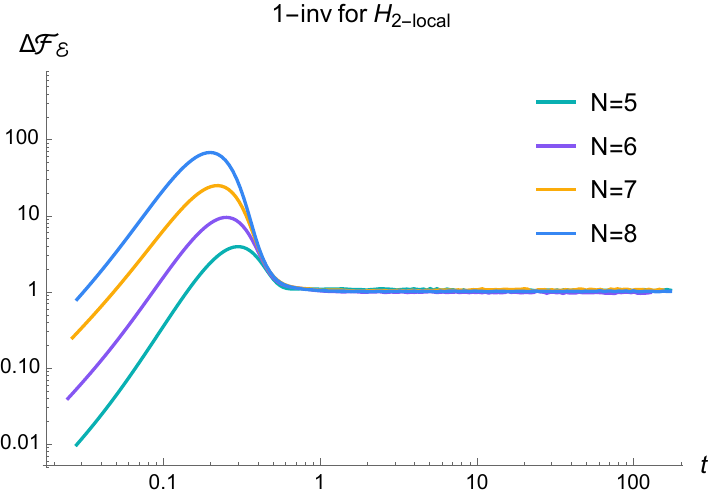} \quad \includegraphics[width=0.45\linewidth]{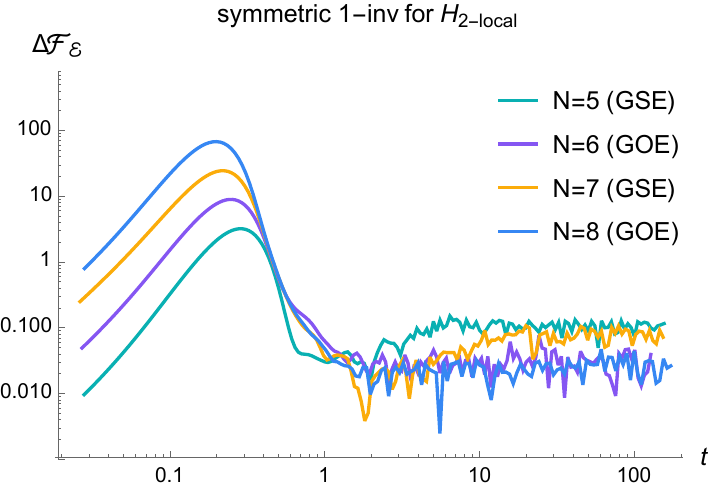}};
\node[fill=white] at (-8.5,2.55) {\scalebox{0.69}{$\CF^{(1)}_{\CE_t}-\CF^{(1)}_{\widetilde \CE_t}$}};
\node[fill=white] at (-0.32,-2.6) {\scriptsize $t$};
\node[fill=white] at (1.16,2.55) {\scalebox{0.69}{$\CF^{(1)}_{\CE_t}-\CF^{(1)}_{\widetilde \CE^{\rm sym}_t}$}};
\node[fill=white] at (8.82,-2.6) {\scriptsize $t$};
\end{tikzpicture}
\caption{We plot $\CF^{(1)}_{\CE_t} - \CF^{(1)}_{\widetilde{\CE}_t}$ and $\CF^{(1)}_{\CE_t} - \CF^{(1)}_{\widetilde{\CE}_t^{\rm sym}}$ for the random 2-local model with $N=5,6,7$ and $8$ spins, where the invariant ensemble $\widetilde{\CE}_t^{\rm sym}$ is taken to be either $\widetilde{\CE}_t^{O}$ or $\widetilde{\CE}_t^{Sp}$ depending on the symmetry class. We observe in the left plot that the system is not unitarily 1-invariant due to a non-zero late-time floor, but when accounting for symmetry as in the right plot, the system achieves approximate 1-invariance at late times.}
\label{fig:1invH2L}
\end{figure}

\subsubsection*{Random Nonlocal 2-body}
The next Pauli spin model we consider is a system of $N$ spins with all-to-all interactions via random couplings, summed over all possible 2-body Pauli interactions
\begin{equation}
H_\text{2-local} = \sum_{i,j,\alpha,\beta} J_{ij,\alpha\beta}\, \sigma^\alpha_i \sigma^\beta_j\,.
\label{eq:HR2L}
\end{equation}
Here, each $J_{ij,\alpha\beta}$ is an i.i.d.\ Gaussian random variable with zero mean and variance $1/N$, and we sum $i$ and $j$ over all pairs of sites and sum $\alpha$ and $\beta$ over local Paulis at each site. First, we note that $H_{\text{2-local}}$ realizes time-reversal symmetry and commutes with the antiunitary operator
\begin{equation}
\label{eq:Top1}
T = \prod_{j=1}^N (i\sigma^y_j) K\,,
\end{equation}
where $K$ is the antiunitary complex conjugation operator. As $T$ reverse the sign of single-site Paulis under conjugation, i.e., $T \sigma_i^\alpha T^{-1} = -\sigma_i^\alpha$, the 2-body Hamiltonian above commutes with $T$. Note that $T^2=1$ if the number of spins $N$ is even, and thus $H_\text{2-local}$ belongs to the GOE symmetry class.  Also, $T^2=-1$ if the number of spins $N$ is odd, meaning $H_\text{2-local}$ belongs to the GSE symmetry class.

Therefore, when computing $\CF^{(1)}_{\CE_t}-\CF^{(1)}_{\widetilde\CE_t}$, instead of using the expression for $\CF^{(1)}_{\widetilde\CE_t}$ assuming unitary invariance (see Eqn.~\eqref{eq:FP1inv}), we should instead expect orthogonal or symplectic invariance (depending on the number of spins $N$) and use Eqn.~\eqref{eq:OinvFP} or Eqn.~\eqref{eq:SpinvFP} accordingly.

In Figure~\ref{fig:1invH2L}, we plot the difference in frame potentials to show approximate unitary 1-invariance and symmetric 1-invariance at late times. As in the previous example, the difference in the frame potentials increases at early times, and with a magnitude that increases with $N$. In the left plot, we observe the presence of a late-time floor in the unitary invariance at a small but non-zero value, indicating a lack of 1-invariance. Once the symmetry is accounted for using the appropriate symmetrically invariant ensemble, in the right plot we find that the late time floor goes to zero and the ensembles display approximate orthogonal or symplectic invariance at late times.

We further note that a 4-local analog of $H_{2\text{-local}}$ is also time-reversal invariant, and appears to achieve orthogonal/symplectic $1$-invariance at late times better than the $2$-local case (see \cite{NHJthesis} for some numerics demonstrating this). We believe that this is because, in some appropriate sense, the $4$-local Hamiltonian is more ``chaotic'' than the $2$-local ensemble.

\subsubsection*{2-point functions and form factors}

Recall that there is an exact relation between the average of 2-point functions over the basis of Pauli strings and the 2-point spectral form factor \cite{ChaosRMT}
\begin{equation}
\frac{1}{d^2}\sum_{A\in P} \vev{A(t)A} = \frac{\CR_2(t)}{d^2}\,.
\end{equation}
This expression is true for any ensemble of Hamiltonians (even if that ensemble 
consists of a single Hamiltonian). We previously argued that approximate unitary 1-invariance implies that the variance of 2-point functions is small in Section \ref{sec:kinvcorr}, and that approximate 1-invariance for time-reversal symmetric systems implies a larger variance in 2-point correlators. 

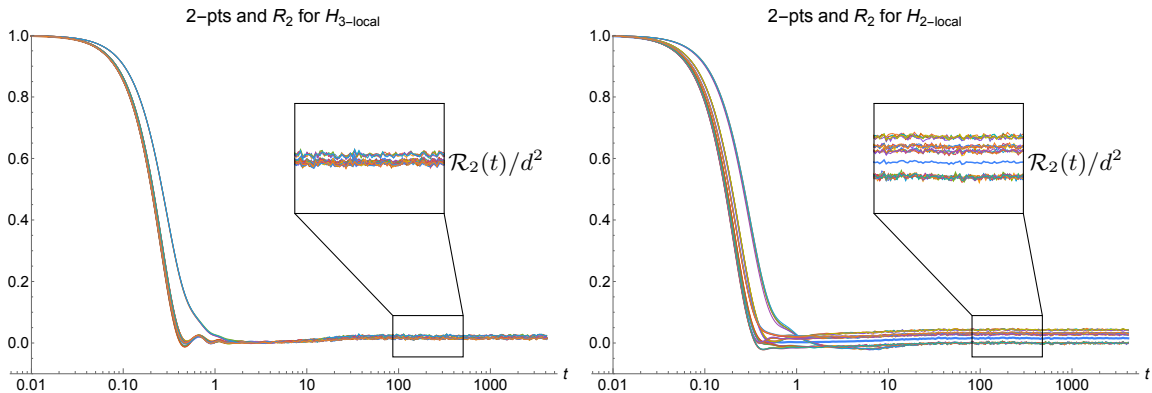
\begin{figure}
\centering
\begin{tikzpicture}[scale=0.55]
\node[anchor=east] at (0,0) {\includegraphics[width=0.48\linewidth]{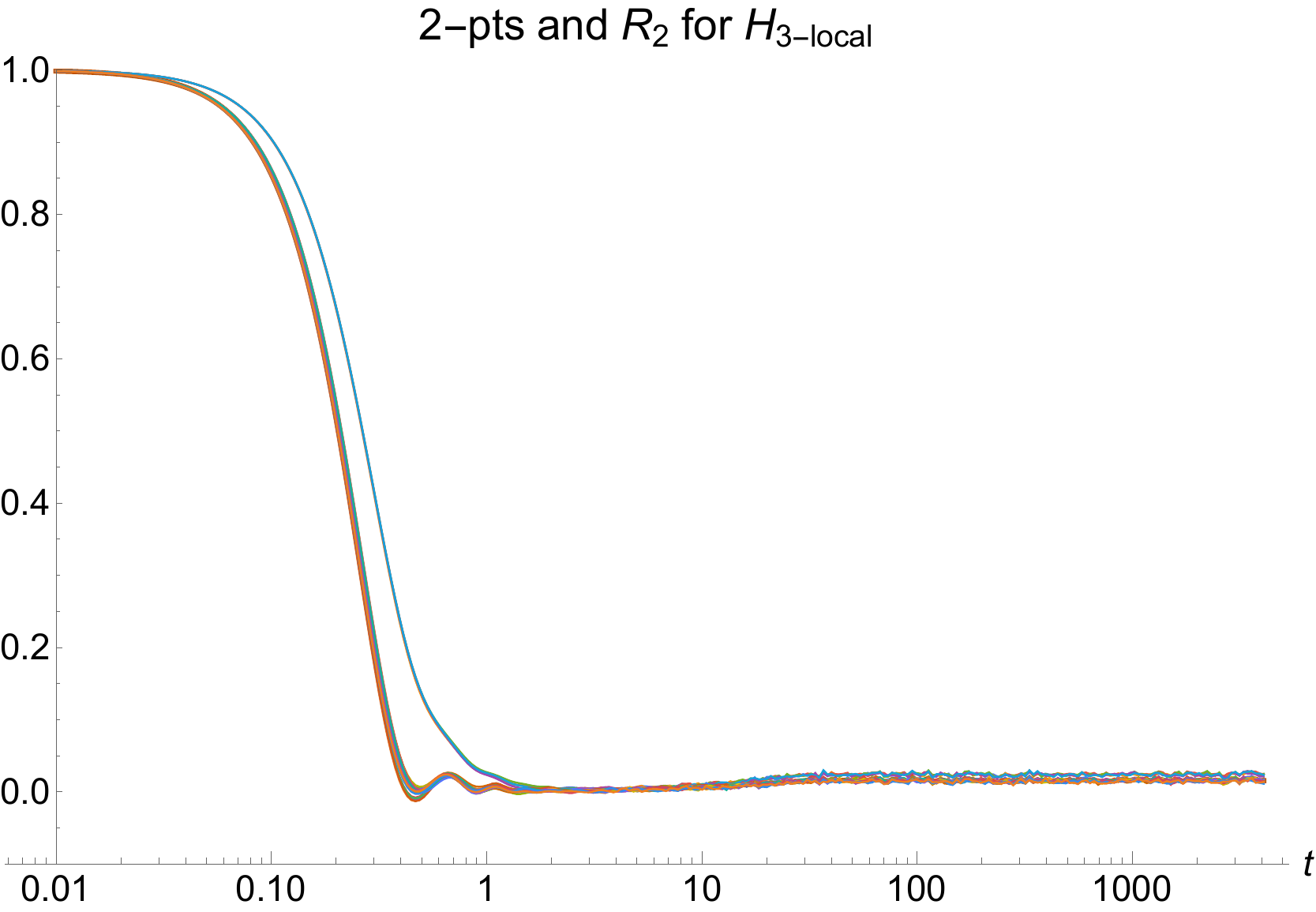}}; \node[anchor=west] at (0,0) {\includegraphics[width=0.48\linewidth]{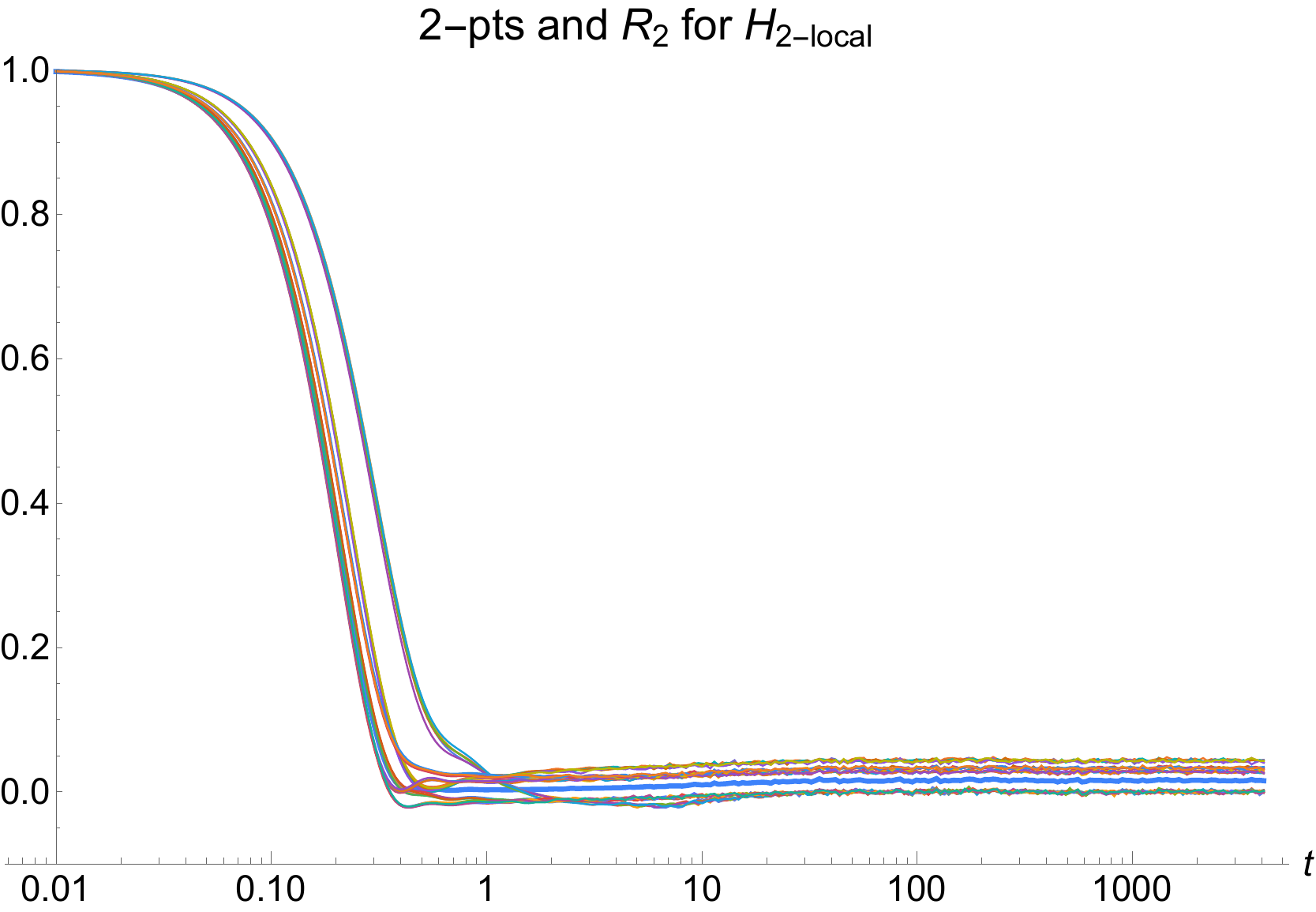}};
\node[anchor=east] at (-3,1) {\framebox[2cm]{\includegraphics[width=2cm]{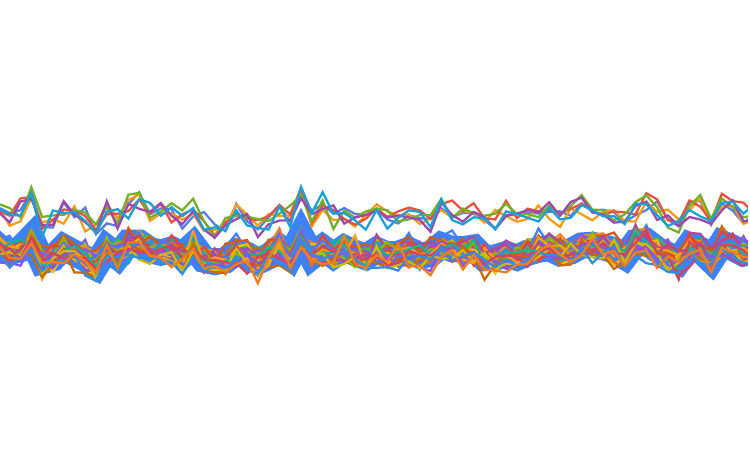}}};
\draw (-2.8,-2.8) rectangle (-4.5,-3.8);
\draw (-2.8,-2.8) -- (-3.26,-0.33);
\draw (-4.5,-2.8) -- (-6.87,-0.33);
\node[anchor=east] at (11,1) {\framebox[2cm]{\includegraphics[width=2cm]{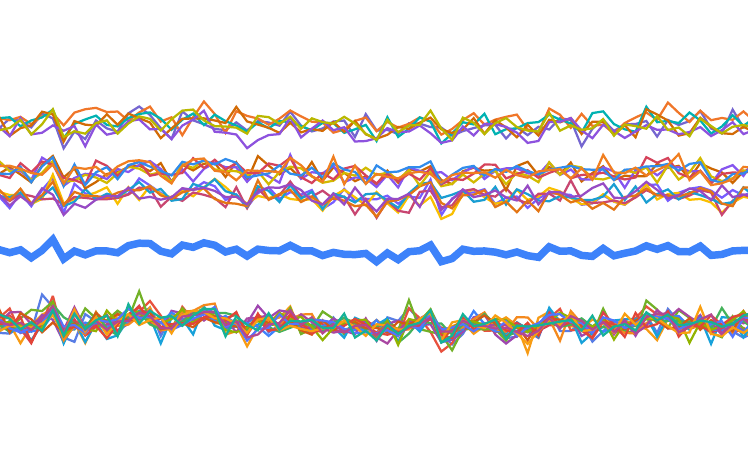}}};
\draw (11.2,-2.8) rectangle (9.5,-3.8);
\draw (11.2,-2.8) -- (10.74,-0.33);
\draw (9.5,-2.8) -- (7.13,-0.33);
\node at (-2,0.9) {{\footnotesize $\CR_2(t)/d^2$}};
\node at (12,0.9) {{\footnotesize $\CR_2(t)/d^2$}};
\end{tikzpicture}
\caption{We plot the 2-point functions of the form $\vev{A(t)A}$ for randomly chosen Pauli operators $A$ alongside the spectral form factor $\CR_2(t)$ for the random 2-local and 3-local Hamiltonians in Eqns.~\eqref{eq:HR2L} and \eqref{eq:HR3L}. The form factor, depicted by the thick blue line, exactly equals the average of all 2-point functions, but the variance of the 2-point functions is a lot smaller in the absence of time-reversal symmetry (on the left).}
\label{fig:2ptsR2}
\end{figure}

In Figure~\ref{fig:2ptsR2}, we plot 2-point functions of the form $\vev{A(t)A}$ for different Pauli operators $A$ alongside the form factor $\CR_2(t)$ for the random nonlocal 2-body and 3-body Hamiltonians described above. As there are $4^N$ Pauli operators, we randomly chose 8 Pauli operators of each weight (\ie from weight 1 up to weight $N$ Paulis), where the weight of a Pauli operator is the number of non-identity operators in the tensor product. For each plot in Figure~\ref{fig:2ptsR2}, we randomly sampled 300 random Hamiltonians with $N=8$ spins and computed 64 different 2-point functions $\vev{A(t)A}$, \ie 8 randomly chosen Pauli operators for weights 1 to 8, and where each 2-point function is averaged over the ensemble of Hamiltonians. 
 
We observe that the average of this subset of all 2-point functions is indistinguishably close to the form factor. Moreover, we see that late time correlators are tightly clustered around $\CR_2(t)$, but in the case of the 2-local system the correlators spread out around the form factor in bands of varying Pauli weights. 

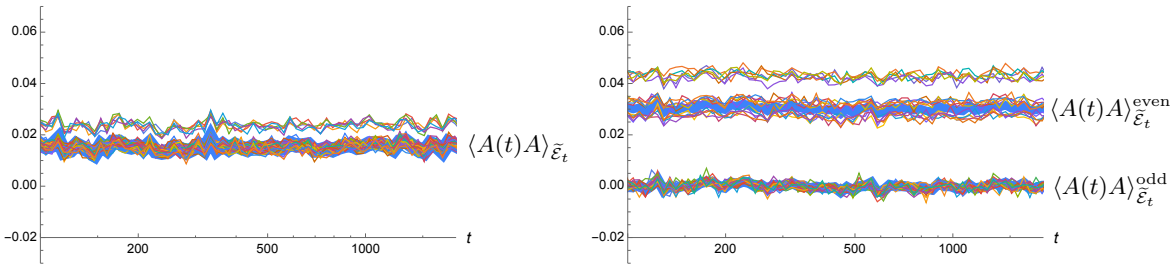
\begin{figure}
\centering
\begin{tikzpicture}[scale=0.55]
\node[anchor=east] at (-2,0) {\includegraphics[width=0.4\linewidth]{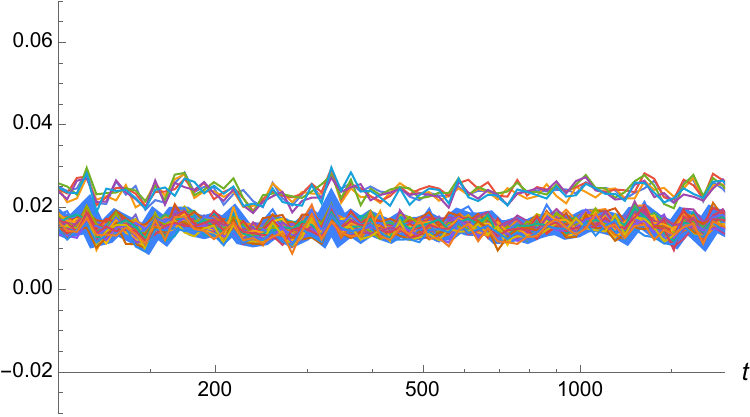}};
\node[anchor=west] at (0.4,0) {\includegraphics[width=0.4\linewidth]{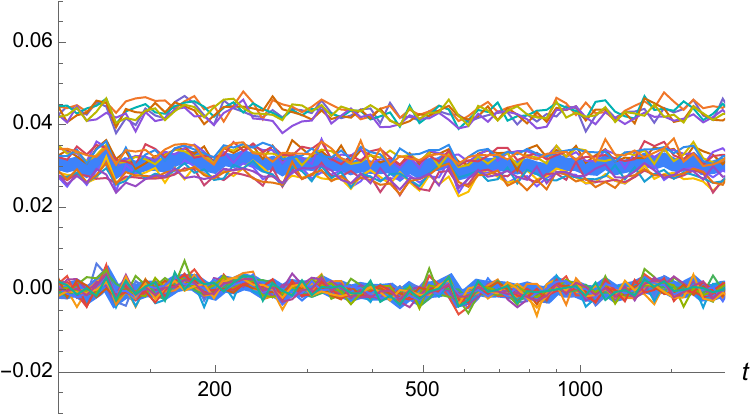}};
\node at (-1.1,-0.3) {{\footnotesize $\vev{A(t)A}_{\widetilde \CE_t}$}};
\node at (13.2,0.55) {{\footnotesize $\vev{A(t)A}_{\widetilde \CE_t}^{\rm even}$}};
\node at (13.2,-1.3) {{\footnotesize $\vev{A(t)A}_{\widetilde \CE_t}^{\rm odd}$}};
\end{tikzpicture}
\caption{We plot the late time floor of the 2-point functions $\vev{A(t)A}$ of varying weight Pauli operators on top of the invariant 2-point functions (thick blue lines) computed in terms of the spectral form factor $\CR_2(t)$ for the random 2-local and 3-local Hamiltonians.}
\label{fig:inv2ptsR2}
\end{figure}

One of the more interesting properties of $1$-invariance is it implies that 2-point functions $\langle A(t) A \rangle_{\mathcal{E}_t}$ are close to their spectrally decoupled versions $\langle A(t) A \rangle_{\widetilde{\mathcal{E}}_t} \approx (\CR_2(t)/d^2) \langle A A \rangle$.  For the 3-local system, which becomes unitarily invariant, all 2-point functions should become close to $\vev{A(t)A}_{\widetilde \CE_t} = \frac{\CR_2(t)-1}{d^2-1}$. For time-reversal symmetric systems, the invariant 2-point function depends on if the operator is even or odd, but again are expressions in terms of $\CR_2$ (given in Eqn.~\eqref{eq:Oinv2pt}). In Figure \ref{fig:inv2ptsR2}, we zoom in on the late time floor of the 2-point functions $\vev{A(t)A}$ for our 3-local and 2-local Hamiltonians, again for Pauli operators $A$ of all different weights (randomly sampling operators of each weight). Plotted on top of the 2-point functions are the invariant 2-point functions computed in terms of $\CR_2(t)$. The late time 2-point functions are mostly closely clustered around their respective invariant counterparts.

We note that in Figure \ref{fig:2ptsR2}, the correlation functions of single site Pauli operators decay slower than correlation functions of multi-site Paulis.  In the left panel of Figure~\ref{fig:inv2ptsR2}, for the 3-local system the band above the approximately $1$-invariant 2-point functions consists of 2-point functions of single site Paulis.  In the right panel of Figure~\ref{fig:inv2ptsR2}, for the 2-local system the top band consists of correlation functions of 2-site Pauli operators.  It would be interesting to better understand these features.

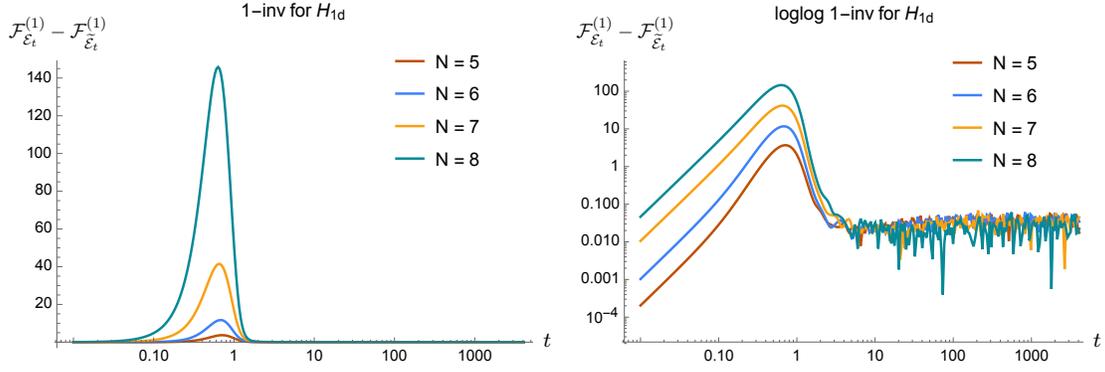
\begin{figure}
\centering
\begin{tikzpicture}[scale=0.8,baseline=-0.4cm]
\node at (0,0) {\includegraphics[width=0.45\linewidth]{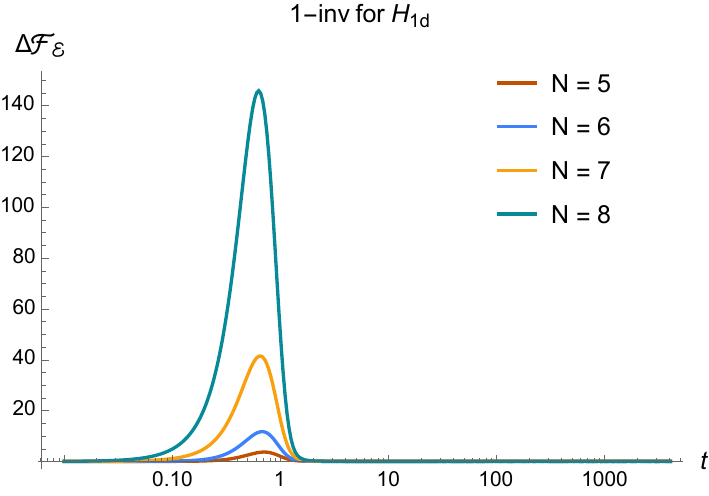}\quad \includegraphics[width=0.45\linewidth]{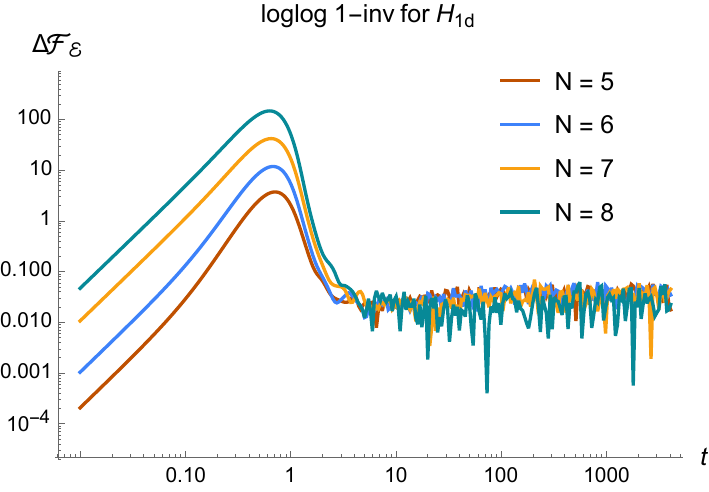}};
\node[fill=white] at (-8.45,2.5) {\scalebox{0.69}{$\CF^{(1)}_{\CE_t}-\CF^{(1)}_{\widetilde \CE_t}$}};
\node[fill=white] at (0.98,2.5) {\scalebox{0.69}{$\CF^{(1)}_{\CE_t}-\CF^{(1)}_{\widetilde \CE_t}$}};
\node[fill=white] at (-0.32,-2.6) {\scriptsize $t$};
\node[fill=white] at (8.82,-2.6) {\scriptsize $t$};
\end{tikzpicture}
\caption{Here we plot the unitary 1-invariance for the 1d local Hamiltonian in Eqn.~\eqref{eq:Hlocal} and observe a small value for the late time floor.}
\label{fig:kinvHloc}
\end{figure}

\subsubsection*{Geometrically local spin systems}
Above we considered disordered spin systems with random few-body interactions that are not geometrically local.  Here we consider more physical, geometrically local systems, namely a 1d system of $N$ Pauli spins with all allowed nearest-neighbor and next-to-nearest-neighbor interactions
\begin{equation}
H_{\rm 1d} = \sum_{i,\alpha, \beta} K_{i,\alpha \beta} \,\sigma_i^\alpha \sigma_{i+1}^\beta + \sum_{i,\alpha, \beta, \gamma} J_{i, \alpha \beta \gamma} \, \sigma_i^\alpha \sigma_{i+1}^\beta \sigma_{i+2}^\gamma\,.
\label{eq:Hlocal}
\end{equation}
We have permitted 3-body interactions in order to break time-reversal symmetry, opting to do so with 3-body terms instead of 1-body terms to avoid regimes where the system might many-body localize.  As usual, we take each $J_{i, \alpha \beta \gamma}$ and $K_{i, \alpha \beta}$ to be Gaussian random variables.  In Figure~\ref{fig:kinvHloc}, we generate an ensemble of 300 random Hamiltonians and plot $\CF^{(1)}_{\CE_t}-\CF^{(1)}_{\widetilde\CE_t}$ for $N=5,6,7,8$ spins, observing a decay to approximate 1-invariance at late times as the difference in frame potentials becomes small. We note that the floor value is small, but does not appear to approach zero up to numerical precision in contrast with the case of the geometrically local 3-body spin system discussed previously.

\subsection{Approximate symmetry in Pauli spin models}
\label{sec:ApproxPauliExamples}

Now we consider a system with ``approximate'' symmetry.  We return to Hamiltonians with non-local interactions for convenience.  Consider a family of Hamiltonians of the form
\begin{equation}
H_{\rm 1d}^{(\lambda)} = \lambda \sum_{i=1}^{N-1} \sum_{\alpha, \beta = 1}^3 K_{i, \alpha \beta} \,\sigma_i^\alpha \sigma_{i+1}^\beta + (1-\lambda) \sum_{i=1}^{N-2} \sum_{\alpha, \beta, \gamma = 1}^3 J_{i, \alpha \beta \gamma} \, \sigma_i^\alpha \sigma_{i+1}^\beta \sigma_{i+2}^\gamma 
\label{eq:Hlambda}
\end{equation}
with $\lambda$ taken between $0$ and $1$.  Here, $J_{i,\alpha \beta \gamma}$ are i.i.d.~Gaussian random variables with zero mean and variance $1/N^2$, and the $K_{i, \alpha \beta}$ are i.i.d.~Gaussian random variables with zero mean and variance $1/N$.  Notice that the 2-local terms in the Hamiltonian are time-reversal invariant with respect to the $T$ operator in Eqn.~\eqref{eq:Top1}. For $\lambda = 1$ our Hamiltonian is either GOE or GSE class, depending on the number of spins $N$.  On the other hand, for $\lambda = 0$, we are only left with the 3-local terms which do not possess any symmetry, and hence render the Hamiltonian GUE class.  Thus as $\lambda$ runs from $0$ to $1$, we can explore the effects of time-reversal symmetry breaking on approximate 1-invariance. 

\begin{figure}
\centering
\begin{tikzpicture}[scale=0.8,baseline=-0.4cm]
\node at (0,0) {\includegraphics[height=4.6cm]{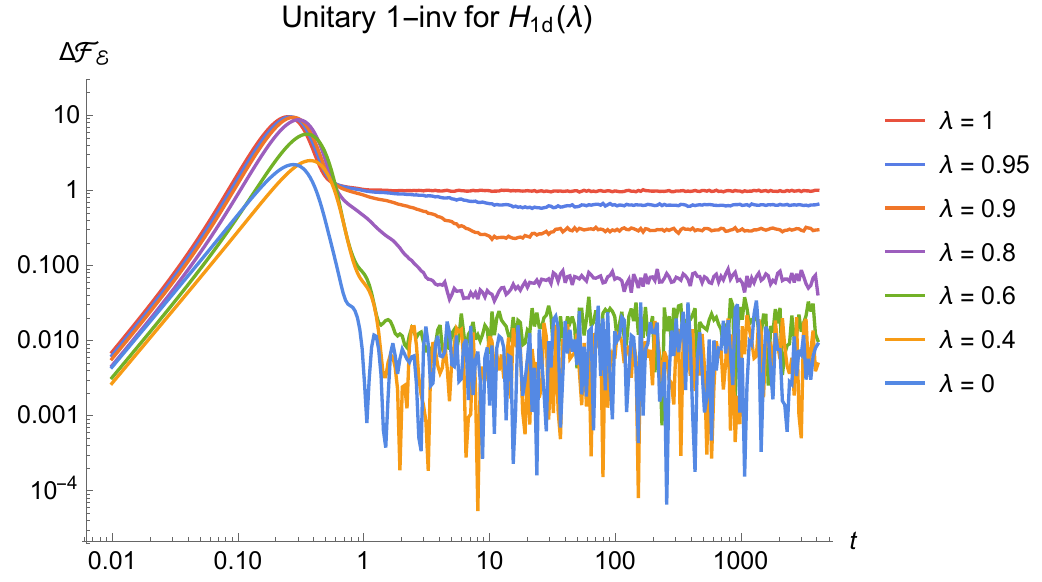}\includegraphics[height=4.6cm]{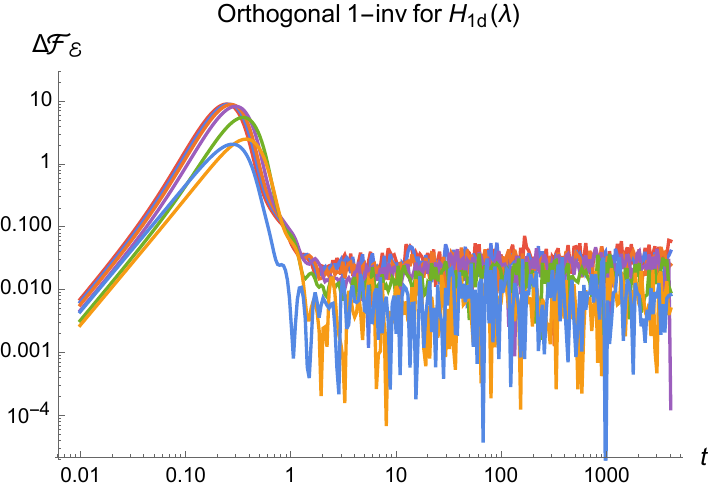}};
\node[fill=white] at (-8.45,2.5) {\scalebox{0.69}{$\CF^{(1)}_{\CE_t}-\CF^{(1)}_{\widetilde \CE_t}$}};
\node[fill=white] at (1.7,2.5) {\scalebox{0.69}{$\CF^{(1)}_{\CE_t}-\CF^{(1)}_{\widetilde \CE^O_t}$}};
\node[fill=white] at (-0.92,-2.45) {\scriptsize $t$};
\node[fill=white] at (9.12,-2.45) {\scriptsize $t$};
\end{tikzpicture}
\caption{Here we plot the approach to unitary and orthogonal 1-invariance for the local Hamiltonian in Eqn.~\eqref{eq:Hlambda}, tuning $\lambda$ to break time reversal symmetry.}
\label{fig:approxsym}
\end{figure}
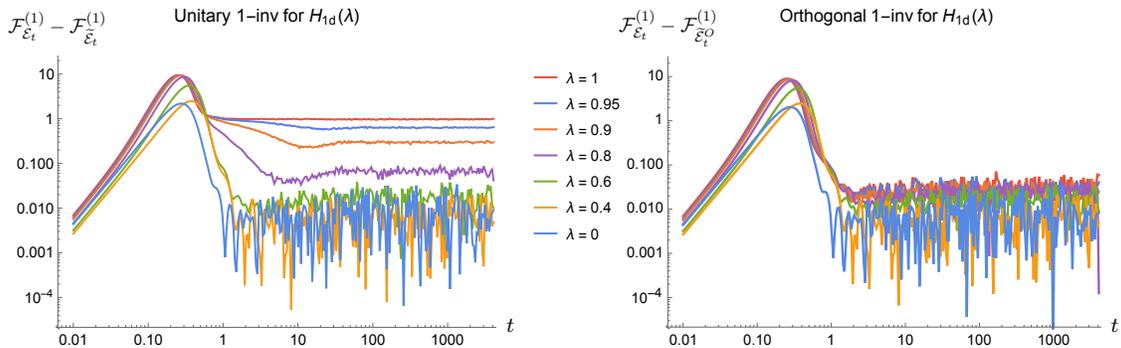

In Figure~\ref{fig:approxsym}, we plot $\mathcal{F}_{\mathcal{E}_t}^{(1)} - \mathcal{F}_{\widetilde{\mathcal{E}}_t}^{(1)}$ and $\mathcal{F}_{\mathcal{E}_t}^{(1)} - \mathcal{F}_{\widetilde{\mathcal{E}}_t^O}^{(1)}$ for $H_{\rm 1d}^{(\lambda)}$ with $N=6$ spins and an ensemble size of 300, tuning $\lambda$ from 0 to 1. We find that as we perturb away from the time-reversal symmetric point $\lambda=1$, the system becomes increasingly unitarily 1-invariant at late times. Approximate orthogonal 1-invariance is achieved for all values of $\lambda$. This is because for $\lambda=1$ the Hamiltonian is GOE class, but as we deform the Hamiltonian to break time-reversal symmetry the Hamiltonian becomes unitarily invariant, and unitary invariance is a sufficient condition for orthogonal invariance. Here orthogonal invariance in the time-reversal broken regime was computed using the full expression in Eqn.~\eqref{eq:OinvFPfull}, evaluated in a $T$-invariant basis. For details on how to construct such a basis, see Appendix~\ref{app:num}.

\label{sec:SYKexamples}
\subsection{SYK models}
We now turn to studying $k$-invariance in versions of the Sachdev-Ye-Kitaev model (SYK), a system of $N$ strongly-interacting all-to-all randomly coupled Majorana fermions $\chi_i$. The SYK model \cite{Kitaev15,SachdevYe, MS_SYK} has garnered much interest as a tractable model of quantum many-body chaos and captures features of low-dimensional black holes, exhibiting an emergent reparametrization invariance and maximal chaos \cite{Kitaev15,MS_SYK}. The Hamiltonian for the $q$-local SYK model is written as
\begin{equation}
H_{\rm SYK} = (i)^{q/2}\!\!\sum_{1 \leq i_1 < i_2 < \cdots < i_q \leq N} \!\! J_{i_1 i_2 \cdots i_q} \,\chi_{i_1}\chi_{i_2} \cdots \chi_{i_q}\,, 
\end{equation}
where the Majorana fermions are Hermitian, $\chi_i = \chi_i^\dagger$, and obey the anticommutation relations $\{\chi_i,\chi_j\} = \delta_{ij}$. We disorder average the Hamiltonian over i.i.d. Gaussian random couplings $J_{i_1 i_2 \cdots i_q}$ with zero mean and variance $\vev{J_{i_1 i_2 \cdots i_q}^2} = \frac{2^{q-1}}{q} \frac{J^2 (q-1)!}{N^{q-1}}$, where $J$ is a positive constant. The theory is solvable in the limit $1\ll \beta J\ll N$, where an emergent conformal symmetry allows one to compute correlation functions of the theory \cite{Kitaev15, MS_SYK}.

For purposes of numerical computations, we can represent the $N$ Majorana operators $\chi_i$ for $i=1,...,N$ in terms of Pauli strings, using the Jordan-Wigner transformation as an intermediate step.  If we have a Hilbert space $\mathcal{H} \simeq \bigotimes_{j=1}^{N/2} \mathbb{C}^2$ with $N/2$ sites $j=1,...,N/2$, we can then define the $N$ Majorana operators using the standard Clifford algebra representation as 
\begin{equation}
\chi_{2j} = \frac{1}{\sqrt{2}} \bigg( \prod_{i=1}^{j-1} \sigma_i^z\bigg) \sigma_{j}^y\,, \qquad \chi_{2j-1} = \frac{1}{\sqrt{2}} \bigg( \prod_{i=1}^{j-1} \sigma_i^z\bigg) \sigma_{j}^x\,,
\end{equation}
which are Hermitian and satisfy the desired anticommutation relation $\{\chi_i, \chi_j\} = \delta_{ij}$. 
Notice that each Majorana operator can be represented by a string of $N/2$ Pauli operators (including $2 \times 2$ identity matrices), and so can be expressed as a $N/2$ by $N/2$ complex Hermitian matrix.  Thus, the Hilbert space of $N$ Majoranas has dimension $d = 2^{N/2}$, although it can be regarded as comprising $N$ sites.

Relevant for our discussion is that SYK has a particle-hole symmetry $P$ which acts non-trivially on the spectrum \cite{You16,BHRMT16,KanazawaRMTSYK17}. Here, $P$ is an anti-linear operator which can be written as complex conjugation times a product of Majorana fermions, specifically
\begin{equation}
\label{eq:PHop1}
P =  2^{N/4} \, K \, \prod_{j=1}^{N/2} \chi_{2j-1}\,.
\end{equation}
$P$ commutes with the Hamiltonian and squares to $\pm 1$.  The $N$-dependence is \cite{BHRMT16}
\begin{equation}
    P^2 = \begin{cases}
    +1 \quad\text{if }N \equiv 0\,\,(\text{mod }8) \\
    +1 \quad\text{if }N \equiv 2\,\,(\text{mod }8) \\
    -1 \quad\text{if }N \equiv 4\,\,(\text{mod }8) \\
    -1 \quad\text{if }N \equiv 6\,\,(\text{mod }8)
    \end{cases}
\end{equation}
For $N$ Majorana fermions, the statistics of the spectrum of $H$ depends on $N\,\,(\text{mod }8)$, which can be understood by the action of $P$ on the charge sectors.

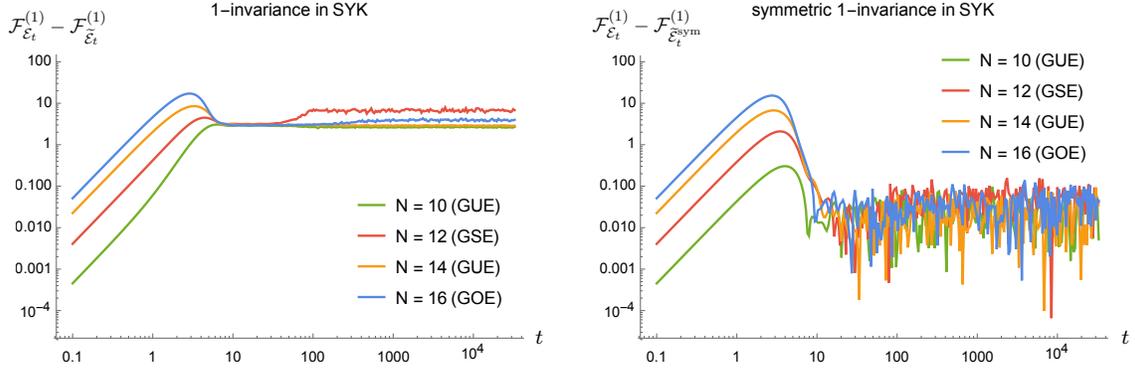
\begin{figure}
\centering
\begin{tikzpicture}[scale=0.8,baseline=-0.4cm]
\node at (0,0) {\includegraphics[width=0.45\linewidth]{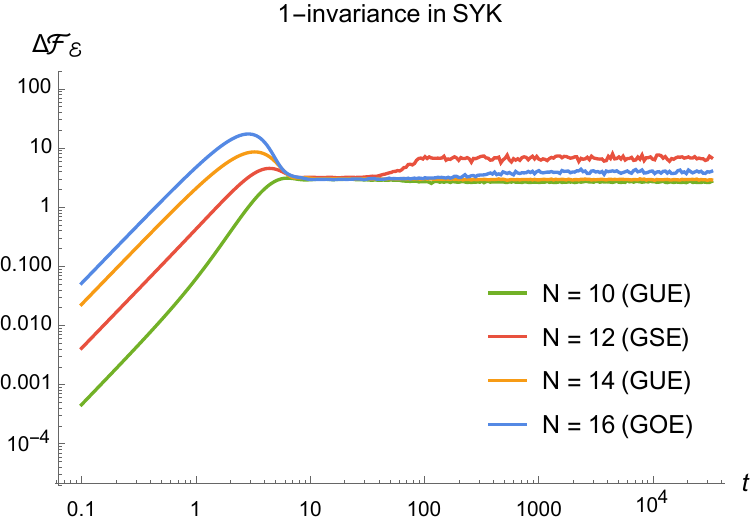}\qquad \includegraphics[width=0.45\linewidth]{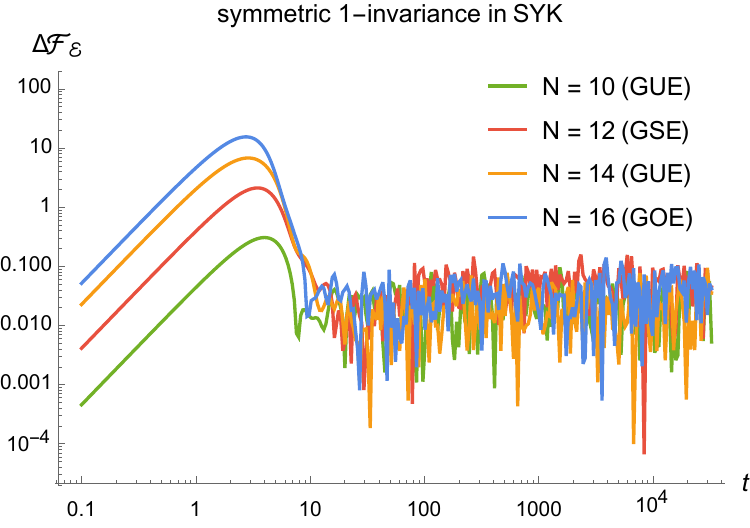}};
\node[fill=white] at (-8.52,2.6) {\scalebox{0.69}{$\CF^{(1)}_{\CE_t}-\CF^{(1)}_{\widetilde \CE_t}$}};
\node[fill=white] at (1.28,2.6) {\scalebox{0.69}{$\CF^{(1)}_{\CE_t}-\CF^{(1)}_{\widetilde \CE^{\rm sym}_t}$}};
\node[fill=white] at (-0.52,-2.55) {\scriptsize $t$};
\node[fill=white] at (9.22,-2.55) {\scriptsize $t$};
\end{tikzpicture}
\caption{Left: $\CF^{(1)}_{\CE_t} - \CF^{(1)}_{\widetilde \CE_t}$, the difference between the frame potential for the $q=4$ SYK model and its unitary-invariant ensemble at different number of Majoranas. Right: the difference between frame potentials where $\widetilde \CE^{\rm sym}_t$ is taken to be the invariant ensemble with respect to appropriate symmetry classes for the given $N$ and $q$.}
\label{fig:kinvsyk}
\end{figure}

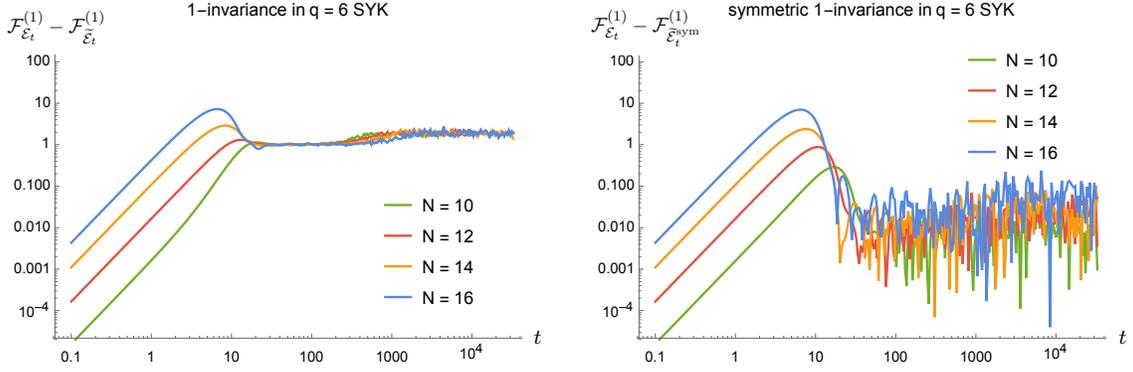
\begin{figure}
\centering
\begin{tikzpicture}[scale=0.8,baseline=-0.4cm]
\node at (0,0) {\includegraphics[width=0.45\linewidth]{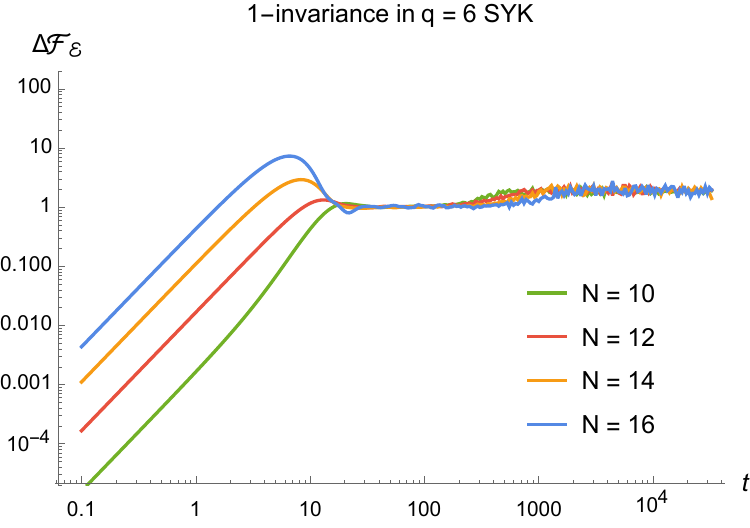}\qquad \includegraphics[width=0.45\linewidth]{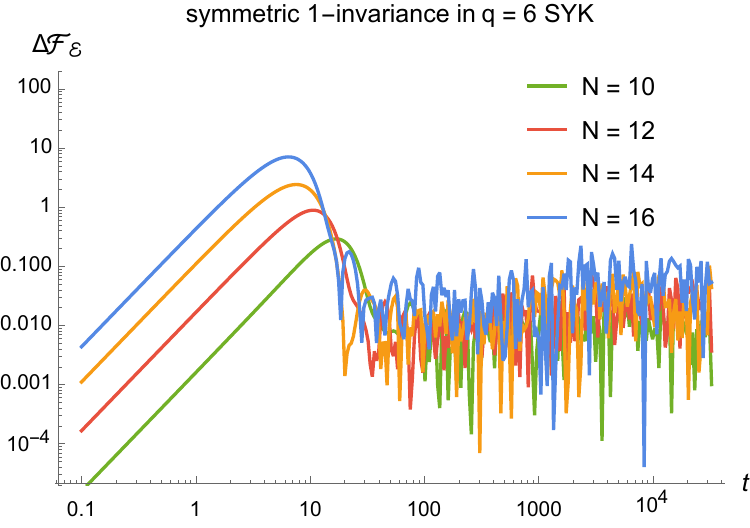}};
\node[fill=white] at (-8.52,2.6) {\scalebox{0.69}{$\CF^{(1)}_{\CE_t}-\CF^{(1)}_{\widetilde \CE_t}$}};
\node[fill=white] at (1.28,2.6) {\scalebox{0.69}{$\CF^{(1)}_{\CE_t}-\CF^{(1)}_{\widetilde \CE^{\rm sym}_t}$}};
\node[fill=white] at (-0.52,-2.55) {\scriptsize $t$};
\node[fill=white] at (9.2,-2.55) {\scriptsize $t$};
\end{tikzpicture}
\caption{Left: $\CF^{(1)}_{\CE_t} - \CF^{(1)}_{\widetilde \CE_t}$, the difference between the frame potential for the $q=6$ SYK model and its unitary-invariant ensemble at different values of $N$. Right: the difference between frame potentials where $\widetilde \CE^{\rm sym}_t$ is taken to be the invariant ensemble with respect to appropriate symmetry classes for the given $N$ and $q$.}
\label{fig:kinvq6syk}
\end{figure}

We understand that for time-reversal invariant systems, the ensemble $\mathcal{E}_t$ may become $k$-invariant with respect to orthogonal or symplectic transformations. For fermionic systems with different charge-parity sectors, the non-trivial block diagonal structure of the Hamiltonian again necessitates a refinement of our definition of $k$-invariance. In what follows, we analyze how to correctly account for the particle-hole symmetry that acts nontrivially on the SYK spectrum and determine the suitable version of symmetric $k$-invariance. By using the appropriate expressions for different values of $N$ and $q$, we then show numerically that SYK achieves approximate $1$-invariance in Figure~\ref{fig:kinvsyk} and Figure~\ref{fig:kinvq6syk}.

\subsubsection*{SYK with $N \equiv 2, 6 \,\, (\text{mod }8)$}
For $N \equiv 2\text{ or }6 \,\,(\text{mod }8)$, $P$ maps the even sector to the odd sector, i.e.~even eigenstates are mapped to odd eigenstates. Therefore two sectors have the same spectra spec$(H_+) = $ spec$(H_-)$, and so the entire spectrum is doubly degenerate. We also note that $P$ does not commute with either $H_+$ or $H_-$ and therefore regardless of whether $P^2=\pm 1$, there is no symmetry within the sectors. This means each sector should have GUE statistics and become unitarily invariant.

The SYK Hamiltonian in this case has a block diagonal form $H = {\rm diag}(H_+,H_-)$, where $H_+=H_-^*$ and so the diagonalizing unitaries $U_+$ and $U_-$ should not be taken as independent. We must take $U_+=U_-^*$ and then the expression for the symmetric 1-invariance in Eqn.~\eqref{eq:symFP} becomes
\begin{align}
\CF_{\widetilde\CE_t}^{(1)} = \int_{U(d/2)} dU \, \int_{\CE_H} \!\!dH_{1}\, dH_{2} \, \Big( 4 \,|\Tr(U e^{i \widetilde{H}_{1} t} U^\dagger e^{-i \widetilde{H}_{2} t}) |^2 \Big)
\end{align}
where since the spectra of $H_+$ and $H_-$ are the same, we use $\widetilde H$ to denote a diagonalized single sector of the Hamiltonian. Integrating, we find
\begin{equation}
\CF_{\widetilde\CE_t}^{(1)} = \frac{4}{{\widetilde d}^2-1} \big( \widetilde \CR_2(t)^2 + \widetilde d\,{}^2 - 2 \widetilde\CR_2(t)\big)
\end{equation}
where $\widetilde d = d/2$ is the dimension of each sector, and the form factor for each sector is simply
\begin{equation}
\widetilde\CR_2(t) = \int D\lambda\sum_{i,j\in \widetilde H} e^{i(\lambda_i-\lambda_j)t}\,,
\end{equation}
where we sum over only the eigenvalues in a sector, \ie only the unique eigenvalues in the spectrum. Note that this is related to the na\"{i}ve spectral form factor $\CR_2(t)$ computed from directly the Hamiltonian by a factor of four, $\CR_2(t) = 4\widetilde\CR_2(t)$. We can write the Haar'ed frame potential for SYK in terms of the full form factor as
\begin{equation}
\CF_{\widetilde\CE_t}^{(1)} = \frac{1}{d^2-4} \big(\CR_2(t)^2 + 4d^2 - 8 \CR_2(t)\big)\,.
\end{equation}

\subsubsection*{SYK with $N \equiv 0\,\,(\text{mod }8)$}
In the case of $N \equiv 0 \,\,(\text{mod }8)$, we have that the particle-hole operator maps the even sector to itself and the odd sector to itself. As $P^2=1$, the operator maps an eigenstate to itself and thus the entire spectrum of the Hamiltonian has no degeneracy. Moreover, the particle-hole operator is a symmetry of each sector and commutes with $H_+$ and $H_-$. As $P^2=1$, we thus expect each sector to have GOE statistics and become orthogonally invariant.

To consider this case, we want to compute the Haar-invariant frame potential where the symmetry of the ensemble is a block diagonal orthogonal matrix of the form
\begin{equation}
    \begin{pmatrix}
    O_+ & \textbf{0} \\ \textbf{0} & O_-
    \end{pmatrix}\,.
\end{equation}
Moreover, the blocks $H_+$ and $H_-$ are independent, so we need to integrate over Haar random $O_+$ and $O_-$ taken independently. The first two terms in Eqn.~\eqref{eq:symFP} for Haar random orthogonal matrices $O_+$ and $O_-$, instead of Haar unitary matrices $U_+$ and $U_-$, give our standard orthogonally-invariant expressions for each sector plus an interaction term:
\begin{equation}
\CF_{\widetilde\CE_t}^{(1)} = \frac{(d_+ + 1) \CR_{2,+}^2 + 2d_+^3 - 4d \CR_{2,+}}{d_+(d_+ +2)(d_+ -1)} + \frac{(d_- +1) \CR_{2,-}^2 + 2d_-^3 - 4d_- \CR_{2,-}}{d_-(d_- +2)(d_- -1)} + \CF^{(1)}_{\rm int}\,.
\end{equation}
Now to compute the interaction term between the two sectors, we simply need to compute the coupled terms of the form
\begin{equation}
\int dO_+ dO_- \Tr(O_+ e^{i\Lambda_{1,+}t} O_+^T e^{-i\Lambda_{2,+}t})\Tr(O_- e^{-i\Lambda_{1,-}t} O_-^T e^{i\Lambda_{2,-}t})\,,
\end{equation}
and using the first orthogonal moment, we obtain
\begin{equation}
\CF^{(1)}_{\rm int} = \frac{2}{d_+ d_-} \big| \big\langle \Tr(e^{iH_+ t}) \Tr(e^{-iH_- t}) \big\rangle_{\CE_H} \big|^2\,.
\end{equation}
In summary, for the SYK model with charge sectors with GOE statistics, we have the invariant frame potential with which to define symmetric 1-invariance
\begin{equation}
\CF_{\widetilde\CE_t}^{(1)} = \CF^{(1)}_{\CE^O_{t,+}} + \CF^{(1)}_{\widetilde{\CE}^O_{t,-}} + \frac{2}{d_+ d_-} \big| \big\langle \Tr(e^{iH_+ t}) \Tr(e^{-iH_- t}) \big\rangle_{\CE_H} \big|^2\,.
\end{equation}

\subsubsection*{SYK with $N \equiv 4\,\,(\text{mod }8)$}
In the case of SYK with $N \equiv 4\,\,(\text{mod }8)$, again the particle-hole operator acts by mapping each charge-parity sector to itself. But as $P^2=-1$, the operator cannot take an eigenstate to itself and thus must map each eigenstate within a sector to another eigenstate within the sector, meaning each charge-parity sector must be doubly degenerate. Furthermore, as $P$ will commute with $H_+$ and $H_-$, and squares to $-1$, each charge sector should have GSE statistics and achieve symplectic invariance at late times.

As the first symplectic moment is the same as the other Haar moments, we simply find the for the SYK model with charge sectors with GSE statistics, we have the invariant frame potential with which to define symmetric 1-invariance
\begin{equation}
\CF_{\widetilde\CE_t}^{(1)} = \CF^{(1)}_{\CE^{Sp}_{t,+}} + \CF^{(1)}_{\CE^{Sp}_{t,-}} + \frac{2}{d_+ d_-} \big| \big\langle \Tr(e^{iH_+ t}) \Tr(e^{-iH_- t}) \big\rangle_{\CE_H} \big|^2\,.
\end{equation}

\subsubsection*{$k$-invariance for $q \equiv 2\,\,(\text{mod }4)$ SYK}
Now we turn to considering $k$-invariance in SYK with $q$-body interactions with $q \equiv 2\,\,(\text{mod }4)$. In this case, the particle-hole operator anticommutes with the Hamiltonian and thus the operator $P$ maps each eigenstate to another eigenstate with the opposite signed energy \cite{KanazawaRMTSYK17}. The Hamiltonian can again be block-diagonalized into charge-parity sectors, where the statistics of the sectors depends on $N$. 

\subsubsection*{SYK with $q \equiv 2 \,\,(\text{mod }4)$ and $N \equiv 0, 4\,\,(\text{mod }8)$}
In these cases the Hamiltonian is block diagonal, and each block is of Bogoliubov-de Gennes (BdG) type \cite{KanazawaRMTSYK17}.  
For the $N \equiv 0 \,\,(\text{mod }8)$ case, $P^2=1$ and is a bosonic operator, which means that the Hamiltonian in each sector is pure imaginary and skew symmetric. This is the extended ensemble BdG class D, meaning the Hamiltonian in each block lives in the tangent space of $O(d)$. There is no special invariance class for these ensembles, and thus we simply achieve unitary 1-invariance in each sector independently.

Similarly, for the $N \equiv 4\,\,(\text{mod }8)$ case, $P^2=-1$ and each sector of the Hamiltonian belongs to BdG class C, meaning that each sector of the Hamiltonian lives in the tangent space of $Sp(d)$. Again, there is no special invariance group and the invariant ensemble is simply given by two independent unitarily invariant sectors.  The corresponding frame potential is:
\begin{align}
\CF_{\widetilde\CE_t}^{(1)} = \frac{\CR_{2,+}^2 + d_+^2 -2 \CR_{2,+}}{d_+^2-1} + \frac{\CR_{2,-}^2 + d_-^2 -2 \CR_{2,-}}{d_-^2-1} + \frac{2}{d_+ d_-} \big| \big\langle \Tr(e^{iH_+ t}) \Tr(e^{-iH_- t}) \big\rangle_{\CE_H} \big|^2\,.
\end{align}

\subsubsection*{SYK with $q \equiv 2 \,\,(\text{mod }4)$ and $N \equiv 2,6 \,\,(\text{mod }8)$}
In this case the particle-hole operator exchanges the two blocks $H_+$ and $H_-$, relating the blocks to one another.  However, the blocks are related with inverted spectra, $H_+ = -H_-^*$, where each block has GUE statistics. As the diagonalizing unitaries for each block are not independent, i.e.~we have $U_+ = U_-^*$, the expression for the 1-invariance can be computed as
\begin{align}
\CF_{\widetilde\CE_t}^{(1)} = \frac{2}{\widetilde d^2-1} \big( \widetilde \CR_2^2 + \widetilde d\,{}^2-2 \widetilde \CR_2\big) + \CF^{(1)}_{\rm int}\,,
\end{align}
where we find a contribution from the $+$ and $-$ sectors, but the interaction term must be computed with identical $U_+$ and $U_-$. Computing this interaction term with the oppositely signed sectors, we find
\begin{equation}
\CF^{(1)}_{\rm int} = \frac{2}{\widetilde d^2 -1} \big( \CR_{2'} \CR_{2'}^* + \CR_{1'} \CR_{1'}^* - \frac{1}{d} \CR_{2'} \CR_{1'}^* - \frac{1}{d} \CR_{2'}^* \CR_{1'} \big)
\end{equation}
where
\begin{equation}
\begin{split}
\CR_{2'} &:=  \big\langle \Tr(e^{iHt})\Tr(e^{iHt}) \big\rangle_{\CE_H} = \int D\lambda \, \sum_{i,j} e^{i(\lambda_i+\lambda_j) t}\,,\\ \CR_{1'} &:= \big\langle \Tr(e^{2iHt})\big\rangle_{\CE_H} = \int D\lambda \, \sum_{j} e^{2 i \lambda_j t}\,.
\end{split}
\end{equation}

\subsection{\ktitle-invariance and self-averaging}

In this paper, we have focused on properties of correlation functions and spectral correlators for ensembles of Hamiltonians.  It is natural to ask how our results interface with ``self-averaging," that is, does our analysis of $k$-invariance for ensembles have anything to say about the dynamical onset of scrambling for a single Hamiltonian? More specifically, if we have an ensemble $\CE_H$ of Hamiltonians whose unitary evolution becomes (approximately) $k$-invariant after some time, then will a single randomly sampled Hamiltonian from $\CE_H$ behave similarly?

Our main diagnostic of $k$-invariance is the $k$-th frame potential. For the ensemble $\CE_H$ consisting of a single Hamiltonian (with unit probability), the frame potential of $\CE_t = \{ e^{-iHt}\}$ is simply computed to be $\CF_{\CE_t}^{(k)} = |\tr(e^{iH t} e^{- i H t})|^{2k} = d^{2k}$. Therefore, the dynamics induced by a single Hamiltonian will not become $k$-invariant for any $k$. While this conclusion is correct, it may be somewhat misguided.  Consider, for instance, the spectral form factor $\CR_2(t) = \vev{|\tr(e^{-iHt})|^2}_{\CE_H}$.  It is well known that the spectral form factor is not self-averaging~\cite{prange1997spectral} after $\op(1)$ times. Moreover, the connected piece of the 2-point spectral form factor is trivially zero for a single Hamiltonian. Nevertheless, one may recover the dip, ramp, and plateau features of the form factor for a single Hamiltonian by averaging over a moving time window \cite{BHRMT16}. 

Turning back to $k$-invariance, one can imagine ``smearing out'' a single Hamiltonian in time so that the frame potential  $\CF_{\CE_t}^{(k)}$ is non-trivial, \ie generating an ensemble of unitary time-evolutions $e^{-iHt}$ for a single $H$ by randomly sampling times in a small window around a fixed time. We anticipate that with such a finite time resolution, then in an appropriate sense the evolution of a single Hamiltonian can manifest $k$-invariance, but we leave a detailed analysis for future work.

\subsection{Comments on \ktitle-invariance in JT gravity}

It is difficult to analytically probe $k$-invariance due to the late timescales involved.  In particular, there are few theories that enable precise, late time computations, for which the results are non-trivial.  One notable exception is Jackiw-Teitelboim (JT) gravity \cite{jackiw1985lower, teitelboim1983gravitation, jackiw1992gauge}.  We consider the version of JT gravity which includes the contributions of higher-genus spacetimes.  Remarkably, this theory of quantum gravity in nearly-AdS$_2$ spacetimes admits a non-perturbative completion in terms of a matrix integral \cite{SaadJT19}.  This means that the completion can be regarded as a theory of disorder-averaged Hamiltonians.  There is a similar story for JT gravity in nearly-dS$_2$ \cite{deSitterQG19} (see also \cite{maldacena2019two}).  We will presently focus on the case of nearly-AdS$_2$ gravity, for which $\Lambda = -1$.

Here, we briefly recount the relation between JT gravity and a disordered ensemble, and then explain the connection with $k$-invariance.  The action for JT gravity on a surface $\mathcal{M}$ with boundaries $\partial \mathcal{M}$ can be written as
\begin{equation}
I[g_{\mu \nu},\phi] = - S_0 \,\chi(\mathcal{M}) - \frac{1}{16 \pi G_N}\int_{\mathcal{M}} d^2 x \, \sqrt{g} \, \phi \,(R + 2\Lambda) - \frac{1}{8 \pi G_N}\int_{\partial \mathcal{M}} dx\,\sqrt{h} \, \phi \, (K-1)\,.
\end{equation}
Here, $g_{\mu \nu}$ is the metric, $\phi$ is a non-dynamical dilaton field, $\chi(\mathcal{M})$ is the Euler characteristic of $\mathcal{M}$, $G_N$ is Newton's constant, $R$ is the Ricci scalar, $\Lambda$ is the cosmological constant, $h$ is the induced metric on $\partial \mathcal{M}$, and $K$ is the induced scalar curvature on $\partial \mathcal{M}$.  For simplicity, we take $S_0 \sim 1/G_N$.  The amplitude of a spacetime with boundary $\partial \mathcal{M}$ is given by the path integral
\begin{equation}
\label{E:amplitude1}
\mathcal{A}(\partial \mathcal{M}) = \sum_{\text{topologies}(\partial \mathcal{M})}\int \mathcal{D}g_{\mu \nu} \, \mathcal{D}\phi \, e^{- I[g_{\mu \nu}, \phi]}\,,
\end{equation}
where the sum over topologies entails summing over all surfaces with boundary $\partial \mathcal{M}$.  Since we are in two dimensions, $\partial \mathcal{M}$ is a union of circles, which are each parameterized by a (renormalized) length.  If there are $n$ such circles so that
$$\partial \mathcal{M} \simeq \underbrace{S^1 \times \cdots \times S^1}_{n}\,,$$
then we can regard $\mathcal{A}(\partial \mathcal{M})$ as a function of $n$ lengths, which we denote by $\beta_1,...,\beta_n$.  Thus we write $\mathcal{A}(\beta_1,...,\beta_n)$.  This amplitude can be written as an asymptotic series in the parameter $e^{- S_0}$, which is in fact a genus expansion of the spacetime(s) ending at the $n$ boundaries.  Explicit details on all of these matters are given in \cite{SaadJT19} for the AdS$_2$ case (and in \cite{deSitterQG19} for the dS$_2$ case; see also \cite{maldacena2019two}).

JT gravity on higher genus topologies can be expressed as a type of matrix integral.  In particular,
\begin{equation}
\label{E:MMeq1}
\mathcal{A}(\beta_1,...,\beta_n) \sim \lim_{d \to \infty} \frac{1}{\mathcal{Z}}\int dH \, e^{- d \, V(H, d)} \, \text{tr}(e^{- \beta_1 H}) \cdots \text{tr}(e^{- \beta_n H})\,,
\end{equation}
where we are integrating over $d \times d$ Hermitian matrices\footnote{In \cite{SaadJT19} and related works, $L$ is often used instead of $d$.} (here, $d$ can be regarded as the dimension of the Hilbert space) for a particular potential $V$, and taking a ``double scaling limit'' where $d \to \infty$.  Note that the potential $V$ also contains an explicit dependence on $d$ (i.e., it does not just depend on $d$ via the matrix $H$), which enables the double scaling limit to be non-trivial.  Also, we have used ``$\sim$'' instead of ``$=$'' in Eqn.~\eqref{E:MMeq1} to denote that the matrix integral agrees with the asymptotic series expansion in $e^{- S_0}$ given by the JT path integral above, but contains additional non-perturbative corrections in $\sim e^{-e^{S_0}}$.  It is intriguing that in JT gravity, the amplitudes for spacetimes are the same as spectral form factors in a certain random matrix ensemble.

Since we are in the double scaling limit, the Hamiltonians comprising the ensemble are formally infinite-dimensional, although each has discrete level spacings.  Nonetheless, amplitudes $\mathcal{A}$ still have an asymptotic expansion in $e^{- S_0}$, which essentially plays the role of $1/d$ as per our other analyses in this paper.  For instance, we will have ``large $e^{S_0}$ factorization'' like
\begin{equation}
\langle Z(\beta_1) \, Z(\beta_2) \rangle = \langle Z(\beta_1) \rangle \langle Z(\beta_2) \rangle \left(1 + O(e^{- S_0}) \right),
\end{equation}
where the averages are taken with respect to the matrix integral.

Now we consider coupling the gravitational theory to matter fields.  At present, it is not clear how to introduce propagating matter fields into the matrix integral in Eqn.~\eqref{E:MMeq1}, but one can more modestly introduce probe matter fields in the path integral in Eqn.~\eqref{E:amplitude1} and compute terms in the $e^{-S_0}$ expansion.  This was carried out in \cite{saad2019latetime} for 1+1D conformal matter with conformal dimension $\Delta$ (see \cite{MertensJTclocks} for earlier work).  For times $t \gg S_0$, it was found that
\begin{equation}
\label{E:OO1}
\langle \mathcal{O}_{\Delta}(t)\, \mathcal{O}_{\Delta}(0)\rangle_{\beta} \sim e^{- S_0}\, f(\Delta) \, [\text{spectral quantity}](\beta, t)\,,
\end{equation}
where the left-hand side is taken with respect to the gravitational path integral at finite temperature $\beta$, and $[\text{spectral quantity}](\beta, t)$ is a purely spectral quantity depending only on $t$ and $\beta$.  This exemplifies spectral decoupling, analogous to $1$-invariance.

In particular, the time-dependence is contained purely in the $[\text{spectral quantity}](\beta, t)$ term, and the data of the operator $\mathcal{O}_\Delta$ (packaged as a function $f(\Delta)$ of the conformal dimension $\Delta$) factors out.  One can check that $\langle \mathcal{O}_{\Delta}(t) \mathcal{O}_{\Delta}(0)\rangle_{\beta}$ does not factorize in this way at early times, and so the form of Eqn.~\eqref{E:OO1} is indeed novel.  The calculational techniques at hand are not sharp enough to determine the precise timescale of the onset of this (approximate) $1$-invariance-like behavior -- we only know that $t_\text{1-inv} \leq O(S_0)$ \cite{MertensJTclocks,saad2019latetime}, which also may hold in related supersymmetric models \cite{mertens2017solving}.

There are some important differences with 1-invariance, however.  The first is that our formulas for $1$-invariance entail  $\text{tr}( \mathcal{O}_{\Delta}(0) \,\mathcal{O}_{\Delta}(0) )$ which is not well-defined for 1+1D conformal matter, and so our prediction for the form of a 2-point function approximately satisfying 1-invariance (after a specified timescale) cannot literally hold.  Relatedly, our derivation of the form of a 2-point function approximately satisfying 1-invariance is no longer valid, since the derivation involves manipulating traces of a product of operators which are not trace class.  Nonetheless, if nearly-AdS$_2$ JT gravity coupled to conformal matter turns out to have a non-perturbative completion in terms of a (multi-)matrix integral, possibly with additional fields, then the 1+1D conformal matter theory may become resolved in some microscopic manner.  In such a scenario, our $k$-invariance formulas may be more precisely realized.

Even in the absence of speculation, Eqn.~\eqref{E:OO1} is already rather suggestive.  The equation advances our point of view that the late time dynamics of correlation functions in highly chaotic many-body theories (such a gravity) are captured by purely spectral quantities.  In our quantum gravity example here, it appears that the fluctuations of spacetime dominate the late time dynamics of matter correlators, since the $[\text{spectral quantity}](\beta, t)$ in Eqn.~\eqref{E:OO1} only knows about the spectrum of the pure gravitational theory.  It would be interesting to understand spectral decoupling in variations of JT gravity with other symmetries, as in \cite{stanford2019JTRMT}, and then find analogs with symmetric $k$-invariance.

Furthermore, it would be appealing if other theories of gravity, beyond JT in nearly-AdS$_2$ and nearly-dS$_2$, can be expressed as disorder averaged theories.  Then perhaps there is some appropriate analog of $k$-invariance in such theories.

\section{Discussion}
\label{sec:Discussion}

We have quantified spectral decoupling by a detailed analysis of $k$-invariance, and its refinement to systems with symmetry.  Systems with quenched disorder which undergo approximate spectral decoupling adopt universal dynamics, captured by the joint eigenvalue statistics of the Hamiltonian ensemble.  In such systems, the late time physics of OTOCs, Keldysh ordered correlation functions, and operator growth are captured by appropriate spectral forms factors.  Thus spectral decoupling provides a bridge between recent diagnostics of quantum many-body chaos in terms of operators and correlation functions, and more traditional diagnostics in terms of random matrix eigenvalue statistics.

Our numerical and analytic evidence suggests that chaotic quantum many-body systems with non-local $q$-body interactions and quenched disorder become approximately $k$-invariant at late times, for $k$ small relative to the number of sites $N$.  The timescale of the onset of approximate $k$-invariance depends on the system, choice of norm, and value of $k$, but appears to be at most $\text{poly}(N)$.  For random quantum circuits, the timescale of $k$-invariance can be quantified more precisely.  Random circuits are exactly $1$-invariant, and become $2$-invariant just before the $2$-design time. It would be desirable to have more precise calculations of the $1$-invariance time in a chaotic system, either by analytical or numerical methods.  Further study of $k$-invariance for higher $k$ would also be interesting.

There are many natural questions and directions suggested by this work.  Foremost is understanding the range and scope of Hamiltonian systems with quenched disorder which experience $k$-invariance, and types of spectral decoupling more broadly.  This requires both more extensive numerical (and possibly analytic) understanding of spectral decoupling in systems with non-local interactions, as well as more detailed study of spectral decoupling in systems with geometrically local interactions.  Quantifying the class of systems which undergo some form of spectral decoupling would provide a sharper boundary between stronger and weaker forms of quantum many-body chaos.

Although we have done an initial investigation of spectral decoupling in disordered many-body systems with geometrically local interactions and found positive results, there is more to be understood.  The number of random variables in the ensemble may play a key role.  Also, there may be a more refined notion of spectral decoupling and $k$-invariance in particular, that is better suited to geometrically local systems. This could enable more precise contact with disordered systems in nature.

\subsection*{Acknowledgments}
We thank {\'A}lvaro Alhambra, Kristan Jensen, Richard Kueng, Thomas Mertens, Phil Saad, Douglas Stanford, Alex Turzillo, and Beni Yoshida for helpful discussions. JC is supported by the Fannie and John Hertz Foundation and the Stanford Graduate Fellowship program. NHJ would like to thank the Aspen Center for Physics, McGill University, and Stanford University for their hospitality during the completion of part of this work.  Research at Perimeter Institute is supported by the Government of Canada through the Department of Innovation, Science and Economic Development Canada and by the Province of Ontario through the Ministry of Research, Innovation and Science.

\appendix
\section{Random unitaries and operator averages}
\label{app:designs}

In this appendix we provide a few definitions for the quantum information theoretic notions used in the paper, as well as derive some identities for averages over Pauli operators.

\subsubsection*{$k$-fold channels and unitary designs}
The $k$-fold channel with respect to an ensemble of unitaries $\CE_U$ of an operator $\op$ was given in Eqn.~\eqref{eq:kfold} for a discrete ensemble. For a continuous ensemble, a subset of the unitary group equipped with a probability measure, the $k$-fold channel is
\begin{equation}
\Phi^{(k)}_{\CE_U} (\op) := \int_{\mathcal{E}_U} dU\, U^{\otimes k}\,\op\, U^\dagger{}^{\otimes k}\,.
\end{equation}
Again, we say that an ensemble $\CE_U$ forms a unitary $k$-design if the $k$-fold channel of $\CE_U$ and the Haar ensemble are equal $\Phi^{(k)}_{\CE_U}(\op) = \Phi^{(k)}_{\mathcal{E}_\text{Haar}} (\op)$ for all $\op$. 

For $p\geq 1$, the superoperator norm of a quantum channel is defined as
\begin{equation}
\|\Phi\|_{p\ra p} := \sup_{\op\neq 0} \frac{\| \Phi(\op)\|_p}{\|\op\|_p}\,,
\end{equation}
where $\|\op\|_p := \big( \Tr |\op|^p \big)^{1/p}$ is the Schatten $p$-norm of the operator $\op$. We can now define the notion of distinguishability of quantum channels with the diamond norm, defined as
\begin{equation}
\|\Phi\|_\diamond := \sup_d \| \Phi \otimes \mathcal{I}_d\|_{1\ra 1}\,,
\end{equation}
where $\mathcal{I}_d$ is the identity channel on an ancilla system of dimension $d$. 

The diamond norm allows us to specify when an ensemble is close to being Haar random. We say that an ensemble of unitaries $\CE_U$ forms an $\epsilon$-approximate $k$-design when the $k$-fold channels satisfy
\begin{equation}
\big\|\Phi^{(k)}_{\CE_U} -  \Phi^{(k)}_{\CE_\text{Haar}} \big\|_\diamond \leq \epsilon\,.
\label{eq:approxk2}
\end{equation}
A review of different operator norms and the various definitions of approximate designs and their relation to one another is given in \cite{LowThesis}. 

In the present paper, we focus on a weaker measure of approximate design called the frame potential, where the $k$-th frame potential for an ensemble $\CE_U$ defined by \cite{Scott08,Gross07}
\begin{equation}
\CF^{(k)}_{\CE_U} := \int_{\CE_U} dU dV\, \big| \Tr (U^\dagger V)\big|^{2k}\,,
\end{equation}
which is lower bounded for any ensemble as $\CF^{(k)}_{\CE_U} \geq \CF^{(k)}_{\CE_\text{Haar}} $, where the Haar value is $\CF^{(k)}_{\CE_\text{Haar}} = k!$ for $k\leq d$.

The difference in frame potentials $\CF^{(k)}_{\CE_U} - \CF^{(k)}_{\mathcal{E}_\text{Haar}}$ corresponds to the 2-norm distance to Haar randomness, specifically \cite{Scott08}
\begin{equation}
\big\| \Phi^{(k)}_{\CE_U} - \Phi^{(k)}_{\CE_\text{Haar}} \big\|_2^2 = \CF^{(k)}_{\CE_U} - \CF^{(k)}_{\CE_\text{Haar}}\,,
\label{eq:kmomop}
\end{equation}
where here we mean the operator 2-norm of the channels as operators acting on $\CH^{\otimes 2k}$, explicitly written as $\widehat\Phi^{(k)}_{\CE_U} $, the $k$-th moment operator of the ensemble:
\begin{equation}
\widehat\Phi^{(k)}_{\CE_U} = \int_{\CE_U} dU\, U^{\otimes k} \otimes U^\dagger{}^{\otimes k}\,.
\end{equation}
The frame potential is related to an approximate unitary $k$-design in Eqn.~\eqref{eq:approxk2}, where the difference in frame potentials bounds the difference in channels as \cite{HaarRQC}
\begin{equation}
\big\| \Phi^{(k)}_{\CE_U} - \Phi^{(k)}_{\CE_\text{Haar}} \big\|_\diamond \leq d^{2k} \big( \CF^{(k)}_{\CE_U} - \CF^{(k)}_{\CE_\text{Haar}} \big)\,.
\end{equation}

\subsubsection*{$k$-invariance}
As we discussed, the notion of an approximate design is suited to capture the ergodicity of ensembles of time-dependent evolutions, such as random circuits or stochastic Hamiltonians, but for evolution by disordered Hamiltonians in Eqn.~\eqref{eq:Htens} we should instead focus on $k$-invariance. We defined $k$-invariance as the distance between the ensemble of time evolutions $\CE_t$ and the unitarily invariant ensemble $\widetilde \CE_t$. The frame potential for $\CE_t$ is lower bounded by that for $\widetilde \CE_t$ as \cite{ChaosRMT}
\begin{equation}
\CF^{(k)}_{\CE_t} - \CF^{(k)}_{\widetilde \CE_t}\geq 0\,.
\end{equation}
Thus the difference in frame potentials quantifies a distance of the ensemble to unitary invariance. Similar to our discussion of approximate designs, we could also consider the distance between the $k$-fold channels over the two ensembles as a stronger notion of $k$-invariance
\begin{equation}
\big\| \Phi^{(k)}_{\CE_t} - \Phi^{(k)}_{\widetilde \CE_t} \big\|_\diamond \leq \epsilon\,,
\end{equation}
but the difference in frame potentials bounds the difference in $k$-fold channels of the ensembles $\CE_t$ and $\widetilde \CE_t$ as
\begin{equation}
\big\| \Phi^{(k)}_{\CE_t} - \Phi^{(k)}_{\widetilde \CE_t} \big\|_\diamond \leq d^{2k} \big( \CF^{(k)}_{\CE_t} - \CF^{(k)}_{\widetilde \CE_t} \big)\,.
\end{equation}
Next, we establish that the difference in frame potentials is related to the 2-norm of the channels, by viewing the channel as an operator on $\CH^{\otimes 2k}$, \ie the $k$-th moment operator of the ensemble in Eqn.~\eqref{eq:kmomop}. The difference in $k$-th moment operators of the two ensembles is
\begin{equation}
\widehat\Phi^{(k)}_{\CE_t} - \widehat\Phi^{(k)}_{\widetilde \CE_t} = \int_{\CE_t} dU \, U^{\otimes k} \otimes U^{\dagger \, \otimes k} - \int_{\widetilde\CE_t} dU \, U^{\otimes k} \otimes U^{\dagger \, \otimes k}\,.
\end{equation}
Computing the 2-norm, and using the invariance of the $\widetilde\CE_t$ ensemble, we find
\begin{equation}
\big\| \Phi^{(k)}_{\CE_t} - \Phi^{(k)}_{\widetilde \CE_t} \big\|_2^2 = \CF^{(k)}_{\CE_t} - \CF^{(k)}_{\widetilde \CE_t}\,.
\end{equation}

\subsubsection*{Haar integrals and random unitaries}
The general expressions for averaging monomials of Haar random unitaries are known. The $k$-th moment of $U(d)$ is given by \cite{Collins02,Collins04}
\begin{equation}
\int dU\, U_{i_1j_1}\ldots U_{i_k j_k} U^\dagger_{\ell_1 m_1}\ldots U^\dagger_{\ell_k m_k} = \sum_{\sigma,\tau\in S_k} \delta_\sigma (\vec \imath\,| \vec m) \delta_\tau (\vec \jmath\,| \vec\ell\, ) \Wg(\sigma^{-1}\tau,d)\,,
\label{eq:Haarint}
\end{equation}
where we sum over elements of the permutation group $S_k$ and have a contraction of indices indexed by a permutation $\sigma$ defined as $\delta_\sigma(\vec \imath\,|\vec\jmath\,) = \delta_{i_1, j_{\sigma(1)}} \cdots \delta_{i_k, j_{\sigma(k)}}$. The weight attributed to a term in the sum, i.e.~a given index contraction, is called the Weingarten function and is a function of permutations $\sigma \in S_k$, which can be computed from the characters of the symmetric group \cite{Collins04}.  The first two moments of the Haar ensemble are
\begin{align}
&\int_{U(d)} dU \, U_{i_1 j_1} U^\dagger_{\ell_1 m_1} = \frac{1}{d} \, \delta_{i_1, m_1} \delta_{j_1, \ell_1} \\ \nn
&\int_{U(d)} dU\, U_{i_1 j_1} U_{i_2 j_2} U_{\ell_1 m_1}^\dagger U^\dagger_{\ell_2 m_2} = \frac{1}{d^2 - 1}\left(\delta_{i_1, m_1} \delta_{i_2, m_2} \delta_{j_1, \ell_1} \delta_{j_2, \ell_2} + \delta_{i_1, m_2} \delta_{i_2, m_1} \delta_{j_1, \ell_2} \delta_{j_2, \ell_1}\right)  \nn
& \qquad \qquad \qquad \qquad \qquad \qquad \quad - \frac{1}{d(d^2 - 1)} \left(\delta_{i_1, m_1} \delta_{i_2, m_2} \delta_{j_1, \ell_2} \delta_{j_2, \ell_1} + \delta_{i_1, m_2} \delta_{i_2, m_1} \delta_{j_1, \ell_1} \delta_{j_2, \ell_2}\right)\,. \label{eq:U2ndmom}
\end{align}

A unitary $k$-design exactly reproduces moments of the Haar measure and captures the above averages. For instance, the Pauli group forms an exact 1-design, which is equivalently expressed as
\begin{equation}
\int_{U(d)} dU\, U\otimes U^\dagger = \frac{1}{d^2} \sum_{A\in P} A\otimes A^\dagger  = \frac{1}{d} \,\Pi_{\rm swap} \,,
\end{equation}
where we sum over Paulis $A$ and $\Pi_{\rm swap}$ is the swap permutation on the tensor factors.

Lastly, we will quickly review integration over other compact Lie groups. The general expression for moments of Haar random orthogonal matrices is \cite{Collins04,CollinsMat09}
\begin{equation}
\int_{O(d)} dO\, O_{i_1 j_1}\ldots O_{i_{2k} j_{2k}} = \sum_{\sigma, \tau \in M_{2k}} \Delta_\sigma (\vec \imath\,) \Delta_\tau (\vec \jmath\,) \Wg^O (\sigma^{-1} \tau, d)\,,
\end{equation}
where we must sum over a subset $M_{2k}$ of the permutation group $S_{2k}$ corresponding to pair partitions, and define an index contraction with respect to a permutation $\sigma\in S_{2k}$ as $\Delta_\sigma(\vec \imath) := \delta_{i_{\sigma(1)},i_{\sigma(2)}} \cdots \delta_{i_{\sigma(2k-1)},i_{\sigma(2k)}} $. The orthogonal Weingarten function $ \Wg^O (\sigma, d)$ also admits an expansion in terms of characters of the symmetric group \cite{CollinsMat09}. Here we explicitly write out the first two moment of the orthogonal group
\begin{align}
& \int_{O(d)} dO \, O_{i_1 j_2} O^T_{j_2 i_2} = \frac{1}{d} \, \delta_{i_1 ,i_2} \delta_{j_1, j_2} \\ \nn
&\int_{O(d)} dO\, O_{i_1 j_1}O_{i_2 j_2}O_{j_3 i_3}^T O^T_{j_4 i_4} \label{eq:O2ndmom}\\
&\qquad = \frac{d+1}{d(d-1)(d+2)} 
\big( \delta_{i_1,i_2} \delta_{i_3,i_4} \delta_{j_1,j_2} \delta_{j_3,j_4} + \delta _{i_1,i_3} \delta_{i_2,i_4} \delta_{j_1,j_3} \delta_{j_2,j_4} + \delta_{i_1,i_4} \delta_{i_2,i_3} \delta_{j_1,j_4} \delta_{j_2,j_3} \big)\nn
&\qquad \quad -\frac{1}{d(d-1)(d+2)} \big(\delta _{i_1,i_3} \delta _{i_2,i_4} \delta _{j_1,j_4} \delta_{j_2,j_3} + \delta _{i_1,i_2} \delta _{i_3,i_4} \delta _{j_1,j_4} \delta_{j_2,j_3} + \delta _{i_1,i_4} \delta _{i_2,i_3} \delta _{j_1,j_3} \delta_{j_2,j_4}\nn
&\hspace*{4cm} +\delta _{i_1,i_2} \delta _{i_3,i_4} \delta _{j_1,j_3} \delta_{j_2,j_4}+\delta _{i_1,i_4} \delta _{i_2,i_3} \delta _{j_1,j_2} \delta_{j_3,j_4}+\delta _{i_1,i_3} \delta _{i_2,i_4} \delta _{j_1,j_2} \delta_{j_3,j_4} \big)\,.\nonumber
\end{align}

The general expression for integration of monomials of Haar random symplectic matrices\footnote{We note that the unitary symplectic group is often defined with even dimension $2d$, as the intersection $Sp(2d,\C) \cap U(2d)$. Here, for convenience and ease in comparing expressions, we denote the symplectic group as $Sp(d)$ keeping in mind that $d$ must be taken to be even.} of dimension $d \times d$ is given by \cite{Collins04,CollinsStolz08}
\begin{equation}
\int_{Sp(d)} dS\, S_{i_1 j_1}\ldots S_{i_{2k} j_{2k}} = \sum_{\sigma, \tau \in M_{2k}} \Delta^J_\sigma (\vec \imath\,) \Delta^J_\tau (\vec \jmath\,) \Wg^{Sp} (\sigma^{-1} \tau, d)\,,
\end{equation}
where $\Delta^J_\sigma$ is the index contraction with respect to a pair partition, similar to that for the orthogonal group except with a symplectic $J$ inserted in the contraction, and $\Wg^{Sp}$ is the symplectic Weingarten function.  Let us consider the canonical symplectic form $J$, defined by
\begin{equation}
J := \begin{bmatrix}
\phantom{-}\textbf{0} & & \textbf{1} \, \\ - \textbf{1} & & \textbf{0} \,
\end{bmatrix}
\end{equation}
where $\textbf{0}$ is the $d/2 \times d/2$ matrix where every entry is zero, and $\textbf{1}$ is the $d/2 \times d/2$ identity matrix.  For $i,j=1,...,d$, we can equivalently write $J_{ij} = \delta_{i+d/2,j} - \delta_{i,j+d/2}$.  The first two moments of the symplectic ensemble are
\begin{align}
&\int_{Sp(d)} dS \, S_{i_1 j_2} S^D_{j_2 i_2} = \frac{1}{d} \, \delta_{i_1, i_2} \delta_{j_1, j_2} \\ \nn
&\int_{Sp(d)} dS\, S_{i_1 j_1} S_{i_2 j_2} S_{j_3 i_3}^D S^D_{j_4 i_4} \label{eq:Sp2ndmom}\\
&\qquad = \frac{d-1}{d(d+1)(d-2)} 
\big( J_{i_1,i_2} J_{i_3,i_4} J_{j_1,j_2} J_{j_3,j_4} + \delta_{i_1,i_3} \delta_{i_2,i_4} \delta_{j_1,j_3} \delta_{j_2,j_4} + \delta_{i_1,i_4} \delta_{i_2,i_3} \delta_{j_1,j_4} \delta_{j_2,j_3} \big)\nn
&\qquad \quad -\frac{1}{d(d+1)(d-2)} \big(\delta _{i_1,i_3} \delta _{i_2,i_4} \delta _{j_1,j_4} \delta_{j_2,j_3} - J _{i_1,i_2} J _{i_3,i_4} \delta _{j_1,j_4} \delta_{j_2,j_3} + \delta _{i_1,i_4} \delta _{i_2,i_3} \delta_{j_1,j_3} \delta_{j_2,j_4}\nn
&\hspace*{4cm} +J _{i_1,i_2} J _{i_3,i_4} \delta_{j_1,j_3} \delta_{j_2,j_4} - \delta_{i_1,i_4} \delta_{i_2,i_3} J _{j_1,j_2} J_{j_3,j_4} + \delta _{i_1,i_3} \delta_{i_2,i_4} J _{j_1,j_2} J_{j_3,j_4} \big)\,.\nonumber
\end{align}
Here, $S^D := J S^T J^{-1}$, so that $S^D S = S S^D = \textbf{1}$.

\subsubsection*{$k$-point spectral form factors}
For convenience, we explicitly define and write out the form factors that appear in many of the second moment calculations in the paper:
\begin{align}
\CR_2(t) &:= \big\langle \Tr(e^{-iHt})\Tr(e^{iHt})\big\rangle_{\mathcal{E}_H} = \int D\lambda \sum_{i,j} e^{i(\lambda_i-\lambda_j) t}\\
\CR_4(t) &:= \big\langle \Tr(e^{-iHt})^2\Tr(e^{iHt})^2\big\rangle_{\mathcal{E}_H} = \int D\lambda \sum_{i,j,k,\ell} e^{i(\lambda_i+\lambda_j -\lambda_k-\lambda_\ell) t}\\
\CR_{4,1}(t) &:= \big\langle \Tr(e^{iHt})^2\Tr(e^{-2iHt})\big\rangle_{\mathcal{E}_H} = \int D\lambda \sum_{i,j,\ell} e^{i(\lambda_i+\lambda_j -2\lambda_\ell) t}\,,\\
\CR_{4,2}(t) &:= \CR_2(2t) = \big\langle \Tr(e^{-2iHt})\Tr(e^{2iHt})\big\rangle_{\mathcal{E}_H} = \int D\lambda \sum_{j,k} e^{2i(\lambda_j -\lambda_k) t}\,.
\label{eq:specfuncs}
\end{align}
Note that $\CR_{4,1}(t)$ is generically complex unless all eigenvalues of $H$ come in pairs $\pm \lambda$.

\section{More on \ktitle-invariance and correlators}
\label{app:OTOCs}

\subsubsection*{$k$-invariant correlators}
As we mentioned in Section~\ref{sec:corrRMT}, the central property of certain random matrix ensembles which allows us to compute correlation functions from spectral quantities is the invariance of the measure. For $k$-invariant Hamiltonians, the same property holds and we can compute correlation functions using the approximate invariance of the measure on our set of Hamiltonians. For instance, the invariant 2-point function $\vev{A(t) B}$ with $A(t) = e^{iHt}Ae^{-iHt} $ and averaged over $H\in \CE_H$. For traceless operators we have
\begin{equation}
\vev{A(t) B}_{\widetilde \CE_t} = \frac{\CR_2(t) - 1}{d^2-1}\vev{AB} \where \CR_2(t) = \big\langle \Tr(e^{iHt}) \Tr(e^{-iHt}) \big\rangle_{\mathcal{E}_H}
\end{equation}
is the 2-point spectral form factor averaged over the ensemble. The above expression only requires the first moment and thus follows from 1-invariance of the ensemble. 

We can also consider generic out-of-time-ordered correlation functions $\vev{A(t)BC(t)D}$, where again we average over $H\in\CE_H$. For traceless operators $A$, $B$, $C$, and $D$ we find the full expression
{\small
\begin{align}
\label{eq:invOTOC}
&\hspace*{-12pt}\vev{A(t) B C(t) D}_{\widetilde \CE_t}=\\
&-\frac{\left( 5 d \CR_4 - (d^2+6) (\CR_{4,1}+\CR^*_{4,1}) + 5d \CR_{4,2}  - 4d (d^2-4) \CR_2 + d^3 (d^2-9)\right) }{d(d^2-1)(d^2-4)(d^2-9)} \vev{AC}\vev{BD} \nn
&+\frac{\left((d^4-8 d^2+6) \CR_{4,1} + (d^2+6) \CR^*_{4,1} - d(d^2-4)(\CR_4 - 4 \CR_2 + \CR_{4,2})\right)}{d(d^2-1)(d^2-4)(d^2-9)}\big(\vev{AD}\vev{BC}+\vev{AB}\vev{CD}\big)\nn
&+\frac{\left( (2d^2-3) \CR_4 -5d \CR_{4,1} - d(d^2-4) \CR^*_{4,1} + (2 d^2-3) \CR_{4,2} - (d^4-d^2-12) \CR_2 +  d^2(d^2-9)\right)}{d^2 (d^2-1)(d^2-4)(d^2-9)}\nn
&\qquad \times \big( \langle ABDC\rangle + \langle ACBD\rangle+\langle ACDB\rangle + \langle ADBC\rangle\big)\nn[4pt]
&+\frac{\left( (d^4-8 d^2+6) (\CR_4-4\CR_2) - d(d^2-4)(\CR_{4,1}+\CR^*_{4,1}) + (d^2+6) \CR_{4,2} + 2d^2(d^2-9)\right)}{d^2 (d^2-1)(d^2-4)(d^2-9)} \vev{ABCD}\nn
&+\frac{\left( (d^2+6) \CR_4 - d (d^2-4) (\CR_{4,1}+\CR^*_{4,1}) + (d^4-8 d^2+6) \CR_{4,2} - 4(d^2+6)\CR_2 - 2 d^2 (d^2-9)\right)}{d^2 (d^2-1)(d^2-4)(d^2-9)}\langle ADCB\rangle\,, \nonumber
\end{align}
}
in terms of the spectral functions defined above.

For OTOCs of the form $\vev{A(t)BA(t)B}$, where both $A$ and $B$ are non-identity Pauli operators and assuming $A\neq B$, we find the above expression reduces to
\begin{equation}
\vev{A(t) B A(t) B}_{\widetilde \CE_t} = \frac{d\CR_4 - 3\CR_{4,1} - 3\CR^*_{4,1} + d \CR_{4,2} - 4d \CR_2-d(d^2-9)}{d(d^2-1)(d^2-9)}\,.
\label{eq:invABOTOC}
\end{equation}
Noting that the spectral functions in Eqn.~\eqref{eq:specfuncs} are upper bounded by their initial time values, we can take the large $d$ limit of the invariant OTOC in Eqn.~\eqref{eq:invOTOC} and find that to leading order in $1/d$
\begin{align}
&\hspace*{-8pt}\vev{A(t) B C(t) D}_{\widetilde \CE_t}\\
&\quad = \frac{\CR_4}{d^4} \vev{ABCD} + \Big( \frac{\CR_{4,1}}{d^3} - \frac{\CR_4}{d^4}\Big) \vev{AB}\vev{CD} +  \Big( \frac{\CR^*_{4,1}}{d^3} - \frac{\CR_4}{d^4}\Big) \vev{AD}\vev{BC} + O\Big(\frac{1}{d^2}\Big)\,. \nonumber 
\end{align}
For the OTOCs in Eqn.~\eqref{eq:invABOTOC} with non-equal Paulis operators $A$ and $B$, this reduces to
\begin{equation}
\vev{A(t) B A(t) B}_{\widetilde \CE_t} \approx \frac{\CR_4(t)}{d^4}\,.
\end{equation}

\subsubsection*{Approximate 2-invariance from OTOCs}
We now review the calculation of the $k=2$ invariant frame potential done in \cite{ChaosRMT}, and present a rederivation in terms of invariant correlation functions. As usual, consider an ensemble $\CE_t$ of unitary time evolutions to time $t$ by an ensemble of disordered Hamiltonians. We can compute the second frame potential for the unitarily invariant ensemble $\widetilde \CE_t$ as
\begin{equation}
\CF^{(2)}_{\widetilde \CE_t} = \int_{\widetilde \CE_H} dHdH' \, \big| \Tr(e^{iHt} e^{-iH't})\big|^4\,.
\end{equation}
As the measure over $\widetilde \CE$ is defined to be unitarily invariant, we take $H \ra UHU^\dagger$ and average over random unitaries using the fourth moment. We then find
{\small
\begin{align}
&\CF_{\widetilde\CE_t}^{(2)} = \frac{1}{d^2 (d^2-1) (d^2-4) (d^2-9)}\bigg((d^4-8 d^2+6) \CR_4^2
+ 4 (d^6-9 d^4+4d^2+24) \CR_2^2+ 4 d^2(d^2-9)\CR_4\nn
&\qquad\quad - 8 d^2 (d^4-11 d^2+18) \CR_2 + (d^4-8 d^2+6) \CR_{4,2}^2 -4 d^2(d^2-9) \CR_{4,2}+ 2(d^4-8d^2+6) \CR_{4,1}\CR^*_{4,1}\nn
&\qquad\quad + (d^2+6)\big(\CR_{4,1}^2+\CR^{*\,2}_{4,1}\big)+ 8 d(d^2-4) \CR_2 \big(\CR_{4,1}+\CR^*_{4,1}\big) - 2d(d^2-4) \big(\CR_4+\CR_{4,2}\big) \big(\CR_{4,1}+\CR^*_{4,1}\big)\nn
&\qquad\quad  - 8(d^2+6) \CR_2 \CR_{4,2}  + 2 (d^2+6) \CR_4 \CR_{4,2} - 8 (d^4-8d^2+6) \CR_2 \CR_4 + 2 d^4 (d^4-12 d^2+27) \bigg)\,,
\label{eq:FP2inv}
\end{align}}
in terms of the spectral functions disorder-averaged over the ensemble $\CE_H$. This expression was derived in Appendix C of \cite{ChaosRMT}.

We want to explore approximate $2$-invariance from the perspective of the constituent 4-point functions. As described in Section~\ref{sec:kinvcorr}, 
we understand the onset of approximate 1-invariance as the decay of generic 2-point functions, and their closeness to the average 2-point function at late times. But as approximate 2-invariance is sufficient for several definitions of scrambling, we should attempt to precisely formulate its onset as the late time behavior of OTOCs.

Recall the for any ensemble of unitaries, we can write the frame potential as an average of OTOCs as
\begin{equation}
\CF_{\CE_t}^{(2)} = \frac{1}{d^2} \sum_{A,B,C,D\in P} \big| \vev{A(t) B C(t) D}_{\CE_t} \big|^2\,,
\end{equation}
where we sum over all Pauli operators, and $\vev{\,\cdot\,}_{\CE_t}$ is the ensemble averaged correlator. We want to evaluate this for the invariant ensemble $\widetilde \CE_t$. First we separately consider the terms in the sum with any of the Paulis being the identity. We find
\begin{equation*}
\CF_{\widetilde\CE_t}^{(2)} = \frac{1}{d^2} \left( 1 + 2(d^2-1) + 4\frac{(\CR_2-1)^2}{d^2-1} + 4(d^2-2)\frac{(\CR_2-1)^2}{d^2-1} + \hspace*{-6pt}\sum_{A,B,C,D\in P'} \hspace*{-6pt}\big| \vev{A(t) B C(t) D}_{\widetilde \CE_t} \big|^2\right),
\end{equation*}
where the first term comes from the OTOC with four consituent identity operators, there is no contribution from OTOCs with three identity operators, the second term comes from OTOCs for which $A=C=\textbf{1}$ or $B=D=\textbf{1}$, the third term comes from the other 4 possible contributions from two of the operators in the OTOC equaling $\textbf{1}$, and the fourth term comes from a single identity operator in the OTOC. The last term is summed over non-identity Paulis $P'$. Here we have used that
\begin{equation}
\sum_{A,B\in P'} \vev{AB} = (d^2-1) \and \sum_{A,B,C\in P'} |\vev{ABC}|^2 = (d^2-1)(d^2-2)\,.
\end{equation}
We explain how to compute these sums in the next part of the Appendix.  Already, we see something interesting. Recall that at the dip time the GUE forms a $k$-design \cite{ChaosRMT}, for which the frame potential equals 2. This can be thought of as the contribution from the OTOCs when $A,C=\textbf{1}$ and when $B,D=\textbf{1}$. At late times, the GUE frame potential is equal to 10. We can already see where part of this arises, since at late times the double and single identity contributions give $2$ and $4$. However, there are still $\sim d^2$ OTOCs that give a late time contribution in the final sum. But this already tells us that almost all of the $(d^2-1)^4$ terms in the remaining sum over non-identity Paulis, when summed over, contribute at subleading order in $1/d$. We knew some of the contribution at late times had to come from 2-point functions and some from 4-point functions as the late time invariant frame potential is
\begin{equation}
\CF^{(2)}_{\widetilde \CE_t} \approx \frac{\CR_4^2}{d^4} + \frac{4\CR_2^2}{d^2} + 2\,.
\end{equation}
It turns out that the only contribution from the sum over $(d^2-1)^4$ non-identity OTOCs is from $\vev{ABCD}$, such that
\begin{equation}
\frac{1}{d^2}\sum_{A,B,C,D\in P'} \big| \vev{A(t) B C(t) D}_{\widetilde \CE_t} \big|^2 \approx \frac{1}{d^2}\frac{\CR_4(t)^2}{d^8} \sum_{A,B,C,D \in P'} |\vev{ABCD}|^2 \approx \frac{\CR_4(t)^2}{d^4}
\end{equation}
which at late times when $\CR_4(t) \approx 2d^2$ contributes at leading order. All other terms in the sum over OTOCs are suppressed by factors of $1/d$. 

The above considerations give us the approximation
\begin{equation*}
\CF_{\widetilde\CE_t}^{(2)} \approx \frac{1}{d^2} \bigg(2 \!\sum_{A,C \in P'} \vev{AC}^2 + 4 \!\!\sum_{A,B,C \in P'} \!\!\frac{\CR_2(t)^2}{d^4}|\vev{ABC}|^2 + \!\!\sum_{A,B,C,D \in P'} \!\!\frac{\CR_4(t)^2}{d^8} |\vev{A B C D}|^2 \bigg)\,.
\end{equation*}
The contributing correlators at the dip time are just $\vev{AC}$ and $\vev{BD}$, i.e.~just the time-independent 2-point functions. At late times, the contributing OTOCs are these two 2-point functions, as well as the four 2-point functions of the form $\vev{A(t)BC(t)}$. Moreover, each of the non-identity 4-point OTOCs gives a contribution at late times from the 4-point form factor as $\CR_4(t)/d^4$. At late times this goes as $\sim 1/d^2$.  Squaring and accounting for the $1/d^2$ out front, we get a $1/d^6$ suppression. Over the $d^8$ correlation functions we sum over, $d^6$ of them contribute at leading order, thus giving the constant contribution to the 2-invariant frame potential. 

But we can also arrive at the answer explicitly. We derived the invariant 4-point function $\vev{A(t) B C(t) D}_{\widetilde \CE_t}$ in Eqn.~\eqref{eq:invOTOC}, expressed in terms of spectral form factors.
Thus it remains to square the quantity in Eqn.~\eqref{eq:invOTOC} and explicitly sum over Paulis $A,B,C,D\in P'$. This requires summing squared 4-point functions over non-identity Paulis, \eg $|\vev{ABCD}|^2$ as well as terms like $\vev{ABCD}\vev{DBCA}$. Making use of some explicit Pauli sums given in the next subsection, we can compute these and simplify the entire expression for the invariant frame potential in terms of the OTOCs. In doing so we recover the expression in for the full 2-invariant frame potential in Eqn.~\eqref{eq:FP2inv}.

\subsubsection*{Some explicit Pauli sums}

Consider a system with $N$ qubits, and let $d = 2^N$.  We would like to compute
\begin{equation}
K_j = \frac{1}{d^2} \sum_{\substack{A_1,...,A_j \in P'}} |\Tr (A_1 \ldots A_j)|^2
\end{equation}
for $j \geq 1$, and where we sum each $A_i$ over the set $P'$ of non-identity strings of Pauli operators.  Let
\begin{equation}
M_j := \frac{1}{4} \sum_{\alpha_1,\ldots,\alpha_j =0}^3 |\text{tr}(\sigma^{\alpha_1} \ldots \sigma^{\alpha_j})|^2\,,
\end{equation}
summing over all single-site Pauli operators (including the identity).  Then $M_j = 4^{j-1}$, and we have the recursion relation
\begin{equation}
K_j = M_j^N - \sum_{i = 0}^{j-1} \binom{j}{i} \, K_i = d^{2(j-1)} - \sum_{i = 0}^{j-1} \binom{j}{i} \, K_i\,,
\end{equation}
where we have used $M_j^N = d^{2(j-1)}$.  Thus we have, for instance,
\begin{align}
&K_1 = 0& &K_2 =  (d^2 - 1)& \nn
&K_3 = (d^2 - 1)(d^2 - 2)& &K_4 = (d^2 - 1)\big((d^2 - 1)(d^2 - 2) + 1\big)&
\end{align}
and so on.

\subsubsection*{Relation to $2k$-point correlation functions}

To get a better understanding of symmetric $k$-invariance, we explain its relation to correlation functions.  Consider the operator
\begin{equation}
S^{(k)} := \int_{\mathcal{E}_t} dU \, (U \otimes U^{\dagger})^{\otimes k} -  \int_{\widetilde{\mathcal{E}}_t^\text{sym}} dU \, (U \otimes U^{\dagger})^{\otimes k}  \,.
\end{equation}
From which it follows that
\begin{equation}
\text{tr}(S^{(k)\,\dagger} S^{(k)}) = \mathcal{F}_{\mathcal{E}_t}^{(k)} - \mathcal{F}_{\widetilde{\mathcal{E}}_t^\text{sym}}^{(k)} \geq 0\,.
\end{equation}
Indeed, the size of $\mathcal{F}_{\mathcal{E}_t}^{(k)} - \mathcal{F}_{\widetilde{\mathcal{E}}_t^\text{sym}}^{(k)}$ captures the degree to which $\mathcal{E}_t$ has become symmetric $k$-invariant (where here we mean symmetric with respect to $G$).

In fact, one can understand this difference, $\mathcal{F}_{\mathcal{E}_t}^{(k)} - \mathcal{F}_{\widetilde{\mathcal{E}}_t^\text{sym}}^{(k)}$, in terms of correlation functions.  Note that $S^{(k)}$ is in the space $\mathcal{B}(\mathcal{H}^{\otimes 2k})$.  Now for any operator $M$ on $\mathcal{B}(\mathcal{H}^{\otimes 2k})$, if $\{\mathcal{O}_i\}_{i=1}^{d^{4k}}$ is an orthogonal basis of operators on $\mathcal{B}(\mathcal{H}^{\otimes 2k})$ satisfying $\frac{1}{d^{2k}}\,\text{tr}(\mathcal{O}_i^\dagger \mathcal{O}_j) = \delta_{ij}$, then
\begin{equation}
M = \frac{1}{d^{2k}}\sum_{i=1}^{d^2} \text{tr}(\mathcal{O}_i^\dagger M) \, \mathcal{O}_i\,.
\end{equation} 
This is just a basis expansion.  Now we construct a useful basis $\{\mathcal{O}_i\}_{i=1}^{d^{4k}}$.  Consider the set of operators $\{A_i\}_{i=1}^{d^2}$ on $\mathcal{B}(\mathcal{H})$, satisfying $\frac{1}{d}\,\text{tr}(A_i^\dagger A_j) = \delta_{ij}$.  For instance, this set could be the generalized Pauli operators, or Pauli strings if the Hilbert space dimension is a power of 2.  Then we construct a basis of $\mathcal{B}(\mathcal{H}^{\otimes 2k})$, namely
\begin{equation}
\label{eq:basiswithperm1}
\left\{(A_{i_1} \otimes A_{i_2} \otimes \cdots \otimes A_{i_{2k}}) W_{\pi_{\text{cyc}}} \right\}_{i_1,i_2,...,i_{2k}=1}^{d^2}\,,
\end{equation}
where $W_{\pi_{\text{cyc}}}$ is the cyclic permutation operator on $\mathcal{H}^{\otimes 2k}$.  The operator $W_{\pi_{\text{cyc}}}$ is unitary, and so $W_{\pi_{\text{cyc}}} W_{\pi_{\text{cyc}}}^\dagger = W_{\pi_{\text{cyc}}}^\dagger W_{\pi_{\text{cyc}}} = \textbf{1}$.  Using the above basis, we can rewrite $S^{(k)}$ as
{\small \begin{align}
S^{(k)} &= \frac{1}{d^{2k}}\sum_{i_1,\ldots,i_k=1}^{d^2}\tr\bigg(\int_{\CE_t} dU \, W_{\pi_{\text{cyc}}}^\dagger\left(A_{i_1}^\dagger U \otimes A_{i_2}^\dagger U^{\dagger} \otimes \cdots \otimes A_{i_{2k-1}}^\dagger U \otimes A_{i_{2k}}^\dagger U^\dagger \right) \\
& \hspace*{3cm}- \int_{\widetilde\CE_t^\text{sym}} dU \, W_{\pi_{\text{cyc}}}^\dagger \left(A_{i_1}^\dagger U \otimes A_{i_2}^\dagger U^{\dagger} \otimes \cdots \otimes A_{i_{2k-1}}^\dagger U \otimes A_{i_{2k}}^\dagger U^\dagger \right) \bigg)  \nn
& \hspace*{8cm} \times (A_{i_1} \otimes A_{i_2} \otimes \cdots \otimes A_{i_{2k-1}} \otimes A_{i_{2k}}) W_{\pi_{\text{cyc}}} \nn
&= \frac{1}{d^{2k-1}}\!\sum_{i_1,\ldots,i_{2k}=1}^{d^2}\! \bigg(\langle A_{i_1}^\dagger(t) A_{i_2}^\dagger(0) \ldots A_{i_{2k-1}}^\dagger(t) A_{i_{2k}}^\dagger(0) \rangle_{\CE_t} -\langle A_{i_1}^\dagger(t) A_{i_2}^\dagger(0) \ldots A_{i_{2k-1}}^\dagger(t) A_{i_{2k}}^\dagger(0) \rangle_{\widetilde\CE_t^\text{sym}} \bigg) \nn
& \hspace*{8cm} \times (A_{i_1} \otimes A_{i_2} \otimes \cdots \otimes A_{i_{2k-1}} \otimes A_{i_{2k}}) W_{\pi_{\text{cyc}}} \nonumber
\end{align}}
where $\langle A_{i_1}^\dagger(t) A_{i_2}^\dagger(0) \ldots A_{i_{2k-1}}^\dagger(t) A_{2k}^\dagger(0) \rangle := \langle \frac{1}{d} \,\text{tr}( A_{i_1}^\dagger(t) A_{i_2}^\dagger(0) \ldots A_{i_{2k-1}}^\dagger(t) A_{2k}^\dagger(0)) \rangle$.  Then we have
\begin{align}
\label{correlationRewritek}
0 &\leq \text{tr}(S^{(k) \, \dagger} S^{(k)}) \\
&= \mathcal{F}_{\mathcal{E}_t}^{(k)} - \mathcal{F}_{\widetilde{\mathcal{E}}_t^\text{sym}}^{(k)} \nn
&= \frac{1}{d^{2k-2}}\sum_{i_1,\ldots,i_{2k}=1}^{d^2} \bigg|\langle A_{i_1}(t) A_{i_2}(0) \ldots A_{i_{2k-1}}(t) A_{i_{2k}}(t) \rangle_{\mathcal{E}_t} - \langle A_{i_1}(t) A_{i_2}(0) \ldots A_{i_{2k-1}}(t) A_{i_{2k}}(t) \rangle_{\widetilde{\mathcal{E}}_t^\text{sym}} \bigg|^2 \nonumber
\end{align}
and so $\mathcal{F}_{\mathcal{E}_t}^{(k)} - \mathcal{F}_{\widetilde{\mathcal{E}}_t^\text{sym}}^{(k)}$ equals the averaged sum of the squared difference of all $2k$-OTOCs.  Therefore, if $\mathcal{E}_t$ becomes approximately symmetric $k$-invariant, all $2k$-OTOCs with respect to $\mathcal{E}_t$ are close to all $2k$-OTOCs with respect to $\widetilde{\mathcal{E}}_t^\text{sym}$.

\section{More on numerics}
\label{app:num}

Here we give some additional details on how we numerically evaluate the frame potential for an ensemble of disordered Hamiltonians. As discussed in \cite{ChaosRMT}, for a finite ensemble $\CE_t$, the $k$-th frame potential
\begin{equation}
\CF^{(k)}_{\CE_t} = \sum_{i,j} p_i p_j \big| \Tr (U^\dagger_i U_j) \big|^{2k}
\end{equation}
receives `diagonal' contributions from the $i=j$ terms. For uniform weights, each diagonal term contributes $d^{2k}/|\CE_H|^2$, and the sum of all such terms gives $d^{2k}/|\CE_H|$. In the infinite ensemble size limit, these terms yield a vanishing contribution. But in performing numerics at finite $|\CE_H|$, they can obscure interesting late time physics. Thus, in our numerics we subtract these diagonal terms and compute a modified frame potential
\begin{equation}
\CF^{(k)}_{\CE_t} = \sum_{i\neq j} p_i p_j \big| \Tr (U^\dagger_i U_j) \big|^{2k}\,,
\end{equation}
with uniform weights $p_i = 1/(|\CE_H|(|\CE_H|-1))$. We numerically evaluate the invariant frame potentials $\CF^{(k)}_{\widetilde \CE_t}$ using the defined invariance of the measure $dH$ to relate the quantity to spectral functions, and simply numerically compute the spectral functions (such as $\CR_2(t)$) for the ensemble of disordered Hamiltonians at some time $t$. 

To characterize 1-invariance numerically, we generate an ensemble of Hamiltonians, compute the frame potential $\CF^{(1)}_{\CE_t}$ for the ensemble of time-evolutions at a time $t$, remove the `diagonal' contributions, and then compute the invariant frame potential $\CF^{(1)}_{\widetilde \CE_t}$ via the spectral form-factor $\CR_2(t)$.  Then we calculate the distance to $1$-invariance $\CF^{(1)}_{\CE_t} - \CF^{(1)}_{\widetilde{\CE}_t}$ as a function of time $t$.

\subsubsection*{Constructing $T$-invariant bases}

Here we describe a procedure for generating a $T$-invariant basis, which can be constructed solely from the antiunitary time-reversal operator $T$ \cite{HaakeChaos}. These are a class of bases in which time-reversal symmetric Hamiltonians commuting with $T$ can be written as real matrices.

Consider a $d\times d$ Hermitian matrix which commutes with the antiunitary operator $T$, with $T^2=1$. 
The procedure is: (i) Generate $d$ linearly independent vectors $\{\ket{\phi_i}\}$. It suffices to take a set of random vectors. (ii) Symmetrize the first vector by defining $\ket{\psi_1} = \ket{\phi_1}+T\ket{\phi_1}$, such that $T\ket{\psi_1} = \ket{\psi_1}$, and then normalize the vector to unit norm $\braket{\psi_1|\psi_1}=1$. (iii) Use Gram-Schmidt orthogonalization and symmetrize with respect to $T$ to construct a set of invariant orthonormal basis vectors. Take $\ket{\phi_2}$ and define a vector orthogonal to $\ket{\psi_1}$ as $\ket{\widetilde\phi_2} = \ket{\phi_2} - \braket{\psi_1|\phi_2}\ket{\psi_1}$. Then $T$-symmetrize and define $\ket{\psi_2} = \ket{\widetilde\phi_2} +T\ket{\widetilde\phi_2}$, and normalize the vector to unit norm.
Repeat this procedure by taking the random vector $\ket{\psi_j}$, orthogonalizing with respect to all $i<j$, symmetrizing so that $T\ket{\psi_j} = \ket{\psi_j}$, and rescaling to unit norm. The result is a $T$-invariant orthonormal basis $\{\ket{\psi_i}\}$. With respect to this basis, the Hermitian matrix will be real, $H=H^*$.

\vspace*{8pt}
\bibliographystyle{utphys}
\bibliography{kinv_chaos}

\providecommand{\href}[2]{#2}\begingroup\raggedright\begin{thebibliography}{10}

\bibitem{BHRMT16}
J.~S. Cotler, G.~Gur-Ari, M.~Hanada, J.~Polchinski, P.~Saad, S.~H. Shenker,
  D.~Stanford, A.~Streicher, and M.~Tezuka, ``{Black Holes and Random
  Matrices},'' \href{http://dx.doi.org/10.1007/JHEP05(2017)118}{{\em JHEP}
  {\bfseries 05} (2017) 118},
\href{http://arxiv.org/abs/1611.04650}{{\ttfamily arXiv:1611.04650 [hep-th]}}.

\bibitem{GarciaSYK16}
A.~M. Garc\'{i}a-Garc\'{i}a and J.~J.~M. Verbaarschot, ``{Spectral and
  thermodynamic properties of the Sachdev-Ye-Kitaev model},''
  \href{http://dx.doi.org/10.1103/PhysRevD.94.126010}{{\em Phys. Rev.}
  {\bfseries D94} (2016) 126010},
\href{http://arxiv.org/abs/1610.03816}{{\ttfamily arXiv:1610.03816 [hep-th]}}.

\bibitem{GarciaSpec17}
A.~M. Garc\'{i}a-Garc\'{i}a and J.~J.~M. Verbaarschot, ``{Analytical Spectral
  Density of the Sachdev-Ye-Kitaev Model at finite N},''
  \href{http://dx.doi.org/10.1103/PhysRevD.96.066012}{{\em Phys. Rev.}
  {\bfseries D96} (2017) 066012},
\href{http://arxiv.org/abs/1701.06593}{{\ttfamily arXiv:1701.06593 [hep-th]}}.

\bibitem{ChaosRMT}
J.~Cotler, N.~Hunter-Jones, J.~Liu, and B.~Yoshida, ``{Chaos, Complexity, and
  Random Matrices},'' \href{http://dx.doi.org/10.1007/JHEP11(2017)048}{{\em
  JHEP} {\bfseries 11} (2017) 048},
\href{http://arxiv.org/abs/1706.05400}{{\ttfamily arXiv:1706.05400 [hep-th]}}.

\bibitem{AltlandErgodicity18}
A.~Altland and D.~Bagrets, ``{Quantum ergodicity in the SYK model},''
  \href{http://dx.doi.org/10.1016/j.nuclphysb.2018.02.015}{{\em Nucl. Phys.}
  {\bfseries B930} (2018) 45},
\href{http://arxiv.org/abs/1712.05073}{{\ttfamily arXiv:1712.05073
  [cond-mat.str-el]}}.

\bibitem{gharibyan2018onset}
H.~Gharibyan, M.~Hanada, S.~H. Shenker, and M.~Tezuka, ``{Onset of Random
  Matrix Behavior in Scrambling Systems},''
  \href{http://dx.doi.org/10.1007/JHEP02(2019)197,
  10.1007/JHEP07(2018)124}{{\em JHEP} {\bfseries 07} (2018) 124},
  \href{http://arxiv.org/abs/1803.08050}{{\ttfamily arXiv:1803.08050
  [hep-th]}}.
[Erratum: JHEP02,197(2019)].

\bibitem{SYKramp18}
P.~Saad, S.~H. Shenker, and D.~Stanford, ``{A semiclassical ramp in SYK and in
  gravity},''
\href{http://arxiv.org/abs/1806.06840}{{\ttfamily arXiv:1806.06840 [hep-th]}}.

\bibitem{galitski2019otoc}
E.~B. {Rozenbaum}, S.~{Ganeshan}, and V.~{Galitski}, ``{Universal level
  statistics of the out-of-time-ordered operator},''
  \href{http://dx.doi.org/10.1103/PhysRevB.100.035112}{{\em Phys. Rev.}
  {\bfseries B100} (2019) 035112},
  \href{http://arxiv.org/abs/1801.10591}{{\ttfamily arXiv:1801.10591
  [cond-mat.dis-nn]}}.

\bibitem{ProsenQC}
P.~Kos, M.~Ljubotina, and T.~Prosen, ``{Many-Body Quantum Chaos: Analytic
  Connection to Random Matrix Theory},''
  \href{http://dx.doi.org/10.1103/PhysRevX.8.021062}{{\em Phys. Rev.}
  {\bfseries X8} (2018) 021062},
  \href{http://arxiv.org/abs/1712.02665}{{\ttfamily arXiv:1712.02665
  [nlin.CD]}}.

\bibitem{ProsenSFF}
B.~Bertini, P.~Kos, and T.~Prosen, ``{Exact Spectral Form Factor in a Minimal
  Model of Many-Body Quantum Chaos},''
  \href{http://dx.doi.org/10.1103/PhysRevLett.121.264101}{{\em Phys. Rev.
  Lett.} {\bfseries 121} (2018) 264101},
  \href{http://arxiv.org/abs/1805.00931}{{\ttfamily arXiv:1805.00931
  [nlin.CD]}}.

\bibitem{FRQC17}
A.~Chan, A.~De~Luca, and J.~T. Chalker, ``{Solution of a minimal model for
  many-body quantum chaos},''
  \href{http://dx.doi.org/10.1103/PhysRevX.8.041019}{{\em Phys. Rev.}
  {\bfseries X8} (2018) 041019},
\href{http://arxiv.org/abs/1712.06836}{{\ttfamily arXiv:1712.06836
  [cond-mat.stat-mech]}}.

\bibitem{FRQC18}
A.~Chan, A.~De~Luca, and J.~T. Chalker, ``{Spectral statistics in spatially
  extended chaotic quantum many-body systems},''
  \href{http://dx.doi.org/10.1103/PhysRevLett.121.060601}{{\em Phys. Rev.
  Lett.} {\bfseries 121} (2018) 060601},
\href{http://arxiv.org/abs/1803.03841}{{\ttfamily arXiv:1803.03841
  [cond-mat.stat-mech]}}.

\bibitem{BGSchaos}
O.~Bohigas, M.~J. Giannoni, and C.~Schmit, ``{Characterization of Chaotic
  Quantum Spectra and Universality of Level Fluctuation Laws},''
  \href{http://dx.doi.org/10.1103/PhysRevLett.52.1}{{\em Phys. Rev. Lett.}
  {\bfseries 52} (1984) 1}.

\bibitem{RMTphys}
T.~Guhr, A.~M{\" u}ller-Groeling, and H.~A. Weidenm{\" u}ller, ``{Random matrix
  theories in quantum physics: Common concepts},''
  \href{http://dx.doi.org/10.1016/S0370-1573(97)00088-4}{{\em Phys. Rept.}
  {\bfseries 299} (1998) 189},
\href{http://arxiv.org/abs/cond-mat/9707301}{{\ttfamily
  arXiv:cond-mat/9707301}}.

\bibitem{HaakeChaos}
F.~Haake, {\em Quantum Signatures of Chaos}.
\newblock Springer, 2010.

\bibitem{LarkinOv69}
A.~I. {Larkin} and Y.~N. {Ovchinnikov}, ``{Quasiclassical Method in the Theory
  of Superconductivity},'' {\em JETP} {\bfseries 28} (1969) 1200.

\bibitem{Kitaev15}
A.~Kitaev, ``{A simple model of quantum holography}.'' Talks given at the KITP,
  Apr.\ 7, 2015 and May 27, 2015.

\bibitem{MSSbound}
J.~Maldacena, S.~H. Shenker, and D.~Stanford, ``{A bound on chaos},''
  \href{http://dx.doi.org/10.1007/JHEP08(2016)106}{{\em JHEP} {\bfseries 08}
  (2016) 106},
\href{http://arxiv.org/abs/1503.01409}{{\ttfamily arXiv:1503.01409 [hep-th]}}.

\bibitem{SachdevYe}
S.~Sachdev and J.~Ye, ``{Gapless spin-fluid ground state in a random quantum
  Heisenberg magnet},''
  \href{http://dx.doi.org/10.1103/PhysRevLett.70.3339}{{\em Phys. Rev. Lett.}
  {\bfseries 70} (1993) 3339}.

\bibitem{MS_SYK}
J.~Maldacena and D.~Stanford, ``{Remarks on the Sachdev-Ye-Kitaev model},''
  \href{http://dx.doi.org/10.1103/PhysRevD.94.106002}{{\em Phys. Rev.}
  {\bfseries D94} (2016) 106002},
\href{http://arxiv.org/abs/1604.07818}{{\ttfamily arXiv:1604.07818 [hep-th]}}.

\bibitem{HaarRQC}
N.~Hunter-Jones, ``{Unitary designs from statistical mechanics in random
  quantum circuits},''
\href{http://arxiv.org/abs/1905.12053}{{\ttfamily arXiv:1905.12053
  [quant-ph]}}.

\bibitem{MehtaRMT}
M.~Mehta, {\em Random Matrices}.
\newblock Pure and Applied Mathematics. Elsevier Science, 2004.

\bibitem{cotler2019locality}
J.~S. Cotler, G.~R. Penington, and D.~H. Ranard, ``{Locality from the
  Spectrum},'' \href{http://dx.doi.org/10.1007/s00220-019-03376-w}{{\em Commun.
  Math. Phys.} {\bfseries 368} (2019) 1267},
  \href{http://arxiv.org/abs/1702.06142}{{\ttfamily arXiv:1702.06142
  [quant-ph]}}.

\bibitem{BrezinZee}
E.~Br{\'e}zin and A.~Zee, ``Universality of the correlations between
  eigenvalues of large random matrices,''
  \href{http://dx.doi.org/https://doi.org/10.1016/0550-3213(93)90121-5}{{\em
  Nucl. Phys. B} {\bfseries 402} (1993) 613}.

\bibitem{BrezinHikami1}
E.~Br\'ezin and S.~Hikami, ``Spectral form factor in a random matrix theory,''
  \href{http://dx.doi.org/10.1103/PhysRevE.55.4067}{{\em Phys. Rev.} {\bfseries
  E55} (1997) 4067}, \href{http://arxiv.org/abs/cond-mat/9608116}{{\ttfamily
  arXiv:cond-mat/9608116}}.

\bibitem{Dankert09}
C.~Dankert, R.~Cleve, J.~Emerson, and E.~Livine, ``Exact and approximate
  unitary 2-designs and their application to fidelity estimation,''
  \href{http://dx.doi.org/10.1103/PhysRevA.80.012304}{{\em Phys. Rev.}
  {\bfseries A80} (2009) 012304},
  \href{http://arxiv.org/abs/quant-ph/0606161}{{\ttfamily
  arXiv:quant-ph/0606161}}.

\bibitem{Zhu15}
H.~Zhu, ``Multiqubit clifford groups are unitary 3-designs,''
  \href{http://dx.doi.org/10.1103/PhysRevA.96.062336}{{\em Phys. Rev. A}
  {\bfseries 96} (2017) 062336},
  \href{http://arxiv.org/abs/1510.02619}{{\ttfamily arXiv:1510.02619
  [quant-ph]}}.

\bibitem{Kueng15}
R.~{Kueng} and D.~{Gross}, ``{Qubit stabilizer states are complex projective
  3-designs},'' \href{http://arxiv.org/abs/1510.02767}{{\ttfamily
  arXiv:1510.02767 [quant-ph]}}.

\bibitem{Webb15}
Z.~Webb, ``{The Clifford group forms a unitary 3-design},'' {\em Quantum Info.
  Comput.} {\bfseries 16} (2016) 1379,
  \href{http://arxiv.org/abs/1510.02769}{{\ttfamily arXiv:1510.02769
  [quant-ph]}}.

\bibitem{LowThesis}
R.~A. {Low}, ``{Pseudo-randomness and Learning in Quantum Computation},''
  \href{http://arxiv.org/abs/1006.5227}{{\ttfamily arXiv:1006.5227
  [quant-ph]}}. PhD Thesis, 2010.

\bibitem{Gross07}
D.~{Gross}, K.~{Audenaert}, and J.~{Eisert}, ``{Evenly distributed unitaries:
  On the structure of unitary designs},''
  \href{http://dx.doi.org/10.1063/1.2716992}{{\em J. Math. Phys.} {\bfseries
  48} (2007) 052104}, \href{http://arxiv.org/abs/quant-ph/0611002}{{\ttfamily
  arXiv:quant-ph/0611002}}.

\bibitem{Scott08}
A.~J. {Scott}, ``{Optimizing quantum process tomography with unitary
  2-designs},'' \href{http://dx.doi.org/10.1088/1751-8113/41/5/055308}{{\em J.
  Phys.} {\bfseries A41} (2008) 055308},
  \href{http://arxiv.org/abs/0711.1017}{{\ttfamily arXiv:0711.1017
  [quant-ph]}}.

\bibitem{ChaosDesign}
D.~A. Roberts and B.~Yoshida, ``{Chaos and complexity by design},''
  \href{http://dx.doi.org/10.1007/JHEP04(2017)121}{{\em JHEP} {\bfseries 04}
  (2017) 121},
\href{http://arxiv.org/abs/1610.04903}{{\ttfamily arXiv:1610.04903
  [quant-ph]}}.

\bibitem{ChaosSUSY}
N.~Hunter-Jones and J.~Liu, ``{Chaos and random matrices in supersymmetric
  SYK},'' \href{http://dx.doi.org/10.1007/JHEP05(2018)202}{{\em JHEP}
  {\bfseries 05} (2018) 202},
\href{http://arxiv.org/abs/1710.08184}{{\ttfamily arXiv:1710.08184 [hep-th]}}.

\bibitem{CVFP2019}
Q.~Zhuang, T.~Schuster, B.~Yoshida, and N.~Y. Yao, ``Scrambling and complexity
  in phase space,'' \href{http://dx.doi.org/10.1103/PhysRevA.99.062334}{{\em
  Phys. Rev.} {\bfseries A99} (2019) 062334},
  \href{http://arxiv.org/abs/1902.04076}{{\ttfamily arXiv:1902.04076
  [quant-ph]}}.

\bibitem{Chenu2019}
A.~Chenu, J.~Molina-Vilaplana, and A.~del Campo, ``{Work Statistics, Loschmidt
  Echo and Information Scrambling in Chaotic Quantum Systems},''
  \href{http://dx.doi.org/10.22331/q-2019-03-04-127}{{\em Quantum} {\bfseries
  3} (2019) 127}, \href{http://arxiv.org/abs/1804.09188}{{\ttfamily
  arXiv:1804.09188 [quant-ph]}}.

\bibitem{HL08}
A.~W. {Harrow} and R.~A. {Low}, ``{Random Quantum Circuits are Approximate
  2-designs},'' \href{http://dx.doi.org/10.1007/s00220-009-0873-6}{{\em Commun.
  Math. Phys.} {\bfseries 291} (2009) 257},
  \href{http://arxiv.org/abs/0802.1919}{{\ttfamily arXiv:0802.1919
  [quant-ph]}}.

\bibitem{BHH12}
F.~G.~S.~L. {Brand{\~a}o}, A.~W. {Harrow}, and M.~{Horodecki}, ``{Local Random
  Quantum Circuits are Approximate Polynomial-Designs},''
  \href{http://dx.doi.org/10.1007/s00220-016-2706-8}{{\em Commun. Math. Phys.}
  {\bfseries 346} (2016) 397}, \href{http://arxiv.org/abs/1208.0692}{{\ttfamily
  arXiv:1208.0692 [quant-ph]}}.

\bibitem{Nakata16}
Y.~Nakata, C.~Hirche, M.~Koashi, and A.~Winter, ``{Efficient Quantum
  Pseudorandomness with Nearly Time-Independent Hamiltonian Dynamics},''
  \href{http://dx.doi.org/10.1103/PhysRevX.7.021006}{{\em Phys. Rev.}
  {\bfseries X7} (2017) 021006},
\href{http://arxiv.org/abs/1609.07021}{{\ttfamily arXiv:1609.07021
  [quant-ph]}}.

\bibitem{Onorati17}
E.~Onorati, O.~Buerschaper, M.~Kliesch, W.~Brown, A.~H. Werner, and J.~Eisert,
  ``{Mixing properties of stochastic quantum Hamiltonians},''
  \href{http://dx.doi.org/10.1007/s00220-017-2950-6}{{\em Commun. Math. Phys.}
  {\bfseries 355} (2017) 905},
\href{http://arxiv.org/abs/1606.01914}{{\ttfamily arXiv:1606.01914
  [quant-ph]}}.

\bibitem{Cramer12}
M.~{Cramer}, ``{Thermalization under randomized local Hamiltonians},''
  \href{http://dx.doi.org/10.1088/1367-2630/14/5/053051}{{\em New J. Phys.}
  {\bfseries 14} (2012) 053051},
  \href{http://arxiv.org/abs/1112.5295}{{\ttfamily arXiv:1112.5295
  [quant-ph]}}.

\bibitem{BrandaoThermal12}
F.~G. S.~L. Brand{\~a}o, P.~\'{C}wikli\'{n}ski, M.~Horodecki, P.~Horodecki,
  J.~K. Korbicz, and M.~Mozrzymas, ``{Convergence to equilibrium under a random
  Hamiltonian},'' \href{http://dx.doi.org/10.1103/PhysRevE.86.031101}{{\em
  Phys. Rev.} {\bfseries E86} (2012) 031101},
  \href{http://arxiv.org/abs/1108.2985}{{\ttfamily arXiv:1108.2985
  [quant-ph]}}.

\bibitem{Vinayak12}
{Vinayak} and M.~{{\v{Z}}nidari{\v{c}}}, ``{Subsystem dynamics under random
  Hamiltonian evolution},''
  \href{http://dx.doi.org/10.1088/1751-8113/45/12/125204}{{\em J. Phys.}
  {\bfseries A45} (2012) 125204},
  \href{http://arxiv.org/abs/1107.6035}{{\ttfamily arXiv:1107.6035
  [quant-ph]}}.

\bibitem{Masanes13}
L.~Masanes, A.~J. Roncaglia, and A.~Ac\'{\i}n, ``Complexity of energy
  eigenstates as a mechanism for equilibration,''
  \href{http://dx.doi.org/10.1103/PhysRevE.87.032137}{{\em Phys. Rev.}
  {\bfseries E87} (2013) 032137},
  \href{http://arxiv.org/abs/1108.0374}{{\ttfamily arXiv:1108.0374
  [quant-ph]}}.

\bibitem{NakataThermal}
Y.~Nakata and T.~J. Osborne, ``Thermal states of random quantum many-body
  systems,'' \href{http://dx.doi.org/10.1103/PhysRevA.90.050304}{{\em Phys.
  Rev.} {\bfseries A90} (2014) 050304},
  \href{http://arxiv.org/abs/1407.6136}{{\ttfamily arXiv:1407.6136
  [quant-ph]}}.

\bibitem{ChaosChannels}
P.~Hosur, X.-L. Qi, D.~A. Roberts, and B.~Yoshida, ``{Chaos in quantum
  channels},'' \href{http://dx.doi.org/10.1007/JHEP02(2016)004}{{\em JHEP}
  {\bfseries 02} (2016) 004},
\href{http://arxiv.org/abs/1511.04021}{{\ttfamily arXiv:1511.04021 [hep-th]}}.

\bibitem{HaydenPreskill}
P.~Hayden and J.~Preskill, ``{Black holes as mirrors: Quantum information in
  random subsystems},''
  \href{http://dx.doi.org/10.1088/1126-6708/2007/09/120}{{\em JHEP} {\bfseries
  09} (2007) 120},
\href{http://arxiv.org/abs/0708.4025}{{\ttfamily arXiv:0708.4025 [hep-th]}}.

\bibitem{YoshidaKitaev17}
B.~Yoshida and A.~Kitaev, ``{Efficient decoding for the Hayden-Preskill
  protocol},''
\href{http://arxiv.org/abs/1710.03363}{{\ttfamily arXiv:1710.03363 [hep-th]}}.

\bibitem{LocalizedShocks}
D.~A. Roberts, D.~Stanford, and L.~Susskind, ``{Localized shocks},''
  \href{http://dx.doi.org/10.1007/JHEP03(2015)051}{{\em JHEP} {\bfseries 03}
  (2015) 051},
\href{http://arxiv.org/abs/1409.8180}{{\ttfamily arXiv:1409.8180 [hep-th]}}.

\bibitem{Aleiner16}
I.~L. Aleiner, L.~Faoro, and L.~B. Ioffe, ``{Microscopic model of quantum
  butterfly effect: out-of-time-order correlators and traveling combustion
  waves},'' \href{http://dx.doi.org/10.1016/j.aop.2016.09.006}{{\em Annals
  Phys.} {\bfseries 375} (2016) 378},
\href{http://arxiv.org/abs/1609.01251}{{\ttfamily arXiv:1609.01251
  [cond-mat.stat-mech]}}.

\bibitem{NahumRQC17}
A.~Nahum, S.~Vijay, and J.~Haah, ``{Operator Spreading in Random Unitary
  Circuits},'' \href{http://dx.doi.org/10.1103/PhysRevX.8.021014}{{\em Phys.
  Rev.} {\bfseries X8} (2018) 021014},
\href{http://arxiv.org/abs/1705.08975}{{\ttfamily arXiv:1705.08975
  [cond-mat.str-el]}}.

\bibitem{VRPS17}
C.~von Keyserlingk, T.~Rakovszky, F.~Pollmann, and S.~Sondhi, ``{Operator
  hydrodynamics, OTOCs, and entanglement growth in systems without conservation
  laws},'' \href{http://dx.doi.org/10.1103/PhysRevX.8.021013}{{\em Phys. Rev.}
  {\bfseries X8} (2018) 021013},
\href{http://arxiv.org/abs/1705.08910}{{\ttfamily arXiv:1705.08910
  [cond-mat.str-el]}}.

\bibitem{SYKopgrowth}
D.~A. Roberts, D.~Stanford, and A.~Streicher, ``{Operator growth in the SYK
  model},'' \href{http://dx.doi.org/10.1007/JHEP06(2018)122}{{\em JHEP}
  {\bfseries 06} (2018) 122},
\href{http://arxiv.org/abs/1802.02633}{{\ttfamily arXiv:1802.02633 [hep-th]}}.

\bibitem{SYKopbeta}
X.-L. Qi and A.~Streicher, ``{Quantum Epidemiology: Operator Growth, Thermal
  Effects, and SYK},'' \href{http://dx.doi.org/10.1007/JHEP08(2019)012}{{\em
  JHEP} {\bfseries 08} (2019) 012},
  \href{http://arxiv.org/abs/1810.11958}{{\ttfamily arXiv:1810.11958
  [hep-th]}}.

\bibitem{RQCsym}
N.~Hunter-Jones, ``{Operator growth in random quantum circuits with
  symmetry},''
\href{http://arxiv.org/abs/1812.08219}{{\ttfamily arXiv:1812.08219
  [quant-ph]}}.

\bibitem{BF12}
W.~Brown and O.~Fawzi, ``{Scrambling speed of random quantum circuits},''
\href{http://arxiv.org/abs/1210.6644}{{\ttfamily arXiv:1210.6644 [quant-ph]}}.

\bibitem{BF13}
W.~{Brown} and O.~{Fawzi}, ``{Decoupling with random quantum circuits},''
  \href{http://dx.doi.org/10.1007/s00220-015-2470-1}{{\em Comm. Math. Phys.}
  {\bfseries 340} (2015) 867}, \href{http://arxiv.org/abs/1307.0632}{{\ttfamily
  arXiv:1307.0632 [quant-ph]}}.

\bibitem{HM18}
A.~{Harrow} and S.~{Mehraban}, ``{Approximate unitary $t$-designs by short
  random quantum circuits using nearest-neighbor and long-range gates},''
  \href{http://arxiv.org/abs/1809.06957}{{\ttfamily arXiv:1809.06957
  [quant-ph]}}.

\bibitem{NahumRQC16}
A.~Nahum, J.~Ruhman, S.~Vijay, and J.~Haah, ``{Quantum Entanglement Growth
  Under Random Unitary Dynamics},''
  \href{http://dx.doi.org/10.1103/PhysRevX.7.031016}{{\em Phys. Rev.}
  {\bfseries X7} (2017) 031016},
\href{http://arxiv.org/abs/1608.06950}{{\ttfamily arXiv:1608.06950
  [cond-mat.stat-mech]}}.

\bibitem{RQCstatmech}
T.~Zhou and A.~Nahum, ``{Emergent statistical mechanics of entanglement in
  random unitary circuits},''
  \href{http://dx.doi.org/10.1103/PhysRevB.99.174205}{{\em Phys. Rev.}
  {\bfseries B99} (2019) 174205},
\href{http://arxiv.org/abs/1804.09737}{{\ttfamily arXiv:1804.09737
  [cond-mat.stat-mech]}}.

\bibitem{SrednickiETH}
M.~Srednicki, ``Chaos and quantum thermalization,''
  \href{http://dx.doi.org/10.1103/PhysRevE.50.888}{{\em Phys. Rev.} {\bfseries
  E50} (1994) 888}, \href{http://arxiv.org/abs/cond-mat/9403051}{{\ttfamily
  arXiv:cond-mat/9403051}}.

\bibitem{DeutschETH}
J.~M. Deutsch, ``Quantum statistical mechanics in a closed system,''
  \href{http://dx.doi.org/10.1103/PhysRevA.43.2046}{{\em Phys. Rev.} {\bfseries
  A43} (1991) 2046}.

\bibitem{ChaosETH}
L.~D'Alessio, Y.~Kafri, A.~Polkovnikov, and M.~Rigol, ``From quantum chaos and
  eigenstate thermalization to statistical mechanics and thermodynamics,''
  \href{http://dx.doi.org/10.1080/00018732.2016.1198134}{{\em Adv. Phys.}
  {\bfseries 65} (2016) 239}, \href{http://arxiv.org/abs/1509.06411}{{\ttfamily
  arXiv:1509.06411 [cond-mat.stat-mech]}}.

\bibitem{Srednicki98}
M.~{Srednicki}, ``{The approach to thermal equilibrium in quantized chaotic
  systems},'' \href{http://dx.doi.org/10.1088/0305-4470/32/7/007}{{\em J. Phys.
  A: Math. Gen.} {\bfseries 32} (1999) 1163},
  \href{http://arxiv.org/abs/cond-mat/9809360}{{\ttfamily
  arXiv:cond-mat/9809360}}.

\bibitem{Dymarsky18}
A.~Dymarsky, ``{Bound on Eigenstate Thermalization from Transport},''
\href{http://arxiv.org/abs/1804.08626}{{\ttfamily arXiv:1804.08626
  [cond-mat.stat-mech]}}.

\bibitem{dymarsky2019new}
A.~Dymarsky and H.~Liu, ``{New characteristic of quantum many-body chaotic
  systems},'' \href{http://dx.doi.org/10.1103/PhysRevE.99.010102}{{\em Phys.
  Rev.} {\bfseries E99} (2019) 010102},
  \href{http://arxiv.org/abs/1702.07722}{{\ttfamily arXiv:1702.07722
  [cond-mat.stat-mech]}}.

\bibitem{RigolOffDiag}
R.~Mondaini and M.~Rigol, ``{Eigenstate thermalization in the two-dimensional
  transverse field Ising model. II. Off-diagonal matrix elements of
  observables},'' \href{http://dx.doi.org/10.1103/PhysRevE.96.012157}{{\em
  Phys. Rev.} {\bfseries E96} (2017) 012157},
  \href{http://arxiv.org/abs/1705.08058}{{\ttfamily arXiv:1705.08058
  [cond-mat.stat-mech]}}.

\bibitem{DysonSym}
F.~J. {Dyson}, ``{The Threefold Way. Algebraic Structure of Symmetry Groups and
  Ensembles in Quantum Mechanics},''
  \href{http://dx.doi.org/10.1063/1.1703863}{{\em J. Math. Phys.} {\bfseries 3}
  (1962) 1199}.

\bibitem{Zirnbauer10}
M.~R. {Zirnbauer}, ``{Symmetry Classes},''
  \href{http://arxiv.org/abs/1001.0722}{{\ttfamily arXiv:1001.0722 [math-ph]}}.

\bibitem{AltlandZirnbauer}
A.~Altland and M.~R. Zirnbauer, ``Nonstandard symmetry classes in mesoscopic
  normal- superconducting hybrid structures,''
  \href{http://dx.doi.org/10.1103/PhysRevB.55.1142}{{\em Phys. Rev.} {\bfseries
  B55} (1997) 1142}, \href{http://arxiv.org/abs/cond-mat/9602137}{{\ttfamily
  arXiv:cond-mat/9602137}}.

\bibitem{NHJthesis}
N.~Hunter-Jones, \href{http://dx.doi.org/10.7907/BHZ5-HV76}{{\em {Chaos and
  Randomness in Strongly-Interacting Quantum Systems}}}.
\newblock PhD thesis, California Institute of Technology, 2018.

\bibitem{You16}
Y.-Z. {You}, A.~W.~W. {Ludwig}, and C.~{Xu}, ``{Sachdev-Ye-Kitaev model and
  thermalization on the boundary of many-body localized fermionic
  symmetry-protected topological states},''
  \href{http://dx.doi.org/10.1103/PhysRevB.95.115150}{{\em Phys. Rev.}
  {\bfseries B95} (2017) 115150},
  \href{http://arxiv.org/abs/1602.06964}{{\ttfamily arXiv:1602.06964
  [cond-mat.str-el]}}.

\bibitem{KanazawaRMTSYK17}
T.~Kanazawa and T.~Wettig, ``{Complete random matrix classification of SYK
  models with $\mathcal{N}=0$, $1$ and $2$ supersymmetry},''
  \href{http://dx.doi.org/10.1007/JHEP09(2017)050}{{\em JHEP} {\bfseries 09}
  (2017) 050},
\href{http://arxiv.org/abs/1706.03044}{{\ttfamily arXiv:1706.03044 [hep-th]}}.

\bibitem{prange1997spectral}
R.~E. {Prange}, ``{The Spectral Form Factor Is Not Self-Averaging},''
  \href{http://dx.doi.org/10.1103/PhysRevLett.78.2280}{{\em Phys. Rev. Lett.}
  {\bfseries 78} (1997) 2280},
  \href{http://arxiv.org/abs/chao-dyn/9606010}{{\ttfamily
  arXiv:chao-dyn/9606010}}.

\bibitem{jackiw1985lower}
R.~Jackiw, ``{Lower Dimensional Gravity},''
\href{http://dx.doi.org/10.1016/0550-3213(85)90448-1}{{\em Nucl. Phys.}
  {\bfseries B252} (1985) 343}.

\bibitem{teitelboim1983gravitation}
C.~Teitelboim, ``{Gravitation and Hamiltonian Structure in Two Space-Time
  Dimensions},''
\href{http://dx.doi.org/10.1016/0370-2693(83)90012-6}{{\em Phys. Lett.}
  {\bfseries B126} (1983) 41}.

\bibitem{jackiw1992gauge}
R.~Jackiw, ``{Gauge theories for gravity on a line},''
  \href{http://dx.doi.org/10.1007/BF01017075}{{\em Theor. Math. Phys.}
  {\bfseries 92} (1992) 979},
\href{http://arxiv.org/abs/hep-th/9206093}{{\ttfamily arXiv:hep-th/9206093}}.

\bibitem{SaadJT19}
P.~Saad, S.~H. Shenker, and D.~Stanford, ``{JT gravity as a matrix integral},''
\href{http://arxiv.org/abs/1903.11115}{{\ttfamily arXiv:1903.11115 [hep-th]}}.

\bibitem{deSitterQG19}
J.~Cotler, K.~Jensen, and A.~Maloney, ``{Low-dimensional de Sitter quantum
  gravity},'' \href{http://dx.doi.org/10.1007/JHEP06(2020)048}{{\em JHEP}
  {\bfseries 06} (2020) 048}, \href{http://arxiv.org/abs/1905.03780}{{\ttfamily
  arXiv:1905.03780 [hep-th]}}.

\bibitem{maldacena2019two}
J.~Maldacena, G.~J. Turiaci, and Z.~Yang, ``{Two dimensional Nearly de Sitter
  gravity},''
\href{http://arxiv.org/abs/1904.01911}{{\ttfamily arXiv:1904.01911 [hep-th]}}.

\bibitem{saad2019latetime}
P.~Saad, ``{Late Time Correlation Functions, Baby Universes, and ETH in JT
  Gravity},''
\href{http://arxiv.org/abs/1910.10311}{{\ttfamily arXiv:1910.10311 [hep-th]}}.

\bibitem{MertensJTclocks}
A.~Blommaert, T.~G. Mertens, and H.~Verschelde, ``{Clocks and Rods in
  Jackiw-Teitelboim Quantum Gravity},''
  \href{http://dx.doi.org/10.1007/JHEP09(2019)060}{{\em JHEP} {\bfseries 09}
  (2019) 060},
\href{http://arxiv.org/abs/1902.11194}{{\ttfamily arXiv:1902.11194 [hep-th]}}.

\bibitem{mertens2017solving}
T.~G. Mertens, G.~J. Turiaci, and H.~L. Verlinde, ``{Solving the Schwarzian via
  the Conformal Bootstrap},''
  \href{http://dx.doi.org/10.1007/JHEP08(2017)136}{{\em JHEP} {\bfseries 08}
  (2017) 136},
\href{http://arxiv.org/abs/1705.08408}{{\ttfamily arXiv:1705.08408 [hep-th]}}.

\bibitem{stanford2019JTRMT}
D.~Stanford and E.~Witten, ``{JT Gravity and the Ensembles of Random Matrix
  Theory},''
\href{http://arxiv.org/abs/1907.03363}{{\ttfamily arXiv:1907.03363 [hep-th]}}.

\bibitem{Collins02}
B.~Collins, ``{Moments and cumulants of polynomial random variables on unitary
  groups, the Itzykson-Zuber integral, and free probability},''
  \href{http://dx.doi.org/10.1155/S107379280320917X}{{\em Int. Math. Res. Not.}
  {\bfseries 2003} (2003) 953},
  \href{http://arxiv.org/abs/math-ph/0205010}{{\ttfamily
  arXiv:math-ph/0205010}}.

\bibitem{Collins04}
B.~{Collins} and P.~{{\'S}niady}, ``{Integration with Respect to the Haar
  Measure on Unitary, Orthogonal and Symplectic Group},''
  \href{http://dx.doi.org/10.1007/s00220-006-1554-3}{{\em Commun. Math. Phys.}
  {\bfseries 264} (2006) 773},
  \href{http://arxiv.org/abs/math-ph/0402073}{{\ttfamily
  arXiv:math-ph/0402073}}.

\bibitem{CollinsMat09}
B.~Collins and S.~Matsumoto, ``{On some properties of orthogonal Weingarten
  functions},'' \href{http://dx.doi.org/10.1063/1.3251304}{{\em J. Math. Phys.}
  {\bfseries 50} (2009) 113516},
  \href{http://arxiv.org/abs/0903.5143}{{\ttfamily arXiv:0903.5143 [math-ph]}}.

\bibitem{CollinsStolz08}
B.~Collins and M.~Stolz, ``Borel theorems for random matrices from the
  classical compact symmetric spaces,''
  \href{http://dx.doi.org/10.1214/07-AOP341}{{\em Ann. Probab.} {\bfseries 36}
  (2008) 876}, \href{http://arxiv.org/abs/math/0611708}{{\ttfamily
  arXiv:math/0611708 [math.PR]}}.

\end{thebibliography}\endgroup


\providecommand{\href}[2]{#2}\begingroup\raggedright\endgroup

\end{document}